\documentclass[twocolumn,times]{aastex7}

\usepackage{natbib}
\usepackage{epsfig,amsmath,amssymb}
\usepackage{graphicx}
\usepackage{xcolor}
\usepackage{bm}
\usepackage{caption}

\def\CII{[\ion{C}{2}]}
\def\kms{km\,s$^{-1}$}
\defcitealias{Neeleman2021}{N21}

\begin{document}

\title{\Large  XQR30: ALMA Observations Reveal the Diverse Structure of Cold Gas within $\bm{z \gtrsim 6}$ Quasar Host Galaxies.}

\correspondingauthor{Marcel Neeleman}
\email{mneeleman@nrao.edu}

\author[0000-0002-9838-8191]{Marcel Neeleman}
\affiliation{National Radio Astronomy Observatory, 520 Edgemont Road, Charlottesville, VA 22903, USA}
\email{mneeleman@nrao.edu}

\author[0000-0001-8582-7012]{Sarah E. I. Bosman}
\affiliation{Institut f\"{u}r Theoretische Physik, Universit\"{a}t Heidelberg, Philosophenweg 12, D-69120 Heidelberg, Germany}
\affiliation{Max-Planck-Institut f\"{u}r Astronomie, K\"{o}nigstuhl 17, D-69117 Heidelberg, Germany}
\email{walter@mpia.de}

\author[0000-0003-3693-3091]{Valentina D'Odorico}
\affiliation{INAF -- Osservatorio Astronomico di Trieste, Via G. B. Tiepolo 11, I-34131 Trieste, Italy}
\affiliation{IFPU -- Institute for Fundamental Physics of the Universe, via Beirut 2, I-34151 Trieste, Italy}
\email{valentina.dodorico@inaf.it}

\author[0000-0001-9024-8322]{Bram P. Venemans}
\affiliation{Leiden Observatory, Leiden University, PO Box 9513, 2300 RA Leiden, The Netherlands}
\email{venemans@strw.leidenuniv.nl}

\author[0000-0003-4793-7880]{Fabian Walter}
\affiliation{Max-Planck-Institut f\"{u}r Astronomie, K\"{o}nigstuhl 17, D-69117 Heidelberg, Germany}
\email{walter@mpia.de}

\author[0000-0002-2931-7824]{Eduardo Ba\~{n}ados}
\affiliation{Max-Planck-Institut f\"{u}r Astronomie, K\"{o}nigstuhl 17, D-69117 Heidelberg, Germany}
\email{banados@mpia.de}

\author[0000-0003-2344-263X]{George D. Becker}
\affiliation{Department of Physics \& Astronomy, University of California, Riverside, CA 92521, USA}
\email{georgeb@ucr.edu}

\author[0000-0002-4314-021X]{Manuela Bischetti}
\affiliation{Dipartimento di Fisica, Sezione di Astronomia, Universit\'{a} di Trieste, Via G. B. Tiepolo 11, I-34131 Trieste, Italy}
\affiliation{INAF -- Osservatorio Astronomico di Trieste, Via G. B. Tiepolo 11, I-34131 Trieste, Italy}
\email{manuela.bischetti@inaf.it}

\author[0000-0002-6719-380X]{Stefano Carniani}
\affiliation{Scuola Normale Superiore, Piazza dei Cavalieri, I-56126 Pisa, Italy}
\email{stefano.carniani@sns.it}

\author[0000-0002-2115-5234]{Stefano Cristiani}
\affiliation{INAF -- Osservatorio Astronomico di Trieste, Via G. B. Tiepolo 11, I-34131 Trieste, Italy}
\affiliation{IFPU -- Institute for Fundamental Physics of the Universe, via Beirut 2, I-34151 Trieste, Italy}
\affiliation{INFN -- National Institute for Fundamental Physics, via Valerio 2, I-34127 Trieste, Italy}
\email{stefano.cristiani@inaf.it}

\author[0000-0002-2662-8803]{Roberto Decarli}
\affiliation{INAF -- Osservatorio di Astrofisica e Scienza dello Spazio di Bologna, via Gobetti 93/3, I-40129 Bologna, Italy}
\email{roberto.decarli@inaf.it}

\author[0000-0003-2895-6218]{Anna-Christina Eilers}
\affiliation{Department of Physics, Massachusetts Institute of Technology, Cambridge, MA 02139, USA}
\affiliation{MIT Kavli Institute for Astrophysics and Space Research, 77 Massachusetts Avenue, Cambridge, MA 02139, USA}
\email{eilers@mit.edu}

\author[0000-0002-4227-6035]{Chiara Feruglio}
\affiliation{INAF -- Osservatorio Astronomico di Trieste, Via G. B. Tiepolo 11, I-34131 Trieste, Italy}
\affiliation{IFPU -- Institute for Fundamental Physics of the Universe, via Beirut 2, I-34151 Trieste, Italy}
\email{chiara.feruglio@inaf.it}

\author[0000-0001-5829-4716]{Girish Kulkarni}
\affiliation{Tata Institute of Fundamental Research, Homi Bhabha Road, Mumbai 400005, India}
\email{kulkarni@theory.tifr.res.in}

\author[0000-0001-9372-4611]{Samuel Lai}
\affiliation{Space \& Astronomy, Commonwealth Scientific and Industrial Research Organisation (CSIRO), P. O. Box 1130, Bentley, WA 6102, Australia}
\email{Samuel.Lai@anu.edu.au}

\author[0000-0002-5941-5214]{Chiara Mazzucchelli}
\affiliation{Instituto de Estudios Astrof\'{\i}sicos, Facultad de Ingenier\'{\i}a y Ciencias, Universidad Diego Portales, Avenida Ej\'{e}rcito Libertador 441, Santiago, Chile}
\email{chiara.mazzucchelli@mail.udp.cl}

\author[0000-0001-5492-4522]{Romain A. Meyer}
\affiliation{Department of Astronomy, University of Geneva, Chemin Pegasi 51, 1290 Versoix, Switzerland}
\email{Romain.Meyer@unige.ch}

\author[0000-0002-4544-8242]{Jan-Torge Schindler}
\affiliation{Hamburger Sternwarte, Universität Hamburg, Gojenbergsweg 112, D-21029 Hamburg, Germany}
\email{jan-torge.schindler@uni-hamburg.de}

\begin{abstract}
We present  0\farcs25 ($\approx$1.5 kpc) ALMA observations of the \CII\ and dust continuum emission from the host galaxies of 24 luminous $z \gtrsim 6$ quasars. Together with archival data from 17 quasars, this constitutes a nearly complete sample of all quasars with $J$-band magnitudes of $J_{\rm AB} < 20.0$ and redshifts between 5.8 and 6.6, which are the limits defined in the VLT/XSHOOTER legacy survey, XQR-30. For the full sample of 41 quasars, we find that 39 (95 \%) of the quasar host galaxies are detected in \CII\  emission with \CII\ luminosities ranging between (0.3 and 9.2) $\times 10^9 L_\odot$, and 39 (95 \%)  of the quasar host galaxies are detected in continuum emission with continuum flux densities ranging between 0.22 and 16.3 mJy. We detect eight \CII-emitting companion galaxies, which is consistent with the expected number of galaxies based on clustering results. We find that $z \gtrsim 6$ quasar host galaxies have a range of morphologies, and by defining a new parameter, the kinematic position angle alignment parameter, $\Delta$PA$_{\rm kin-CII}$, we show that most quasar host galaxies deviate from a standard disk model. This suggests that \CII\ emission probes structures beyond simple disk rotation, such as bars, tidally disturbed gas, etc. We show that the dust emission is more compact than the \CII\ emission independent of the morphology of the galaxy. Finally, we provide rough dynamical mass estimates within twice the \CII\ half-light radius for the majority of the galaxies, and show that the average dynamical mass within the inner few kpc is consistent with estimates of the average mass in stars, gas and the SMBH. 
\end{abstract}


\section{Introduction}
\label{sec:intro}

Massive galaxies are expected to form in strong galaxy overdensities, where conditions are more conducive to the rapid buildup of galaxies through mergers and gas accretion \citep{Efstathiou1988,Volonteri2006,Costa2014}. The most massive galaxies are believed to form and grow most rapidly \citep{Cattaneo2006}, giving rise to massive quiescent galaxies at $z  > 4$ \citep[e.g.,][]{DeGraaff2025}. This makes it crucial to observe them at the earliest times $z \gtrsim 6$, in order to understand the environment and physical conditions within these galaxies that facilitate this rapid growth and evolution. However, the number density of massive, high-redshift galaxies is small, and only very large area surveys have the necessary volume to detect these galaxies \citep[e.g.,][]{Casey2014}. An alternative approach to finding massive galaxies is to study high-redshift quasars. Quasars are luminous active galactic nuclei \footnote{In this paper, we will refer to the compact object and its immediate surrounding as a quasar or quasi-stellar object (QSO) whereas the surrounding galaxy is referred to as the quasar host galaxy.}. Their emission is bright enough that optical and near-infrared surveys, such as the Sloan Digital Sky Survey (SDSS), Pan-STARRS and VIKING can detect them out to $z \gtrsim 7$ \citep[e.g.,][]{Fan2001,Banados2018,Wang2021}. Since quasar luminosity is correlated to the mass of the supermassive black hole \citep[e.g.,][]{DeRosa2011,Mazzucchelli2017b,Farina2022}, which in turn is correlated with the mass of the galaxy host \citep[e.g.,][]{Ferrarese2000,Gebhardt2000}, we can use luminous high-redshift quasars as an efficient way to pinpoint the location of massive, star-forming high-redshift galaxies. 

Currently there are over 300 $z \gtrsim 6$ luminous quasars known \citep{Fan2023}, and this number is ever increasing as more telescopes are surveying the sky. However, studying the galaxies that host luminous quasars remains difficult, in part because the quasar outshines the host galaxy by orders of magnitude in the rest-frame optical and near-infrared \citep[e.g.,][]{Mechtley2012}. This has recently changed. Observations with the James Webb Space Telescope (JWST) have proven effective at detecting the stellar light from the host galaxies of luminous quasars at $z \gtrsim 6$ \citep{Ding2023,Yue2023,Stone2024,Yue2024,Ding2025,Marshall2025a,Marshall2025b,Onoue2025,Stone2025}. 
Complementary to these observations are those taken at millimeter wavelengths, because millimeter observations trace the dust and cold gas within the quasar host galaxy: the former through measurements of the redshifted dust continuum emission, and the latter through measurements of fine-structure line emission, particularly from ionized carbon, \CII\ \citep[e.g.,][]{Carilli2013,Decarli2025}. Millimeter observations have several advantages: 1) the host galaxy dominates the emission at these wavelengths, 2) the emission is largely unaffected by dust obscuration, and 3) millimeter interferometry allows for improved resolutions that can resolve the interstellar medium (ISM) of these galaxies. As a result, the Atacama Large Millimeter/sub-millimeter Array (ALMA) has observed more than a hundred $z \gtrsim 6$ quasars \citep[e.g.,][]{Wang2013,Decarli2018,Izumi2018,Venemans2020,Wang2024,Bouwens2025}. 

There are several important results that have come from studying the host galaxies of quasars with ALMA. Spatially unresolved observations have shown that the dust continuum and \CII\ emission of quasar host galaxies is bright with total far-infrared luminosities similar to the most massive galaxies in our local universe \citep[e.g.,][]{Bertoldi2003,Maiolino2005,Wang2013,Decarli2018}. This confirmed the assumption that luminous quasars reside in massive galaxies. These unresolved observations have also revealed a large number of companion galaxies \citep{Decarli2017,Willott2017,Neeleman2019}, which is consistent with the picture that these galaxies reside in galaxy overdensities. Resolving the \CII\ and dust emission in more extended ALMA configurations ($\sim$0\farcs3) have revealed that the dust emission is more compact than the \CII\ emission \citep[e.g.,][]{Venemans2020,Wang2024}, although both show extended emission out to $\sim10$ kpc \citep{Novak2020,Bischetti2024,Bischetti2025}. Using the kinematic information from the \CII\ emission line, these resolved observations also revealed the diverse morphology and kinematics of the host galaxies with some galaxies exhibiting signs of rotation and/or merger activity while others showing no sign of ordered motions \citep{Neeleman2021,Tripodi2022,Wang2024}. Finally, using the most extended configurations with ALMA that probe scales less than 400 pc at $z \sim 6$ ($<$0\farcs08), these observations have started to reveal detailed ISM conditions such as dust temperature gradients within the galaxy \citep{Walter2022,Meyer2025b}, the origin of the \CII\ emission \citep{Meyer2025a}, evidence for kinematics beyond simple disk rotation \citep{Yue2021,Neeleman2023}, alignment of the \CII\ and dust emission with the radio jet \citep{Walter2025}, and the location of the quasar within its host galaxy \citep{Meyer2023,Wilde2026}. 

In addition to pinpointing the position of the massive host galaxy, the rest-frame ultraviolet and optical light from the quasar can also provide detailed information on the properties of the supermassive black hole (SMBH), in particular the mass of the SMBH that is powering the quasar. These masses are derived from the strength and width of emission lines in the spectra of quasars, in particular \ion{Mg}{2} and H$\beta$. Local studies have shown a correlation between the mass of the bulge and the mass of the SMBH over many orders of magnitude \citep{Ferrarese2000,Gebhardt2000,Kormendy2013}. By studying this correlation at high redshift, one can constrain how this correlation evolved over cosmic time \citep{Habouzit2022}. Numerous observations have studied the black hole -- galaxy mass correlation using $z \gtrsim 6$ quasars \citep[e.g.,][]{Wang2013,Willott2015,Venemans2016,Shao2017,Decarli2018,Izumi2018,Pensabene2020,Neeleman2021,Wang2024}. Generally, these observations reveal that luminous quasars fall above the local correlation, although this is partly driven by an observational bias from selecting only luminous quasars \citep{Lauer2007,Li2022,Silverman2025}. Nevertheless, the high quasar luminosity makes these objects ideal laboratories to study feedback and joint evolution to understand what drives the growth of these SMBHs.

In this paper, we study a sample of 41 quasars, 40 of which are drawn from the enlarged XQR-30 sample. All quasars in the enlarged XQR-30 sample are luminous with $J$-band magnitudes of $J_{\rm AB} < 20.0$ and have redshifts between 5.8 and 6.6 \citep{DOdorico2023}, and have been observed with VLT/XSHOOTER \citep{Vernet2011} at a relatively high signal-to-noise ratio (S/N), providing detailed information on the quasar as well as any absorption in its line of sight. Details of the survey are given in \citet{DOdorico2023}, and we refer to this paper for a summary of all of the science that can be done with these data. Of particular use for this paper is the determination of the masses for the SMBHs, which were derived in \citet{Mazzucchelli2023}. In this paper, we present ALMA observations of the dust continuum and \CII\ line emission at a mean resolution of 0\farcs25 ($\approx$1.5 kpc). These resolved observations enable a study of the rest-frame far-infrared (FIR) properties of the host galaxies of the quasars. 

The paper is organized as follows. We present the sample in Section \ref{sec:sample}, and report on the ALMA observations and data reduction in Section \ref{sec:observations}. Derivation of the \CII\ emission line properties, dust continuum properties and kinematics are presented in Section \ref{sec:analysis}. This section also contains a discussion on the companion galaxies found within the ALMA data. Finally the results are discussed in Section \ref{sec:discussion} and summarized in Section \ref{sec:conclusions}. In this paper, we will adopt a standard flat cosmology with $H_0 = 70$ km s$^{-1}$ Mpc$^{-1}$ and $\Omega_{\rm m} = 0.3$. In this cosmology 1$''$ corresponds to 5.7 kpc at $z = 6$.

\section{Sample}
\label{sec:sample}

In this paper, we explore the \CII\ emission line and underlying FIR continuum emission from the host galaxies of the quasars that are part of the enlarged XQR-30 sample \citep{DOdorico2023}, For properties of the quasar and associated references, we refer the reader to \citet{DOdorico2023}. ALMA newly observed 23 XQR-30 quasars as part of three distinct ALMA projects:  2019.1.00111.S (PI: B. Venemans), 2021.1.01018.S (PI: S. Bosman) and 2022.1.00998.S (PI: S. Bosman). We supplement these observations with archival data on 17 additional quasars of this sample \citep[see Table \ref{tab:obs};][]{Decarli2018,Venemans2020,Tripodi2022,Tripodi2024a,Wang2024}. Only two XQR-30 quasars have not been covered with any ALMA observations, SDSS~J0927+2001 and QSO~PSO~J183$-$12. For SDSS~J0927+2001, the available ALMA observations from project 2013.1.00591.S (PI: A. Bolatto) did not cover the redshifted \CII\ emission line and was offset in space providing no useful information on the quasar. QSO~PSO~J183$-$12 was part of the original proposal, but did not get observed. Finally, ALMA project 2021.1.01018.S also covered observations of quasar PSO~J215$-$16 \citep{Morganson2012}, this quasar has a slightly lower redshift than the nominal limits of XQR-30 ($z = 5.783$), but satisfies the luminosity limit, and we will include it in this paper, resulting in a total sample of 41 quasars.

\section{ALMA Observations and Data Reduction}
\label{sec:observations}

\begin{deluxetable*}{llrllclccc}
\tablecaption{Journal of ALMA observations
\label{tab:obs}}
\tabletypesize{\scriptsize}
\tablehead{
\colhead{QSO} &
\colhead{R.A.} &
\colhead{Decl.} &
\colhead{PID} &
\colhead{$\nu_{\rm range}$} &
\colhead{$t_{\rm s}$} &
\colhead{Beam} &
\colhead{rms$_{\rm cont}$} &
\colhead{rms$_{\rm 30MHz}$} &
\colhead{Ref.}\\
\colhead{} &
\colhead{(J2000)} &
\colhead{(J2000)} &
\colhead{} &
\colhead{(GHz)} &
\colhead{(min)} &
\colhead{($'' \times ''$)} &
\colhead{($\mu$Jy bm$^{-1}$)} &
\colhead{(mJy bm$^{-1}$)} &
\colhead{}
}
\startdata
PSO J023$-$02 & 01:32:01.70 & -02:16:03.11 & 2021.1.01018.S & 275.149 --- 290.751 &  38 & 0\farcs24 $\times$ 0\farcs20 & 26 & 0.32 & 1\\
PSO J025$-$11 & 01:40:57.03 & -11:40:59.48 & 2021.1.01018.S & 275.421 --- 291.249 &  37 & 0\farcs23 $\times$ 0\farcs21 & 19 & 0.25 & 1\\
ULAS J0148+0600 & 01:48:37.64 & +06:00:20.06 & 2021.1.01018.S & 254.255 --- 272.990 &  62 & 0\farcs27 $\times$ 0\farcs25 & 16 & 0.27 & 1\\
PSO J029$-$29 & 01:58:04.14 & -29:05:19.25 & 2021.1.01018.S & 255.202 --- 273.552 &  37 & 0\farcs23 $\times$ 0\farcs22 & 22 & 0.32 & 1\\
VST ATLAS J029$-$36 & 01:59:57.97 & -36:33:56.60 & 2021.1.01018.S & 253.812 --- 272.141 &  37 & 0\farcs27 $\times$ 0\farcs21 & 19 & 0.31 & 1\\
PSO J060+24 & 04:02:12.69 & +24:51:24.42 & 2021.1.01018.S & 247.770 --- 266.120 &  77 & 0\farcs34 $\times$ 0\farcs24 & 16 & 0.24 & 1\\
VDES J0408$-$5632 & 04:08:19.23 & -56:32:28.80 & 2019.1.00111.S & 253.425 --- 271.777 &  64 & 0\farcs27 $\times$ 0\farcs24 & 18 & 0.28 & 1\\
PSO J065+01 & 04:23:50.15 & +01:43:24.79 & 2021.1.01018.S & 277.675 --- 293.503 &  40 & 0\farcs25 $\times$ 0\farcs22 & 21 & 0.26 & 1\\
PSO J089$-$15 & 05:59:45.47 & -15:35:00.20 & 2021.1.01018.S & 256.271 --- 274.600 &  37 & 0\farcs25 $\times$ 0\farcs22 & 16 & 0.24 & 1\\
PSO J108+08 & 07:13:46.31 & +08:55:32.65 & 2019.1.00111.S & 256.477 --- 274.904 &  66 & 0\farcs27 $\times$ 0\farcs26 & 21 & 0.33 & 1\\
SDSS J0818+1722 & 08:18:27.40 & +17:22:52.01 & 2021.1.01018.S & 254.291 --- 273.029 &  35 & 0\farcs36 $\times$ 0\farcs25 & 17 & 0.26 & 1\\
SDSS J0836+0054 & 08:36:43.86 & +00:54:53.26 & 2022.1.00998.S & 276.811 --- 292.684 &  39 & 0\farcs29 $\times$ 0\farcs24 & 31 & 0.40 & 1\\
DELS J0923+0402 & 09:23:47.12 & +04:02:54.40 & 2019.1.00111.S & 247.457 --- 267.015 &  40 & 0\farcs32 $\times$ 0\farcs28 & 26 & 0.24 & 1\\
PSO J158$-$14 & 10:34:46.50 & -14:25:15.58 & 2019.1.00111.S & 251.514 --- 270.219 &  72 & 0\farcs27 $\times$ 0\farcs23 & 25 & 0.20 & 1\\
PSO J215$-$16$^{\rm a}$ & 14:20:36.34 & -16:02:30.25 & 2021.1.01018.S & 279.329 --- 294.784 &  37 & 0\farcs27 $\times$ 0\farcs22 & 25 & 0.31 & 1\\
PSO J217$-$16 & 14:28:21.39 & -16:02:43.30 & 2019.1.00111.S & 248.879 --- 268.519 &  72 & 0\farcs28 $\times$ 0\farcs24 & 14 & 0.20 & 1\\
PSO J217$-$07 & 14:31:40.45 & -07:24:43.30 & 2019.1.00111.S & 249.237 --- 267.213 &  97 & 0\farcs28 $\times$ 0\farcs24 & 30 & 0.31 & 1\\
CFHQS J1509$-$1749 & 15:09:41.78 & -17:49:26.80 & 2021.1.01018.S & 249.800 --- 268.235 &  46 & 0\farcs24 $\times$ 0\farcs21 & 15 & 0.21 & 1\\
DELS J1535+1943 & 15:35:32.87 & +19:43:20.10 & 2019.1.00111.S & 255.273 --- 274.788 &  35 & 0\farcs32 $\times$ 0\farcs27 & 17 & 0.23 & 1\\
PSO J239$-$07 & 15:58:50.99 & -07:24:09.59 & 2019.1.00111.S & 250.360 --- 268.687 &  75 & 0\farcs30 $\times$ 0\farcs25 & 17 & 0.26 & 1\\
PSO J242$-$12 & 16:09:45.53 & -12:58:54.11 & 2019.1.00111.S & 276.013 --- 291.833 &  37 & 0\farcs26 $\times$ 0\farcs24 & 21 & 0.25 & 1\\
PSO J308$-$27 & 20:33:55.91 & -27:38:54.60 & 2021.1.01018.S & 277.304 --- 293.131 &  37 & 0\farcs26 $\times$ 0\farcs21 & 21 & 0.27 & 1\\
VDES J2211$-$3206 & 22:11:12.39 & -32:06:12.94 & 2021.1.01018.S & 243.063 --- 261.670 &  37 & 0\farcs24 $\times$ 0\farcs21 & 18 & 0.26 & 1\\
VDES J2250$-$5015 & 22:50:02.01 & -50:15:42.20 & 2021.1.01018.S & 255.486 --- 273.815 &  44 & 0\farcs27 $\times$ 0\farcs23 & 17 & 0.25 & 1\\
\tableline
PSO J007+04 & 00:28:06.56 & +04:57:25.64 & 2017.1.01301.S & 255.293 --- 274.397 &  30 & 0\farcs26 $\times$ 0\farcs22 & 28 & 0.28 & 2\\
PSO J009$-$10 & 00:38:56.52 & -10:25:53.90 & 2017.1.01301.S & 253.650 --- 273.137 &  26 & 0\farcs32 $\times$ 0\farcs22 & 61 & 0.59 & 2\\
SDSS J0100+2802 & 01:00:13.02 & +28:02:25.80 & 2022.1.00300.S & 242.476 --- 260.325 & 454 & 0\farcs20 $\times$ 0\farcs18 & 7.6 & 0.12 & 3\\
 &  &  & 2021.1.01320.S & 244.784 --- 261.222 & 104 &  &  &  & \\
 &  &  & 2021.1.00211.S & 244.007 --- 261.735 & 212 &  &  &  & \\
VST ATLAS J025$-$33 & 01:42:43.72 & -33:27:45.61 & 2017.1.01301.S & 241.834 --- 261.416 &  25 & 0\farcs24 $\times$ 0\farcs21 & 26 & 0.26 & 2\\
VDES J0224$-$4711 & 02:24:26.54 & -47:11:29.40 & 2021.1.00934.S & 250.827 --- 269.812 &  55 & 0\farcs23 $\times$ 0\farcs21 & 14 & 0.18 & 4\\
 &  &  & 2018.1.01188.S & 236.149 --- 255.196 &  14 &  &  &  & \\
PSO J036+03 & 02:26:01.87 & +03:02:59.24 & 2015.1.00399.S & 251.084 --- 269.936 &  77 & 0\farcs15 $\times$ 0\farcs12 & 19 & 0.17 & 2\\
PSO J065$-$26 & 04:21:38.05 & -26:57:15.60 & 2017.1.01301.S & 247.628 --- 266.926 &  25 & 0\farcs29 $\times$ 0\farcs21 & 38 & 0.34 & 2\\
WISEA J0439+1634 & 04:39:47.08 & +16:34:16.01 & 2018.1.00566.S & 235.720 --- 254.822 & 421 & 0\farcs22 $\times$ 0\farcs21 & 8.0 & 0.10 & 4\\
 &  &  & 2018.1.01188.S & 235.777 --- 254.824 &  15 &  &  &  & \\
SDSS J0842+1218 & 08:42:29.43 & +12:18:50.58 & 2016.1.00544.S & 251.050 --- 270.632 &  54 & 0\farcs27 $\times$ 0\farcs23 & 28 & 0.29 & 2\\
SDSS J1030+0524 & 10:30:27.11 & +05:24:55.06 & 2015.1.01115.S & 242.247 --- 261.851 &   9 & 1\farcs07 $\times$ 0\farcs87 & 51 & 0.67 & 5\\
PSO J183+05 & 12:12:26.98 & +05:05:33.49 & 2016.1.00544.S & 239.055 --- 257.926 &  48 & 0\farcs27 $\times$ 0\farcs24 & 35 & 0.32 & 2\\
SDSS J1306+0356 & 13:06:08.26 & +03:56:26.19 & 2017.1.01301.S & 253.071 --- 272.675 &  28 & 0\farcs25 $\times$ 0\farcs24 & 47 & 0.47 & 2\\
ULAS J1319+0950 & 13:19:11.29 & +09:50:51.49 & 2012.1.00240.S & 250.484 --- 270.082 &  48 & 0\farcs27 $\times$ 0\farcs21 & 42 & 0.31 & 2\\
PSO J231$-$20 & 15:26:37.84 & -20:50:00.66 & 2016.1.00544.S & 234.135 --- 253.362 &  44 & 0\farcs21 $\times$ 0\farcs12 & 22 & 0.21 & 2\\
PSO J323+12 & 21:32:33.19 & +12:17:55.26 & 2018.1.00908.S & 233.630 --- 252.119 &  43 & 0\farcs12 $\times$ 0\farcs09 & 29 & 0.27 & 2\\
SDSS J2310+1855 & 23:10:38.89 & +18:55:19.70 & 2019.1.00661.S & 256.063 --- 274.824 & 257 & 0\farcs25 $\times$ 0\farcs22 & 7.4 & 0.09 & 6\\
 &  &  & 2019.1.00080.S & 256.585 --- 274.202 &  89 &  &  &  & \\
PSO J359$-$06 & 23:56:32.45 & -06:22:59.26 & 2017.1.01301.S & 247.263 --- 266.844 &  26 & 0\farcs25 $\times$ 0\farcs21 & 29 & 0.28 & 2
\enddata
\tablerefs{1 - This Work; 2 - \citet{Venemans2020}; 3 - \citet{Tripodi2024a}; 4 - \citet{Wang2024}; 5 - \citet{Decarli2018}; 6 - \citet{Tripodi2022}}
\tablecomments{$^{\rm a}$Quasar is not a part of the extended XQR-30 Sample. $\nu_{\rm range}$ is the maximum frequency range covered by the observations with at least an 8-GHz gap at the center. $t_{\rm s}$ is the approximate time on-source. The beam is the mean beam centered at the redshifted \CII\ frequency. Observations of the sample of quasars from the literature are reported below the horizontal line.}
\end{deluxetable*}

Observations of the 24 quasars that are part of the three projects reported here for the first time took place between March 2021 and April 2023. The mean 80$^{\rm th}$ percentile baseline for these observations was $\approx 500$ m. For all of the observations, each of the spectral windows were configured in the frequency domain mode with 480 channels each with a width of $\approx 3.9$ MHz. Two of the adjacent spectral windows were centered on the expected redshifted \CII\ emission line with a slight overlap in frequency to account for any non-linear behavior in the edge channels. The remaining two spectral windows were set up to measure the rest-frame FIR continuum emission. For each observation at least two different blazars were observed, one to calibrate the bandpass and flux, and the other(s) to calibrate the complex gains. For all of the observations, the phase center was placed at the optical/near-infrared positions of the quasar. Details of the observations, including the on-source time, are listed in Table~\ref{tab:obs}. 

For the 17 quasars that were not part of these three observing programs, we downloaded the archival data in order to obtain a uniform (in spatial and spectral resolution) sample. The 13 quasars that were previously observed by \citet{Venemans2020} have spatial and spectral properties consistent with the above sample, and for these quasars, we use the same data products as reported in that paper. The exception is SDSS~J0100+2802, which had sub-optimal data quality, and for this quasar we used the same procedure as the remaining four quasars not part of the \citet{Venemans2020} sample. For these quasars, we downloaded the archival data products relevant to the spatial scales considered in this paper ($\sim 0\farcs25$). In practice, this implies that we either did not use very high spatial resolution observations or very low spatial resolution data (See Appendix \ref{sec:comb_data}). Data from these five quasars were reduced using the same strategy as the 24 quasars described above, resulting in a sample of 29 quasars that required a full data reduction.

The ALMA observations were reduced using the ALMA pipeline \citep{Hunter2023}, which is part of the Common Astronomy Software Applications \citep[\texttt{CASA};][]{CASA2022} package. Because the data spans many different ALMA cycles, different versions of \texttt{CASA} and the ALMA pipeline were used depending on the recommendation for each cycle. In general, the products from the pipeline were calibrated very well, and we only had to manually excise a minimal number of baselines and antennas using the flagging routines within \texttt{CASA}. Several of the quasars had strong enough continuum to self-calibrate the data. Using the standard routines within \texttt{CASA}, we applied several rounds of phase-only self-calibration, which for very bright sources was followed by a round of amplitude and phase self-calibration. The resulting calibrations were then applied to the data for that quasar. The continuum was subtracted using the \texttt{CASA} task, \texttt{uvcontsub} using zeroth order polynomials.

Quasar SDSS~J1030+0524 only has low-resolution observations ($\sim 1"$) available \citep{Decarli2018}, and we use these data for the integrated properties, but they are not used in the rest of the analysis described in this paper. For the remaining quasars, we checked that the uv-distribution of these combined data sets resulted in uniform sensitivity across all spatial scales \citep{Petry2024}. If this was not the case, we applied flagging to oversampled spatial scales using custom Python routines in order to create a uniform spatial distribution, resulting in a dirty beam that had a volume within 10 \% of the volume of the clean beam for all spatial scales of interest. This obviates the need to apply residual scaling corrections to our data \citep[e.g.,][]{Jorsater1995,Novak2020}. Further details are given in Appendix \ref{sec:comb_data}.

The final data sets were imaged using the \texttt{CASA} task, \texttt{tclean}. In order to optimize the resolution versus the requested sensitivity, we tested several different weighting schemes, and found that the standard Briggs weighting scheme with a robust value of 0.5 provided the best trade-off between sensitivity and the effect of side lobes on the volume of the dirty beam \citep[e.g.][]{ALMA2025}. We applied this weighting scheme to all of the quasars in our sample. To create a relatively spatially uniform data set, we applied a uv-taper to some data sets, and we imaged all of the quasars with the same pixel scale using either the multi-scale or H\"{o}gbom deconvolver, depending on the size of the emission region \citep{ALMA2025}. For each quasar three products were created: 1) a dust continuum image generated from all the channels that do not show any line emission, 2) a continuum-subtracted spectral cube of the two spectral windows covering the redshifted \CII\ emission line of the quasars at a spectral resolution of 30\,MHz (32  to 36 \kms) and a \CII\ image created by imaging the \CII\ emission line using a channel width of 1.2 $\times$ the full width at half maximum (FWHM) centered on the \CII\ line \citep{Novak2020,Venemans2020}. An initial FWHM of the emission line was estimated from the spectral cube centered on the position of the quasar by integrating the emission within a region with a radius of 1$''$. This FWHM measurement was then further refined using the procedure described in Section \ref{sec:ana_cii}. For the quasars that did not show any \CII\ emission, an image was created with a width 200 \kms, which is approximately the minimum FWHM of the \CII\ emission line for the quasars that were detected (see Section \ref{sec:ana_cii}). The resulting spatial resolutions and sensitivities are listed in Table \ref{tab:obs}. We note that PSO~J036+03, PSO~J231$-$20 and PSO~J323+12 were observed at a higher spatial resolution than the other quasars in the sample, which have spatial resolutions between  0\farcs21 and 0\farcs30. We include these observations here for completeness, and we have checked that none of the conclusions in the paper are affected by the inclusion of these quasars.

\section{Analysis}
\label{sec:analysis}
In this section, we analyze the data described in Section~\ref{sec:observations}. For completeness, we show both the new data as well as the literature sample that has been homogenized to the same spatial and spectral resolution as the main sample. The \CII\ emission line properties are analyzed in Section~\ref{sec:ana_cii}, the dust continuum emission properties are analyzed in Section~\ref{sec:ana_cont} and the morphology and kinematics of the galaxies are analyzed in Section~\ref{sec:ana_kin}. We also report on a search for companion galaxies in Section \ref{sec:companions}.

\subsection{[C II] Emission Line Properties}
\label{sec:ana_cii}

\begin{figure*}[!t]
\includegraphics[width=\textwidth]{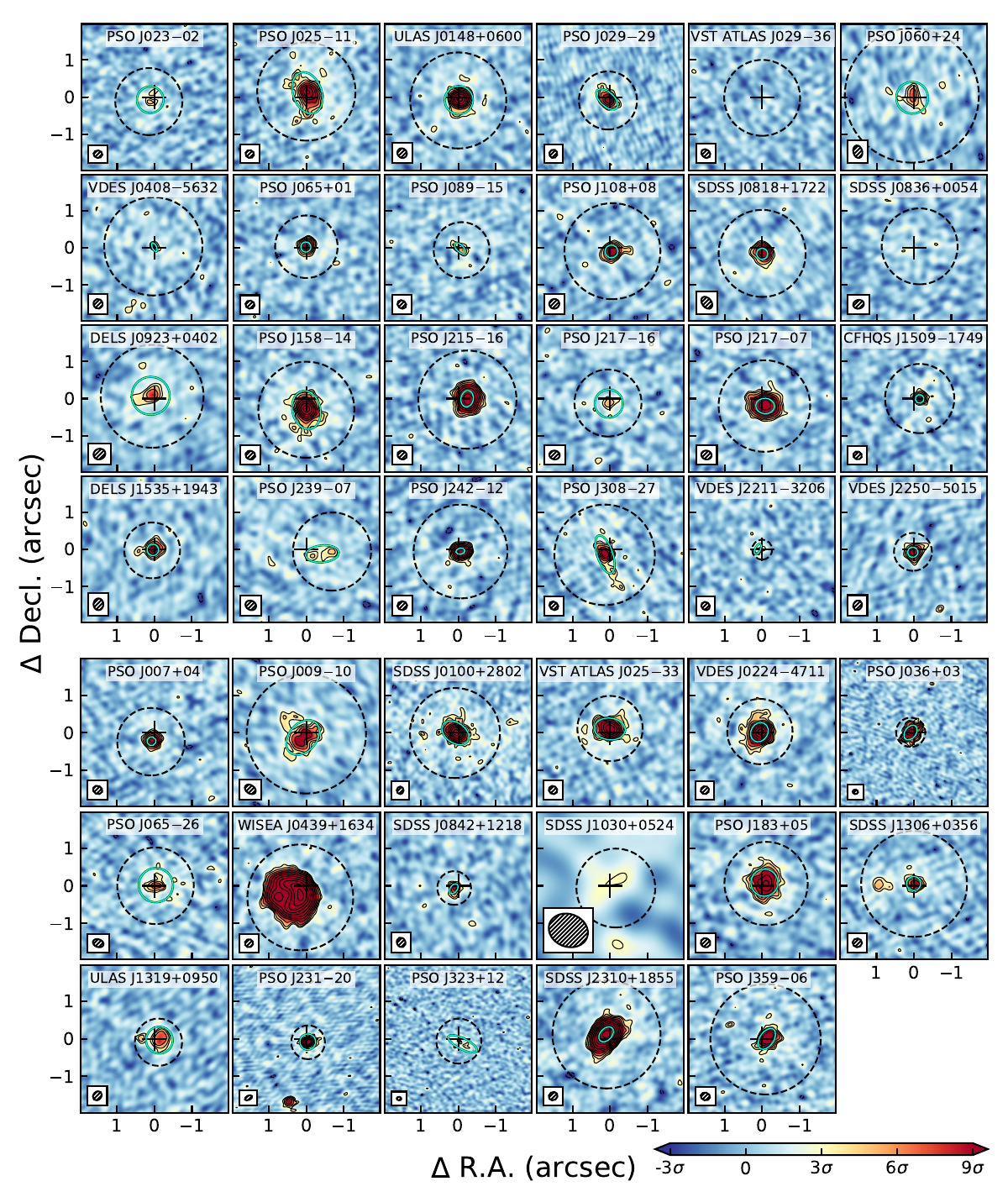}
\caption{\CII\ images of the quasar host galaxies in the enlarged XQR30 sample. The sample is subdivided by the 24 galaxies that were newly observed as part of this survey, and those with archival data on the \CII\ emission line. The contours start at 3$\sigma$ and increase in powers of $\sqrt{2}$. The beam of the observations is shown in the bottom left insert of each panel. The black cross marks the optical position of the quasar, and the black dashed circle is the region used for analysis of the \CII\ emission line. The cyan ellipse marks the beam-deconvolved half-light region of the \CII\ emission, which is plotted for those galaxies with robust size constraints.
\label{fig:ciimap}}
\end{figure*}

\begin{figure*}[!t]
\includegraphics[width=\textwidth]{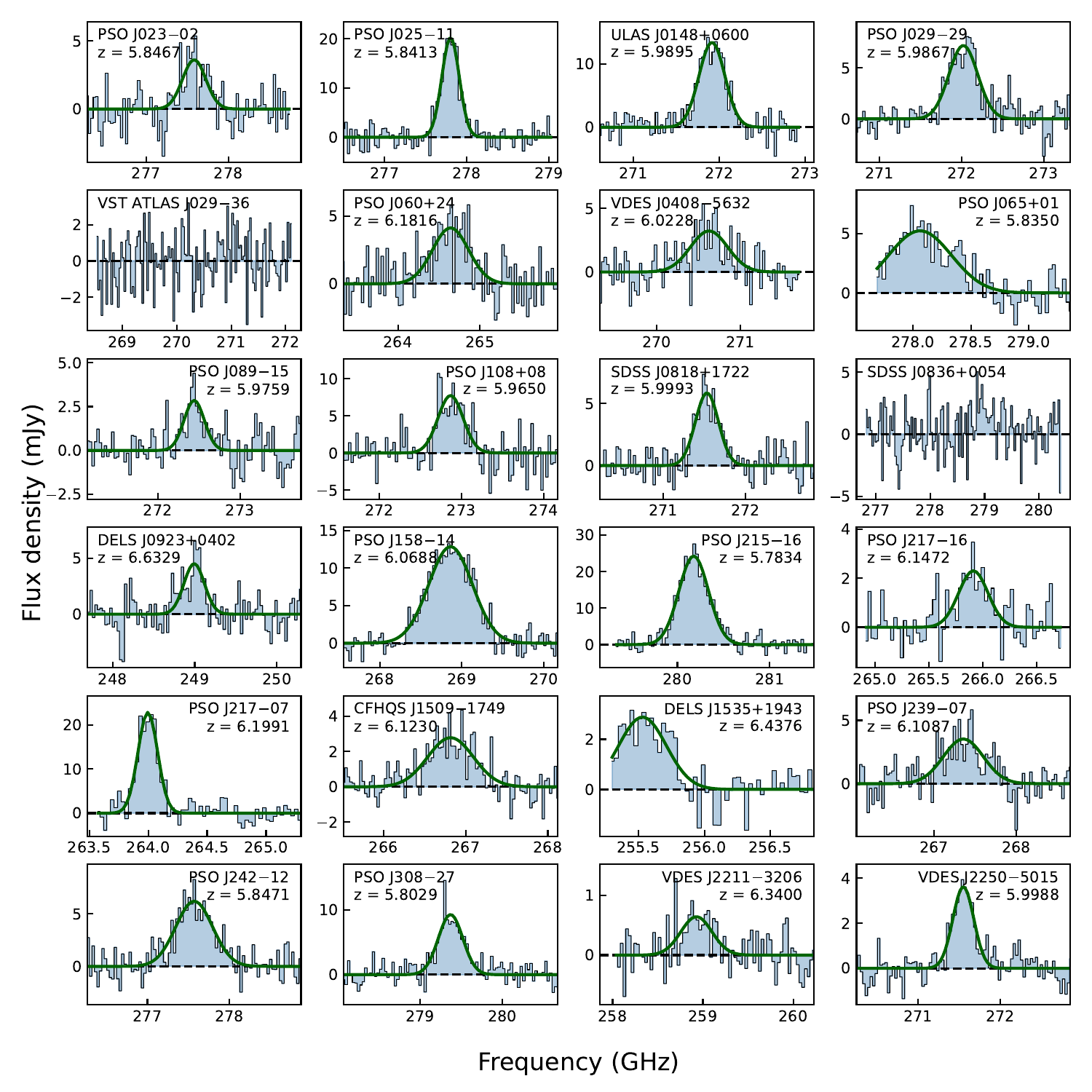}
\caption{Continuum-subtracted spectra of the \CII\ emission line for the quasar host galaxies in the enlarged XQR30 sample. The redshift for the \CII\ emission line was derived from a Gaussian fit to the spectrum (green solid line). The sample is subdivided by the 24 galaxies that were newly observed as part of this survey, and the 17 galaxies with archival \CII\ data.
\label{fig:spectra}}
\end{figure*}

\begin{figure*}[!t]
\ContinuedFloat
\includegraphics[width=\textwidth]{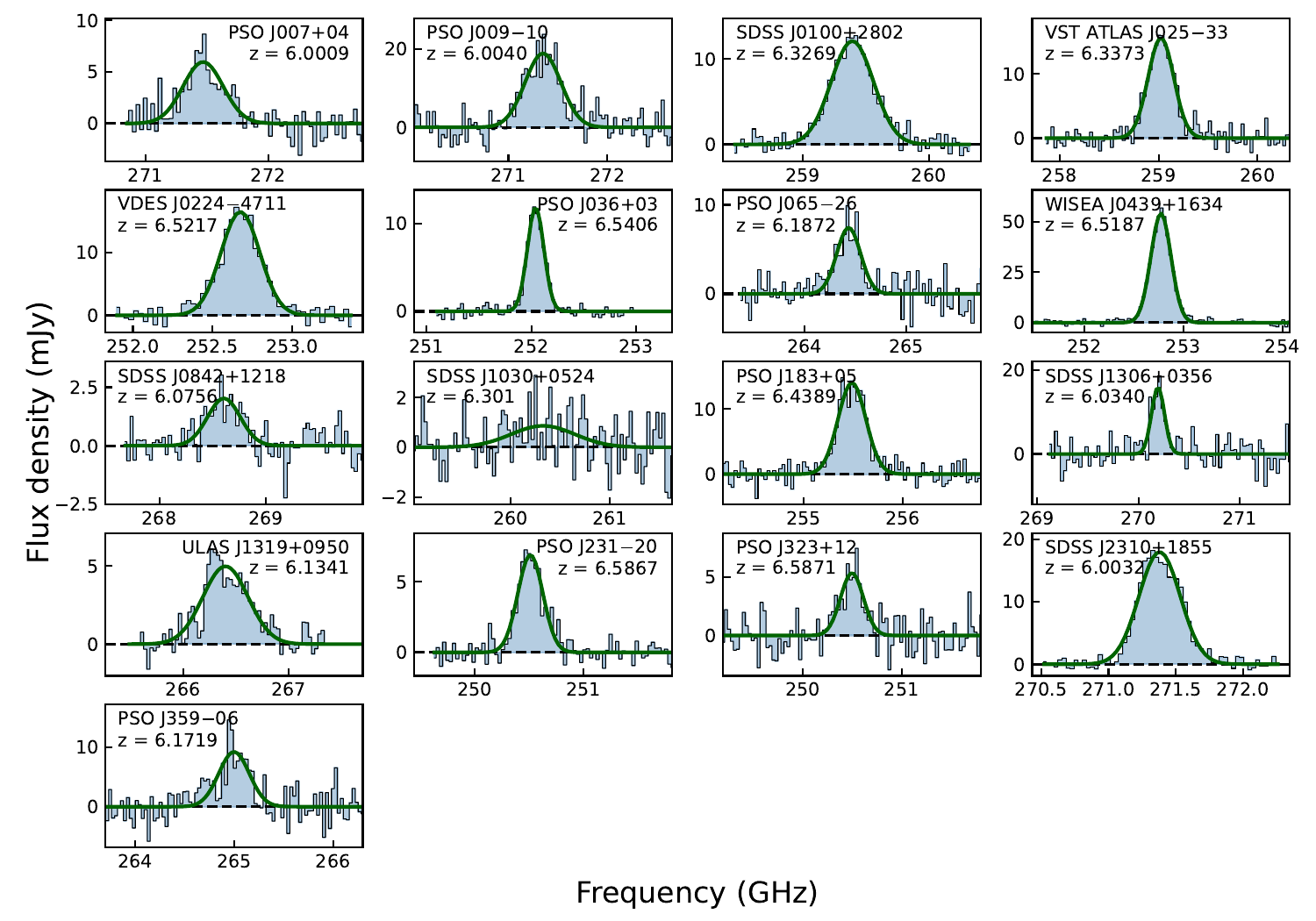}
\caption{Continued.}
\end{figure*}

\begin{deluxetable*}{llrlllllc}
\tablecaption{\CII\ properties of the sample
\label{tab:cii}}
\tabletypesize{\footnotesize}
\tablehead{
\colhead{QSO} &
\colhead{R.A.$_{\rm CII}$} &
\colhead{Decl.$_{\rm CII}$} &
\colhead{$z_{\rm CII}$} &
\colhead{$F_{\rm CII}$} &
\colhead{FWHM$_{\rm CII}$} &
\colhead{$r_{1/2,\rm CII}$} &
\colhead{$L_{\rm CII}$} &
\colhead{S/N$_{\rm CII}$}\\
\colhead{} &
\colhead{(J2000)} &
\colhead{(J2000)} &
\colhead{} &
\colhead{(Jy \kms)} &
\colhead{(\kms)} &
\colhead{(kpc)} &
\colhead{($10^9 L_\odot$)} &
\colhead{}
}
\startdata
PSO J023$-$02 & 01:32:01.707 & $-$02:16:03.18 & 5.8467(8) & $1.42\pm0.35$ & $130\pm 38$ & $2.07\pm0.80$ & $1.29\pm0.32$ & 6.6\\
PSO J025$-$11 & 01:40:57.028 & $-$11:40:59.37 & 5.84134(15) & $5.82\pm0.66$ & $227\pm 15$ & $3.33\pm0.18$ & $5.28\pm0.60$ & 31.1\\
ULAS J0148+0600 & 01:48:37.639 & +06:00:19.98 & 5.9895(3) & $5.50\pm0.67$ & $397\pm 34$ & $1.74\pm0.49$ & $5.17\pm0.63$ & 37.5\\
PSO J029$-$29 & 01:58:04.147 & $-$29:05:19.31 & 5.9867(4) & $3.36\pm0.47$ & $364\pm 42$ & $2.04\pm0.66$ & $3.16\pm0.44$ & 18.4\\
VST ATLAS J029$-$36 & \nodata & \nodata & \nodata & $<$0.35 & 200 & \nodata & $<$0.34 & \nodata\\
PSO J060+24 & 04:02:12.694 & +24:51:24.40 & 6.1816(12) & $2.55\pm0.56$ & $204\pm 50$ & $2.40\pm0.64$ & $2.52\pm0.55$ & 8.2\\
VDES J0408$-$5632 & 04:08:19.230 & $-$56:32:28.78 & 6.0228(15) & $2.00\pm0.52$ & $266\pm 81$ & $0.76\pm0.60$ & $1.89\pm0.50$ & 5.8\\
PSO J065+01 & 04:23:50.151 & +01:43:24.81 & 5.8350(6) & $3.59\pm0.47$ & $517\pm 56$ & $0.79\pm0.11$ & $3.25\pm0.42$ & 23.0\\
PSO J089$-$15 & 05:59:45.467 & $-$15:35:00.21 & 5.9759(6) & $0.92\pm0.20$ & $198\pm 45$ & $1.08\pm0.28$ & $0.87\pm0.18$ & 7.6\\
PSO J108+08 & 07:13:46.308 & +08:55:32.55 & 5.9650(5) & $3.21\pm0.54$ & $363\pm 60$ & $0.98\pm0.14$ & $3.01\pm0.51$ & 17.3\\
SDSS J0818+1722 & 08:18:27.400 & +17:22:51.85 & 5.9993(4) & $2.21\pm0.35$ & $397\pm 60$ & $0.86\pm0.13$ & $2.08\pm0.33$ & 17.4\\
SDSS J0836+0054 & \nodata & \nodata & \nodata & $<$0.43 & 200 & \nodata & $<$0.38 & \nodata\\
DELS J0923+0402 & 09:23:47.127 & +04:02:54.47 & 6.6329(7) & $1.80\pm0.35$ & $217\pm 47$ & $2.75\pm0.43$ & $1.97\pm0.38$ & 9.0\\
PSO J158$-$14 & 10:34:46.500 & $-$14:25:15.89 & 6.0688(3) & $9.37\pm1.01$ & $669\pm 33$ & $2.96\pm0.35$ & $8.99\pm0.97$ & 55.1\\
PSO J215$-$16 & 14:20:36.324 & $-$16:02:30.25 & 5.78336(19) & $10.28\pm1.14$ & $353\pm 21$ & $1.30\pm0.09$ & $9.19\pm1.02$ & 63.6\\
PSO J217$-$16 & 14:28:21.393 & $-$16:02:43.44 & 6.1472(8) & $0.79\pm0.20$ & $372\pm 95$ & $2.13\pm0.73$ & $0.77\pm0.19$ & 6.7\\
PSO J217$-$07 & 14:31:40.445 & $-$07:24:43.50 & 6.19910(14) & $5.20\pm0.61$ & $238\pm 18$ & $1.41\pm0.10$ & $5.15\pm0.61$ & 40.3\\
CFHQS J1509$-$1749 & 15:09:41.770 & $-$17:49:26.81 & 6.1230(11) & $2.15\pm0.39$ & $573\pm106$ & $0.66\pm0.15$ & $2.09\pm0.38$ & 14.2\\
DELS J1535+1943 & 15:35:32.873 & +19:43:20.08 & 6.4376(7) & $1.38\pm0.21$ & $528\pm 78$ & $0.85\pm0.16$ & $1.44\pm0.22$ & 13.7\\
PSO J239$-$07 & 15:58:50.961 & $-$07:24:09.71 & 6.1087(11) & $2.40\pm0.43$ & $437\pm 87$ & $2.55\pm0.52$ & $2.32\pm0.42$ & 7.4\\
PSO J242$-$12 & 16:09:45.526 & $-$12:58:54.16 & 5.8471(6) & $3.79\pm0.55$ & $518\pm 67$ & $0.71\pm0.10$ & $3.44\pm0.50$ & 31.2\\
PSO J308$-$27 & 20:33:55.921 & $-$27:38:54.74 & 5.8029(4) & $3.64\pm0.54$ & $419\pm 58$ & $3.11\pm0.38$ & $3.27\pm0.48$ & 17.4\\
VDES J2211$-$3206 & 22:11:12.399 & $-$32:06:12.93 & 6.3400(14) & $0.30\pm0.10$ & $347\pm117$ & $0.85\pm0.51$ & $0.31\pm0.10$ & 4.1\\
VDES J2250$-$5015 & 22:50:02.014 & $-$50:15:42.28 & 5.9988(3) & $1.27\pm0.17$ & $331\pm 38$ & $0.77\pm0.12$ & $1.20\pm0.16$ & 16.2\\
\tableline
PSO J007+04 & 00:28:06.565 & +04:57:25.41 & 6.0009(4) & $2.66\pm0.38$ & $364\pm 45$ & $0.68\pm0.11$ & $2.51\pm0.36$ & 20.5\\
PSO J009$-$10 & 00:38:56.526 & $-$10:25:54.03 & 6.0040(5) & $9.40\pm1.38$ & $464\pm 61$ & $3.00\pm0.33$ & $8.87\pm1.31$ & 17.9\\
SDSS J0100+2802 & 01:00:13.026 & +28:02:25.80 & 6.32687(17) & $5.79\pm0.62$ & $416\pm 18$ & $2.21\pm0.45$ & $5.90\pm0.63$ & 53.8\\
VST ATLAS J025$-$33 & 01:42:43.721 & $-$33:27:45.51 & 6.33730(17) & $5.66\pm0.63$ & $382\pm 21$ & $2.29\pm0.43$ & $5.79\pm0.64$ & 35.8\\
VDES J0224$-$4711 & 02:24:26.547 & $-$47:11:29.39 & 6.52166(11) & $5.81\pm0.61$ & $285\pm 11$ & $1.47\pm0.18$ & $6.19\pm0.65$ & 59.4\\
PSO J036+03 & 02:26:01.876 & +03:02:59.24 & 6.54059( 9) & $2.72\pm0.29$ & $214\pm 10$ & $1.19\pm0.09$ & $2.91\pm0.31$ & 32.5\\
PSO J065$-$26 & 04:21:38.047 & $-$26:57:15.58 & 6.1872(4) & $2.36\pm0.39$ & $306\pm 49$ & $2.57\pm0.57$ & $2.33\pm0.39$ & 9.9\\
WISEA J0439+1634 & 04:39:47.091 & +16:34:15.71 & 6.51872( 3) & $16.13\pm1.62$ & $285\pm  3$ & \nodata & $4.99\pm0.50$$^{\rm a}$ & 210.9\\
SDSS J0842+1218 & 08:42:29.438 & +12:18:50.50 & 6.0756(7) & $0.82\pm0.17$ & $502\pm108$ & $0.84\pm0.20$ & $0.79\pm0.16$ & 8.7\\
SDSS J1030+0524 & 10:30:27.093 & +05:24:55.25 & 6.301(4) & $0.74\pm0.31$ & $138\pm 68$ & \nodata$^{\rm b}$ & $0.75\pm0.32$ & 3.6\\
PSO J183+05 & 12:12:26.976 & +05:05:33.57 & 6.4389(3) & $5.68\pm0.68$ & $387\pm 32$ & $1.98\pm0.17$ & $5.94\pm0.71$ & 27.4\\
SDSS J1306+0356 & 13:06:08.261 & +03:56:26.25 & 6.0340(3)$^{\rm c}$ & $2.69\pm0.52$ & $200\pm 39$ & $1.07\pm0.17$ & $2.56\pm0.50$ & 11.2\\
ULAS J1319+0950 & 13:19:11.282 & +09:50:51.46 & 6.1341(5) & $2.99\pm0.40$ & $540\pm 58$ & $2.01\pm0.60$ & $2.92\pm0.39$ & 15.6\\
PSO J231$-$20 & 15:26:37.837 & $-$20:50:00.74 & 6.5867(2) & $2.38\pm0.28$ & $251\pm 20$ & $1.14\pm0.38$ & $2.58\pm0.30$ & 24.8\\
PSO J323+12 & 21:32:33.182 & +12:17:55.13 & 6.5871(7) & $1.77\pm0.41$ & $180\pm 44$ & $2.32\pm0.62$ & $1.92\pm0.44$ & 7.5\\
SDSS J2310+1855 & 23:10:38.898 & +18:55:19.82 & 6.00323( 8) & $7.37\pm0.75$ & $431\pm 11$ & $1.34\pm0.09$ & $6.95\pm0.71$ & 168.5\\
PSO J359$-$06 & 23:56:32.442 & $-$06:22:59.26 & 6.1719(6) & $4.11\pm0.66$ & $305\pm 49$ & $1.76\pm0.36$ & $4.04\pm0.65$ & 20.3
\enddata
\tablecomments{All upper limits are 2$\sigma$ upper limits. For $z_{\rm CII}$, the number in parentheses is the uncertainty on the last number(s). $^{\rm a}$The \CII\ luminosity for WISEA~J0439+1634 has been adjusted by a magnifaction factor of 3.44 \citep{Yue2021}. $^{\rm b}$The low resolution data for this quasar does not provide an accurate constraint on the size of the \CII\ emission. $^{\rm c}$The redshift of this target is determined from the combined spectrum of the quasar host galaxy and its nearby merging companion. This results in a slightly discrepant redshift compared to \citet{Venemans2020}.}
\end{deluxetable*}

To estimate the velocity-integrated \CII\ flux density, $F_{\rm CII}$, we adopt the curve-of-growth analysis described in \citet{Novak2020} where we define the center of the emission as the pixel with the maximum flux in the \CII\ images (Fig.~\ref{fig:ciimap}). To measure $F_{\rm CII}$, we use a velocity range of $2\times$ the FWHM, which will include $>99$~\% of the flux, if the emission has a Gaussian spectral profile. Theoretically, the flux density measurement should asymptotically approach the true flux density with increasing radii of the region used for the analysis. However, in practice, noise fluctuations result in a minimum radius for which the flux density is consistent within the uncertainties with the flux density at large radii. For each quasar, we measure this radius, and use this region to perform all of our subsequent analysis (dashed circles in Fig.~\ref{fig:ciimap}). We note that this approach will include emission from faint nearby merging companion galaxies, which will affect both $F_{\rm CII}$ and the redshift. This is most notable in J1306+0356, where the redshift determination is discrepant compared to \citet{Venemans2020}. Only sources where the curve-of-growth analysis yields a maximum signal-to-noise ratio (S/N) $>$ 3 are reported as detections\footnote{The maximum S/N is the maximum of all of the S/N measurements obtained during the curve-of-growth analysis, which almost always comes from a smaller region than what is used for the flux measurement.}.  For the two quasars not detected in \CII, VST~ATLAS~J029$-$36 and SDSS~J0836+0054, a circular region with a radius of 1$''$ centered on the optical position of the quasar is used. In all cases we add a 10~\% uncertainty in quadrature to the measured uncertainty of $F_{\rm CII}$ from the uncertainty in absolute flux calibration of ALMA \citep{ALMA2025}. All fluxes have also been corrected for the primary beam attenuation, and are listed in Table~\ref{tab:cii}. 

The continuum-subtracted spectrum for each quasar is shown in Figure \ref{fig:spectra}. We estimate the redshift of the \CII\ emission line from a Gaussian fit to the spectrum. Although most of the profiles are nearly Gaussian (a notable exception is SDSS~J2310+1855), we opt not to derive the FWHM from this fit, but instead derive it from the data by finding the points where the flux drops below half the peak amplitude of the emission for the first time. This approach is independent of the shape of the emission line. Uncertainties on these measurements are calculated using the scaling relations for Gaussian line fitting derived in \citet{Lenz1992}. From these measurements, we calculate the \CII\ luminosity, $L_{\rm CII}$, using standard equations \citep[e.g.,][]{Carilli2013,Decarli2025}. For the only confirmed lensed quasar of the sample, WISEA~J0439+1634, we have corrected the \CII\ luminosity by a lensing factor of 3.44 as derived in \citet{Yue2021}. Most quasars have $L_{\rm CII}$ between $\approx 10^9$ and $10^{10} L_{\rm \odot}$. Our least luminous \CII\ detection is for quasar VDES~J2211$-$3206 with a luminosity of $(0.31 \pm 0.10) \times 10^9 L_{\rm \odot}$. The maximum S/N for this \CII\ emission line was 4.1. We note that quasar SDSS~J1030$+$0524 is only detected at 3.6$\sigma$ significance, which is the reason it was not reported as a detection in \citet{Decarli2018} who used a 5$\sigma$ threshold for detections.

To estimate the spatial extent and position of the \CII\ emission line, we fit the \CII\ emission map shown in Figure~\ref{fig:ciimap} with a 2D S\'{e}rsic profile that is convolved with the beam. The median parameters and their associated uncertainties were calculated using an MCMC routine that is publicly available as part of the python package \texttt{QubeFit} \citep[hereafter, \citetalias{Neeleman2021}]{Neeleman2021}. Details of the fitting process are given in Appendix \ref{sec:sizes}. In short, we convolve the S\'{e}rsic model with the ALMA synthesized beam and then uses a standard chi-squared likelihood function and the \texttt{emcee} package \citep{Foreman-Mackey2013} to explore the parameter space. All seven parameters of the S\'{e}rsic model (x and y coordinates of the origin, amplitude, S\'{e}rsic index, half-light radius, position angle of the major axis and ellipticity) are kept free during the fitting. In cases where the S\'{e}rsic index was unconstrained, we fitted the emission with a 2D Gaussian, which is the typical profile used in previous studies \citep[e.g.,][]{Venemans2020,Wang2024}. The resultant median half-light ellipse for each of the quasars is shown in Figure~\ref{fig:ciimap}. This is the deconvolved shape, and this ellipse therefore provides a rough estimate of the true shape and size of the \CII\ emission. The half light radius for the \CII\ emission, $r_{\rm 1/2,CII}$, for each quasar is listed in Table~\ref{tab:cii} and Appendix \ref{sec:sizes} and ranges between 0\farcs12 (0.7~kpc) and 0\farcs57 (3.3~kpc). We note that the extent of more diffuse \CII\ emission can be much larger, especially in galaxies that show disturbed morphologies. However, the half light radii provide a good size estimate for the bulk of the \CII\ emission.

\subsection{Dust Continuum Emission Properties}
\label{sec:ana_cont}

\begin{figure*}[!t]
\includegraphics[width=\textwidth]{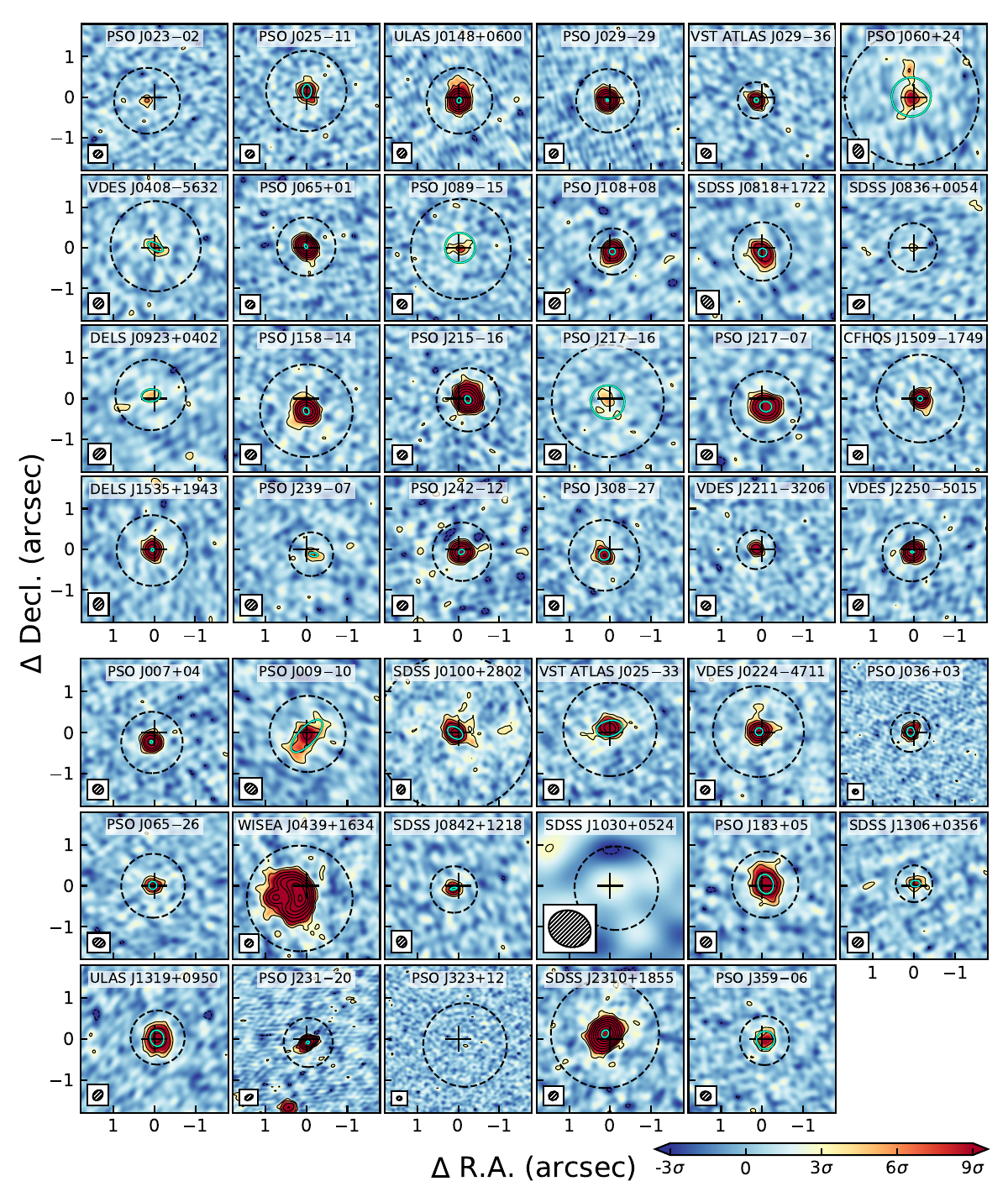}
\caption{Dust continuum images of the quasar host galaxies in the enlarged XQR30 sample. The sample is subdivided by the 24 galaxies that were newly observed as part of this survey, and those with archival data on the dust continuum. The contours start at 3$\sigma$ and increase in powers of 2. The synthesized beam of the observations is shown in the bottom left insert of each panel. The black cross marks the optical position of the quasar, and the black dashed circle is the region used for analysis of the dust continuum emission. The cyan ellipse marks the beam-deconvolved half-light region of the dust continuum emission, which is plotted for those galaxies with robust size constraints.
\label{fig:contmap}}
\end{figure*}

\begin{deluxetable*}{llrclllcc}
\tablecaption{Dust continuum properties of the sample
\label{tab:cont}}
\tabletypesize{\footnotesize}
\tablehead{
\colhead{QSO} &
\colhead{R.A.$_{\rm cont}$} &
\colhead{Decl.$_{\rm cont}$} &
\colhead{Frequency} &
\colhead{$S_{\rm cont}$} &
\colhead{$r_{1/2,\rm cont}$} &
\colhead{$L_{\rm TIR}$} &
\colhead{$L_{\rm TIR}$ ref} &
\colhead{S/N$_{\rm cont}$}\\
\colhead{} &
\colhead{(J2000)} &
\colhead{(J2000)} &
\colhead{(GHz)} &
\colhead{(mJy)} &
\colhead{(kpc)} &
\colhead{($10^{12} L_\odot$)} &
\colhead{} &
\colhead{}
}
\startdata
PSO J023$-$02 & 01:32:01.713 & $-$02:16:03.20 & 283 & $0.41\pm0.17$ & $<$0.87 & $1.1^{+1.0}_{-0.8}$ & 1 & 5.2\\
PSO J025$-$11 & 01:40:57.029 & $-$11:40:59.33 & 283 & $1.60\pm0.22$ & $1.00\pm0.25$ & $4.2^{+3.6}_{-2.8}$ & 1 & 27.7\\
ULAS J0148+0600 & 01:48:37.639 & +06:00:19.98 & 264 & $3.73\pm0.38$ & $0.39\pm0.09$ & $10.2^{+8.7}_{-6.7}$ & 1 & 107.5\\
PSO J029$-$29 & 01:58:04.146 & $-$29:05:19.33 & 264 & $4.93\pm0.51$ & $0.20\pm0.09$ & $13.4^{+11.5}_{-8.8}$ & 1 & 112.1\\
VST ATLAS J029$-$36 & 01:59:57.982 & $-$36:33:56.67 & 263 & $0.84\pm0.10$ & $0.31\pm0.10$ & $2.3^{+2.0}_{-1.5}$ & 1 & 24.4\\
PSO J060+24 & 04:02:12.695 & +24:51:24.42 & 257 & $0.87\pm0.18$ & $2.70\pm0.52$ & $2.5^{+2.2}_{-1.7}$ & 1 & 9.5\\
VDES J0408$-$5632 & 04:08:19.227 & $-$56:32:28.77 & 263 & $0.49\pm0.14$ & $1.19\pm0.24$ & $1.3^{+1.2}_{-0.9}$ & 1 & 7.0\\
PSO J065+01 & 04:23:50.151 & +01:43:24.82 & 286 & $5.80\pm0.59$ & $0.34\pm0.09$ & $15.1^{+12.9}_{-9.9}$ & 1 & 138.7\\
PSO J089$-$15 & 05:59:45.467 & $-$15:35:00.20 & 265 & $0.55\pm0.15$ & $2.11\pm0.77$ & $1.5^{+1.3}_{-1.0}$ & 1 & 6.8\\
PSO J108+08 & 07:13:46.306 & +08:55:32.56 & 266 & $1.70\pm0.19$ & $0.45\pm0.11$ & $4.6^{+3.9}_{-3.0}$ & 1 & 43.7\\
SDSS J0818+1722 & 08:18:27.399 & +17:22:51.90 & 264 & $0.90\pm0.11$ & $0.62\pm0.18$ & $2.5^{+2.1}_{-1.6}$ & 1 & 23.3\\
SDSS J0836+0054 & 08:36:43.862 & +00:54:53.28 & 285 & $0.24\pm0.13$ & $<$2.29 & $0.6^{+0.6}_{-0.5}$ & 1 & 3.0\\
DELS J0923+0402 & 09:23:47.126 & +04:02:54.48 & 257 & $0.57\pm0.15$ & $1.27\pm0.46$ & $1.8^{+1.6}_{-1.3}$ & 1 & 4.8\\
PSO J158$-$14 & 10:34:46.500 & $-$14:25:15.89 & 261 & $4.53\pm0.49$ & $0.52\pm0.10$ & $12.6^{+10.8}_{-8.3}$ & 1 & 78.1\\
PSO J215$-$16 & 14:20:36.324 & $-$16:02:30.27 & 287 & $13.30\pm1.34$ & $0.59\pm0.09$ & $34.2^{+29.2}_{-22.5}$ & 1 & 239.1\\
PSO J217$-$16 & 14:28:21.394 & $-$16:02:43.39 & 259 & $0.48\pm0.13$ & $2.32\pm0.56$ & $1.4^{+1.2}_{-1.0}$ & 1 & 5.5\\
PSO J217$-$07 & 14:31:40.444 & $-$07:24:43.50 & 258 & $5.13\pm0.54$ & $0.84\pm0.09$ & $14.8^{+12.6}_{-9.7}$ & 1 & 68.8\\
CFHQS J1509$-$1749 & 15:09:41.770 & $-$17:49:26.80 & 259 & $1.50\pm0.19$ & $0.35\pm0.09$ & $4.2^{+3.6}_{-2.8}$ & 1 & 44.6\\
DELS J1535+1943 & 15:35:32.873 & +19:43:20.09 & 265 & $0.95\pm0.13$ & $0.20\pm0.10$ & $2.9^{+2.5}_{-1.9}$ & 1 & 26.3\\
PSO J239$-$07 & 15:58:50.978 & $-$07:24:09.72 & 260 & $0.22\pm0.06$ & $0.59\pm0.28$ & $0.6^{+0.5}_{-0.4}$ & 1 & 5.1\\
PSO J242$-$12 & 16:09:45.525 & $-$12:58:54.18 & 284 & $3.31\pm0.35$ & $0.45\pm0.09$ & $8.7^{+7.4}_{-5.7}$ & 1 & 80.8\\
PSO J308$-$27 & 20:33:55.921 & $-$27:38:54.74 & 285 & $1.12\pm0.17$ & $0.76\pm0.27$ & $2.9^{+2.5}_{-1.9}$ & 1 & 19.0\\
VDES J2211$-$3206 & 22:11:12.401 & $-$32:06:12.94 & 252 & $0.51\pm0.08$ & $<$0.39 & $1.5^{+1.3}_{-1.0}$ & 1 & 15.0\\
VDES J2250$-$5015 & 22:50:02.015 & $-$50:15:42.26 & 265 & $1.87\pm0.20$ & $0.25\pm0.10$ & $5.1^{+4.4}_{-3.3}$ & 1 & 59.4\\
\tableline
PSO J007+04 & 00:28:06.565 & +04:57:25.41 & 265 & $2.41\pm0.28$ & $0.19\pm0.10$ & $4.3^{+6.6}_{-1.7}$ & 7 & 45.7\\
PSO J009$-$10 & 00:38:56.519 & $-$10:25:53.98 & 263 & $3.83\pm0.51$ & $2.91\pm0.37$ & $13.6^{+30.6}_{-7.1}$ & 7 & 14.6\\
SDSS J0100+2802 & 01:00:13.026 & +28:02:25.80 & 252 & $1.65\pm0.20$ & $1.13\pm0.21$ & $5.3^{+0.6}_{-0.6}$ & 4 & 52.7\\
VST ATLAS J025$-$33 & 01:42:43.722 & $-$33:27:45.51 & 252 & $2.58\pm0.33$ & $1.79\pm0.52$ & $6.3^{+6.7}_{-2.2}$ & 7 & 32.8\\
VDES J0224$-$4711 & 02:24:26.547 & $-$47:11:29.37 & 253 & $2.42\pm0.27$ & $0.53\pm0.11$ & $33.2^{+22.0}_{-10.0}$ & 6 & 64.3\\
PSO J036+03 & 02:26:01.876 & +03:02:59.26 & 261 & $2.78\pm0.30$ & $0.63\pm0.09$ & $5.3^{+0.7}_{-0.6}$ & 6,7 & 46.1\\
PSO J065$-$26 & 04:21:38.053 & $-$26:57:15.59 & 257 & $1.18\pm0.23$ & $0.57\pm0.14$ & $3.4^{+2.9}_{-2.3}$ & 1 & 11.4\\
WISEA J0439+1634 & 04:39:47.091 & +16:34:15.68 & 245 & $16.25\pm1.63$ & \nodata & $11.3^{+9.5}_{-7.3}$$^{\rm a}$ & 1 & 468.8\\
SDSS J0842+1218 & 08:42:29.438 & +12:18:50.51 & 261 & $0.72\pm0.14$ & $0.50\pm0.15$ & $2.0^{+1.7}_{-1.4}$ & 1 & 11.0\\
SDSS J1030+0524 & \nodata & \nodata & 252 & $<$0.21 & \nodata & $<$1.2 & 1 & \nodata\\
PSO J183+05 & 12:12:26.974 & +05:05:33.53 & 248 & $4.26\pm0.46$ & $1.33\pm0.11$ & $11.9^{+4.0}_{-4.0}$ & 5,6 & 38.2\\
SDSS J1306+0356 & 13:06:08.258 & +03:56:26.24 & 263 & $0.78\pm0.16$ & $0.71\pm0.20$ & $2.1^{+1.9}_{-1.5}$ & 1 & 7.1\\
ULAS J1319+0950 & 13:19:11.287 & +09:50:51.50 & 260 & $3.88\pm0.43$ & $1.12\pm0.10$ & $16.0^{+13.0}_{-6.0}$ & 2 & 31.8\\
PSO J231$-$20 & 15:26:37.837 & $-$20:50:00.75 & 244 & $3.84\pm0.41$ & $0.23\pm0.09$ & $11.0^{+5.5}_{-2.8}$ & 6,7 & 82.4\\
PSO J323+12 & \nodata & \nodata & 243 & $<$1.02 & \nodata & $<$6.5 & 1 & \nodata\\
SDSS J2310+1855 & 23:10:38.899 & +18:55:19.83 & 265 & $8.92\pm0.89$ & $0.58\pm0.09$ & $24.8^{+6.2}_{-5.2}$ & 2,3 & 494.4\\
PSO J359$-$06 & 23:56:32.445 & $-$06:22:59.29 & 257 & $0.94\pm0.15$ & $1.24\pm0.37$ & $2.7^{+2.3}_{-1.8}$ & 1 & 10.9
\enddata
\tablerefs{1. This work, 2. \citet{Carniani2019}, 3. \citet{Tripodi2022}, 4. \citet{Tripodi2023}, 5. \citet{Decarli2023}, 6. \citet{Tripodi2024b}, 7. \citet{Costa2025}}
\tablecomments{All upper limits are 2$\sigma$ upper limits. $^{\rm a}$The total infrared luminosity for WISEA~J0439+1634 has been adjusted by a magnifaction factor of 4.53 \citep{Yue2021}.}
\end{deluxetable*}

Continuum emission is detected at $>3\sigma$ in 39 out of 41 quasars. To estimate the continuum flux emitted from the quasar, we employ the same curve-of-growth analysis as described in Section \ref{sec:ana_cii}. In practice, this approach isolates the quasar from distinct secondary sources (e.g., quasar PSO~J231$-$20). However in cases where the second source is close to the quasar sight line or is connected by extended continuum emission the whole region is used (e.g., quasar PSO~J060+24). These distinct secondary sources only provide a marginal ($<$20~\%) increase in the flux of the quasar. The continuum emission is shown in Figure~\ref{fig:contmap}, including the regions used for the flux density analysis. The continuum flux density for each source is listed in Table~\ref{tab:cont} where the uncertainties take into account a 10~\% uncertainty on the flux due to the absolute flux calibration uncertainty of ALMA.

It is generally assumed that most of the emission arises from the dust within the quasar host galaxies \citep[e.g.,][]{Venemans2020,Tripodi2023}. 
This assumption starts to break down for radio-loud quasars where the continuum emission can have a significant contribution from synchrotron emission at these frequencies \citep{Rojas-Ruiz2021,Khusanova2022,Banados2024,Mazzucchelli2025}. At least one of the quasars, SDSS~J0836+0054, is radio-loud \citep{Wang2007}, but the continuum emission is extremely weak for this quasar. Since the fraction of radio-loud quasars is expected to be around 10 \% \citep{Banados2015}, these quasars will not significantly affect the overall conclusions in this paper. Therefore we will assume that all of the continuum emission arises from dust within the host galaxy of the quasar. However, we note that the far-infrared dust continuum measurements should be taken as upper limits for radio-loud quasars. 

In this paper, we report on only a single continuum measurement. Therefore, we have to assume a fiducial dust temperature and spectral index for the modified black body spectrum in order to estimate a total far-infrared luminosity ($L_{\rm TIR}$; defined as the total luminosity between 8 $\mu$m and 1000 $\mu$m). Here we take the values chosen as in \citet{Venemans2020}, which are based on \citet{Beelen2006} with $T_{\rm dust} = 47 \pm 10$~K and $\beta = 1.6 \pm 0.4$ \citep[see also;][]{Witstok2023,Tripodi2024b,Costa2025}. However, the uncertainties on $L_{\rm TIR}$ in Table ~\ref{tab:cont} take into account the large uncertainties in these values. We note that some of the quasars in the sample have observations at multiple frequencies, which is crucial in order to better sample the dust continuum emission \citep[e.g.,][]{Tripodi2023,Tripodi2024b,Costa2025} and thus constrain $T_{\rm dust}$ and $\beta$. In particular, \citet{Tripodi2024b} derive fitted values for $T_{\rm dust}$ and $\beta$ that are consistent with these fiducial values, although they show that individual quasars can significantly deviate from these average values. Of the 10 quasars that have published multi-frequency dust continuum observations in our sample, only one quasar has a total far-infrared luminosity where the single point measurement significantly deviates from the multi-frequency measurement, VDES~J0224$-$4711 \citep{Tripodi2024b}. This suggests that these single point estimates provide an accurate estimate of the typical far-infrared luminosity, but we emphasize that sampling the dust continuum profile is necessary to provide accurate estimates of $L_{\rm TIR}$ for individual quasars.

To estimate the spatial extent of the dust continuum emission, we use the same approach of fitting a beam-convolved S\'{e}rsic profile that was used to fit the \CII\ emission map (see Section \ref{sec:ana_cii}). Unlike the \CII\ emission line, some of the dust continuum emission remains unresolved in these observations. For those galaxies, we report the 2$\sigma$ upper limit on the half-light radius. Details of the fitting are given in Appendix \ref{sec:sizes} and the half-light radii are shown in Table \ref{tab:cont}. Figure~\ref{fig:contmap} shows the elliptical extent of the half-light radius for each quasar that was resolved in the observations.

\subsection{Galaxy Kinematics}
\label{sec:ana_kin}

\begin{figure*}[!t]
\includegraphics[width=\textwidth]{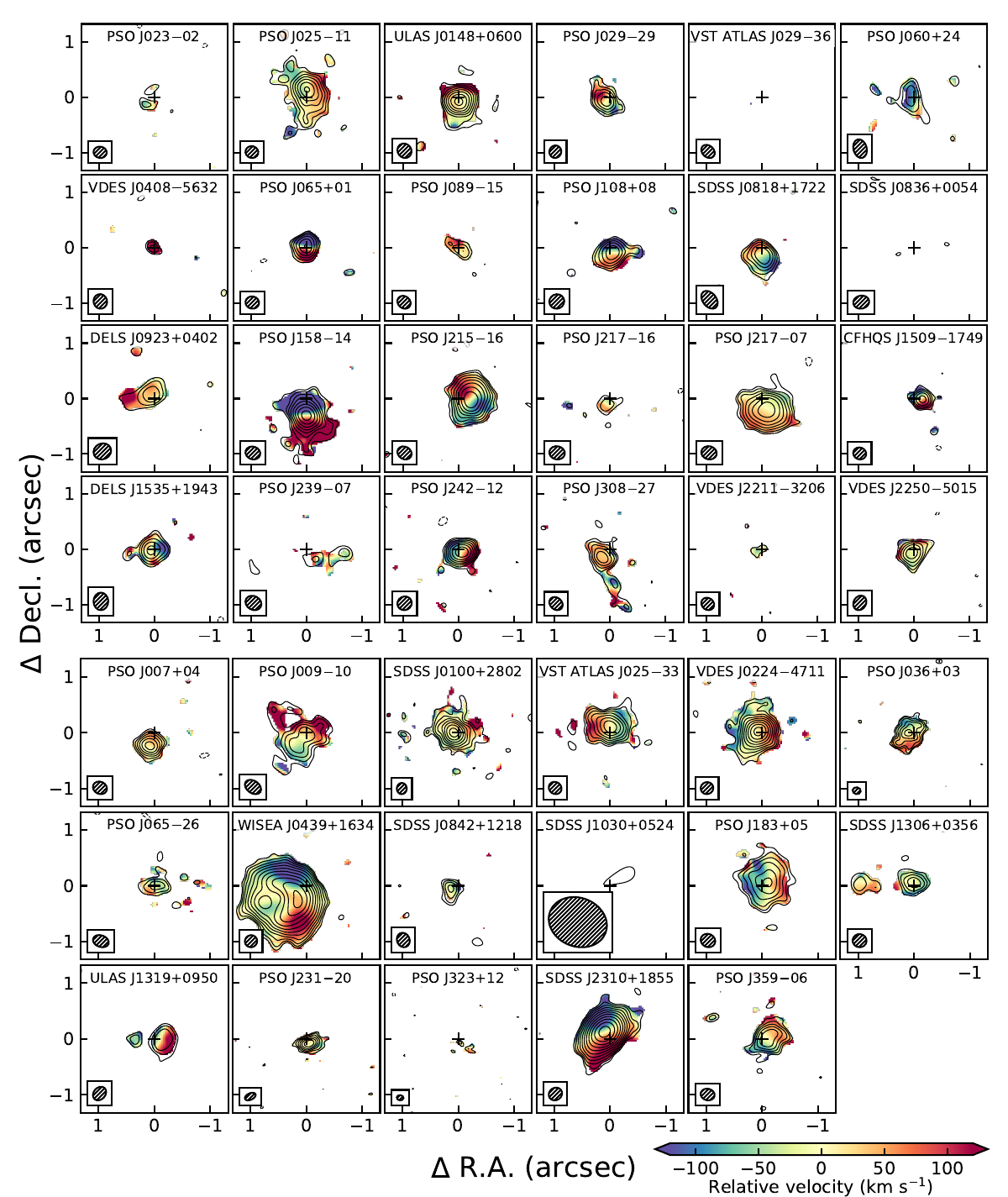}
\caption{Velocity field of the \CII\ emission line subdivided by the 24 galaxies newly observed as part of this survey and the 17 galaxies with archival data. The velocity scale is shown in the bottom right, and is uniform across the sample. We note that some quasar host galaxies (e.g., PSO~J158$-$14 and SDSS~J2310+1855) have velocities extending beyond this range, but for ease of comparison with the rest of the sample, we opted to plot them on the same scale. The contours start at 3$\sigma$ and increase in powers of $\sqrt{2}$. The beam of the observations is shown in the bottom left insert of each panel.
\label{fig:velmap}}
\end{figure*}

\begin{deluxetable*}{lccccccc}
\tablecaption{Kinematics and morphological properties of the sample
\label{tab:kin}}
\tabletypesize{\small}
\tablehead{
\colhead{QSO} &
\colhead{Morphology} &
\colhead{$v_{\rm rot}$} &
\colhead{$\sigma_v$} &
\colhead{$\alpha$} &
\colhead{$i^{\rm a}$} &
\colhead{$M_{\rm dyn}$} &
\colhead{reduced $\chi^2$}\\
\colhead{} &
\colhead{} &
\colhead{(\kms)} &
\colhead{(\kms)} &
\colhead{\degr} &
\colhead{\degr} &
\colhead{$10^{10} M_\odot$} &
\colhead{}
}
\startdata
PSO J023$-$02 & N & \nodata & \nodata & \nodata & \nodata &   1.3 $\pm$   1.5 & \nodata\\
PSO J025$-$11 & Dm & \nodata($>$ 79) & \nodata & \nodata & \nodata & 7$_{-5}^{+7}$ & \nodata\\
ULAS J0148+0600 & B & 110$_{-30}^{+90}$ & 132.7 $\pm$   2.0 & 79 $\pm$ 9 & 20$_{-9}^{+7}$ & 13$_{-11}^{+14}$ & 1.18\\
PSO J029$-$29 & B & 152 $\pm$ 15 & 176 $\pm$ 8 & 40 $\pm$ 3 &  53.6 $\pm$   3.2 & 10$_{-9}^{+11}$ & 0.91\\
VST ATLAS J029$-$36 & \nodata & \nodata & \nodata & \nodata & \nodata & \nodata & \nodata\\
PSO J060+24 & N & \nodata & \nodata & \nodata & \nodata & 4$_{-3}^{+4}$ & \nodata\\
VDES J0408$-$5632 & N & \nodata & \nodata & \nodata & \nodata & 2.0$_{-  2.1}^{+  3.1}$ & \nodata\\
PSO J065+01 & $^*$D & $>$640($>$252) & 145$_{-11}^{+14}$ & 186.7 $\pm$   2.5 & $<$36 & $>$4.9$^{\rm b}$ & 1.07\\
PSO J089$-$15 & N & \nodata & \nodata & \nodata & \nodata & 1.6$_{-  1.4}^{+  1.7}$ & \nodata\\
PSO J108+08 & $^*$Dm & $>$470($>$147) & 139 $\pm$ 11 & 191 $\pm$ 4 & $<$27 & $>$4.0$^{\rm b}$ & 1.10\\
SDSS J0818+1722 & $^*$D & $>$370($>$101) & 107$_{-11}^{+18}$ & 44 $\pm$ 4 & $<$32 & $>$2.0$^{\rm b}$ & 1.16\\
SDSS J0836+0054 & \nodata & \nodata & \nodata & \nodata & \nodata & \nodata & \nodata\\
DELS J0923+0402 & N & \nodata & \nodata & \nodata & \nodata & 5$_{-4}^{+5}$ & \nodata\\
PSO J158$-$14 & Dd & 376 $\pm$ 8($>$238) & 201.1 $\pm$   2.9 & 195.4 $\pm$   0.9 &  47.1 $\pm$   1.1 & 38$_{-5}^{+4}$$^{\rm b}$ & 1.51\\
PSO J215$-$16 & $^*$Dd & $>$580($>$113) & 125.8 $\pm$   2.0 &  43.6 $\pm$   0.9 & $<$16 & $>$4.0$^{\rm b}$ & 1.80\\
PSO J217$-$16 & N & \nodata & \nodata & \nodata & \nodata & 11$_{-10}^{+13}$ & \nodata\\
PSO J217$-$07 & $^*$D & $>$270($>$ 59) &  70.6 $\pm$   1.2 & 151.5 $\pm$   1.9 & $<$13 & $>$1.3$^{\rm b}$ & 1.38\\
CFHQS J1509$-$1749 & D & 680 $\pm$ 210($>$159) & 240$_{-50}^{+70}$ & 189$_{-7}^{+13}$ & 36$_{-15}^{+11}$ & 20 $\pm$ 13$^{\rm b}$ & 1.13\\
DELS J1535+1943 & Bm & 110 $\pm$ 40 & 261$_{-19}^{+24}$ & 97 $\pm$ 7 & 53 $\pm$ 8 & 9 $\pm$ 9 & 1.00\\
PSO J239$-$07 & N & \nodata & \nodata & \nodata & \nodata & 18$_{-16}^{+20}$ & \nodata\\
PSO J242$-$12 & $^*$D & 450$_{-60}^{+90}$($>$140) & 120 $\pm$ 5 & 248.2 $\pm$   2.1 & 28 $\pm$ 5 & 8.2$_{-  2.6}^{+  3.8}$$^{\rm b}$ & 1.22\\
PSO J308$-$27 & Bm & 22 $\pm$ 10 & 89$_{-3}^{+4}$ & 199$_{-5}^{+6}$ & 48$_{-5}^{+4}$ & 21$_{-17}^{+21}$ & 1.00\\
VDES J2211$-$3206 & N & \nodata & \nodata & \nodata & \nodata & 4$_{-4}^{+5}$ & \nodata\\
VDES J2250$-$5015 & B & 110$_{-50}^{+150}$ & 113 $\pm$ 5 & 302 $\pm$ 15 & 22 $\pm$ 14 & 3.2$_{-  2.7}^{+  3.3}$ & 1.05\\
\tableline
PSO J007+04 & B & $>$110 & 136 $\pm$ 4 & 328 $\pm$ 11 & $<$25 & 3.4$_{-  2.8}^{+  3.5}$ & 1.02\\
PSO J009$-$10 & Dm & \nodata($>$152) & \nodata & \nodata & \nodata & 25$_{-20}^{+25}$ & \nodata\\
SDSS J0100+2802 & Bm & \nodata & 164.5 $\pm$   1.6 & \nodata & \nodata & 15$_{-12}^{+15}$ & \nodata\\
VST ATLAS J025$-$33 & D & 145 $\pm$ 8($>$112) & 108.6 $\pm$   2.3 &  81.5 $\pm$   1.7 &  40.7 $\pm$   2.1 & 6.5$_{-  1.1}^{+  1.4}$$^{\rm b}$ & 1.21\\
VDES J0224$-$4711 & D & $>$280($>$ 96) & 127.6 $\pm$   1.4 & 275.9 $\pm$   1.8 & $<$15 & $>$4.4$^{\rm b}$ & 1.42\\
PSO J036+03 & $^*$Dd & 168$_{-26}^{+50}$($>$ 83) &  62.3 $\pm$   1.3 & 188.7 $\pm$   2.4 & 26 $\pm$ 6 & 2.3$_{-  0.7}^{+  1.3}$$^{\rm b}$ & 1.79\\
PSO J065$-$26 & N & \nodata & \nodata & \nodata & \nodata & 9$_{-8}^{+10}$ & \nodata\\
WISEA J0439+1634 & Dl & \nodata($>$115) & \nodata & \nodata & \nodata & \nodata & \nodata\\
SDSS J0842+1218 & N & \nodata & \nodata & \nodata & \nodata & 8$_{-7}^{+9}$ & \nodata\\
SDSS J1030+0524 & N & \nodata & \nodata & \nodata & \nodata & \nodata & \nodata\\
PSO J183+05 & $^*$D & $>$470($>$100) & 140 $\pm$ 5 & 260.3 $\pm$   2.7 & $<$15 & $>$7.0$^{\rm b}$ & 1.18\\
SDSS J1306+0356 & Bm & 10$_{-8}^{+139}$ & 113 $\pm$ 10 & 271$_{-19}^{+166}$ & 44$_{-39}^{+10}$ & 1.6$_{-  1.4}^{+  1.7}$ & 0.90\\
ULAS J1319+0950 & Dm & \nodata($>$124) & \nodata & \nodata & \nodata & 22$_{-19}^{+24}$ & \nodata\\
PSO J231$-$20 & B & 34 $\pm$ 9 & 113.0 $\pm$   2.9 & 84$_{-5}^{+4}$ & 46 $\pm$ 3 & 2.7$_{-  2.3}^{+  3.0}$ & 1.16\\
PSO J323+12 & N & \nodata & \nodata & \nodata & \nodata &   2.8 $\pm$   3.1 & \nodata\\
SDSS J2310+1855 & $^*$Dd & $>$750($>$142) &  89.2 $\pm$   0.5 & 200.49 $\pm$   0.27 & $<$12 & $>$2.9$^{\rm b}$ & 4.57\\
PSO J359$-$06 & D & 126 $\pm$ 8($>$104) & 111 $\pm$ 5 & 316.3 $\pm$   2.9 &  54.2 $\pm$   2.5 & 4.8$_{-  0.8}^{+  1.3}$$^{\rm b}$ & 1.21
\enddata
\tablecomments{The morphological abbreviations are: B--dispersion dominated, D--rotation dominated and N--too low S/N to determine. In addition, the lower case letters describe further characteristics: d--disturbed, m--merger and l--lensed. The asterisk denotes those rotation-dominated galaxies that further satisfy $v_{\rm rot} / \sigma_v > 3$. The value in parenthesis is the strict lower limit to the rotational velocity under the assumption that all of the velocity is perpendicular to the plane of the sky. $^{\rm a}$Inclination uncertainties are the uncertainties derived from the thin disk model fitting and do not include uncertainties from the model assumption. These uncertainties are at least 15\degr. $^{\rm b}$Mass estimates are based on Equation \ref{eq:mdyn}; for the other quasar host galaxies  Equation \ref{eq:mdyn2} is used.}
\end{deluxetable*}

Besides providing a measurement of the \CII\ line flux density, the resolved \CII\ line emission can also be used to study the kinematics of the galaxy. The \CII\ emission line is particularly well suited for studying kinematics, because it is one of the brightest emission lines in the FIR for these galaxies \citep[e.g.,][]{Novak2019}, and the line is optically thin throughout most of the galaxy \citep[e.g.,][]{Goldsmith2012,Guevara2020,Decarli2025}. The latter ensures that the kinematics is weighted by the density of the ionized carbon, which can exist in a range of gas phases. In Figure~\ref{fig:velmap} we show the velocity field of the \CII\ emission line for all of the quasars in the sample. For the velocity field, we adopt the same method as described in \citetalias{Neeleman2021}. In short, we obtain the velocity field from the velocity centroid of a Gaussian fit to the spectrum at each point in the map. As shown in Appendix C of \citetalias{Neeleman2021}, at this resolution and sensitivity, the kinematics derived from this method are more robust compared to the kinematics derived from a standard first-moment image, because the latter is affected by the masking used to generate the image.

Consistent with the results of \citetalias{Neeleman2021}, we observe a wide variety in kinematics among the quasar host galaxies. In order to provide a qualitative assessment of these kinematics, we present a morphological classification in Table~\ref{tab:kin}. The classification is initially based on the existence of a velocity gradient in the velocity field of the quasar host galaxy. As we show below using detailed kinematic modeling, this criterion is an effective way to select galaxies that are rotation-dominated \citep[see also][\citetalias{Neeleman2021}]{Burkert2010}.

Out of the 39 quasar host galaxies in the sample detected in \CII\ emission, we find that 18 galaxies show a distinct velocity gradient and are therefore classified as rotation-dominated, D. We find that 9 galaxies lack a velocity gradient and are labeled as dispersion-dominated, B. Finally, the 12 galaxies that do not have sufficient sensitivity (S/N $<$ 10) are classified as unknown, N. Beyond this initial classification, we further classify those galaxies that show a distinct secondary component in a visual inspection of the data as mergers, m, those that show disturbances not accurately fit with a simple disk model (described below) as disturbed, d, and those galaxies that are lensed, l. This classification can help us compare different properties of the sample between these different galaxy types in Section \ref{sec:discussion}.

It is tempting to classify rotation-dominated systems as disk galaxies, but we caution that other galaxy configurations (e.g., merging protogalactic clumps, tidal features, outflow, inflows) could also produce a velocity gradient, thus the classification of rotation-dominated does not necessarily mean that these are disk galaxies. Similarly, dispersion-dominated systems have kinematics similar to bulges or elliptical galaxies. However, gas is not a collisionless fluid, and requires a source of energy in order to maintain such a velocity distribution, such as shock-driven gas heating and/or turbulence \citep{Alatalo2011,Appleton2023}.

We have fitted the \CII\ emission for all of the galaxies that have been classified as either rotation- or dispersion-dominated using the code \texttt{QubeFit} except for WISEA~J0439+1634. We exclude this lensed quasar because of the added complication of the lensing model, and because it was previously modeled \citep{Yue2021}. We use the same thin disk model that is described in \citetalias{Neeleman2021}, and we refer to this paper for more details. In short, the code fits a nine-parameter model thin disk galaxy to the data using a Markov Chain Monte Carlo approach. This provides realistic uncertainties on the fitted parameters that take into account the parameters covariances (e.g., between the inclination and maximum circular rotation velocity), but we note that the derived uncertainties do not include systematic uncertainties due to choice of model. Results from the fitting code are listed in Table~\ref{tab:kin}. Of the 17 rotation-dominated galaxies that were modeled, 14 galaxies converged on a solution. The three galaxies that did not converged all were classified as mergers. Both PSO~J025$-$11 and PSO~J009$-$10 are good examples of quasar host galaxies that show extended \CII\ emission consistent with tidal material ejected during a merger \citep{Fei2025}. Together with ULAS~J1319+0950 \citep{Venemans2020}, these three quasar host galaxies have sufficiently complex \CII\ distributions that prevent the fitting code from isolating the main galaxy.

For the 14 rotation-dominated galaxies that have been successfully fitted, we find that all galaxies have a rotation velocity to velocity dispersion ratio greater than one ($v_{\rm rot}$/$\sigma_v$ $>$ 1) and are therefore accurately classified as rotation-dominated. In addition, nine galaxies have $v_{\rm rot}$/$\sigma_v$ $>$ 3\footnote{We include PSO~J036+03 because higher resolution observations show this galaxy satisfies this criterion \citep{Neeleman2023}.}. This is the more stringent disk criterion used in \citet{Burkert2010} to determine if a galaxy is a disk \citepalias[see also][]{Neeleman2021}, and we have labeled these galaxies with an asterisk in Table~\ref{tab:kin}. Finally, we find that four quasar host galaxies show significant ($>$6$\sigma$) emission in their residual channel maps, indicting that the simple disk model does not capture the full emission structure of \CII. These galaxies are classified as disturbed, Dd. For one of these galaxies, PSO~J036$+$03, the warp was confirmed in higher resolution observations \citep{Neeleman2023}. It is important to note that these four galaxies are among the brightest \CII\ emitters. This could either be an observational bias, because disturbances can only be seen with sufficient S/N, or it could indicate that these disturbances are driving star-formation and therefore increase the \CII\ emission. This is discussed in more detail in Section \ref{sec:pacontkin}.

A caveat about the kinematic fitting is that the rotation velocity estimate strongly depends on the geometry of the \CII\ emission, in particular the thickness of the disk. Varying the thickness from a thin disk to a spherical distribution would imply about a 40 \% variation in the rotation velocity estimate \citep{Walter1997}. We include this uncertainty in our dynamical mass estimate in Section \ref{sec:mdynmbh}. Regardless of the gas distribution, the velocity gradient seen in the velocity field sets the lower limit of the rotation velocity of the gas. This lower limit is calculated as the average of the absolute value of the 5$^{\rm th}$ and 95$^{\rm th}$ percentile of the velocities in the velocity field, and is given in parenthesis in Table \ref{tab:kin} for all of the rotation-dominated galaxies. 

For the nine galaxies that have been classified as dispersion-dominated, B, we find that eight galaxies are consistent with $v_{\rm rot}$/$\sigma_v$ $\lesssim$ 1. For SDSS~J0100+2802, the merger significantly affects the fitting results of the data and therefore the fitting did not converge\footnote{For this source we list the velocity dispersion measurement from the bulge model in Table \ref{tab:kin}.}. In addition to the thin disk model, we have also fitted the dispersion-dominated galaxies using the bulge model described in \citetalias{Neeleman2021}. We apply this model because the disk model can overfit the data especially in more compact sources. We find that the bulge model yields velocity dispersions that are on-average about 10 \kms\ higher than the disk model while providing a similar fit to the data. This suggests that velocity dispersions could be systematically higher in these systems, but this effect is difficult to separate from actual small velocity gradients within these sources. For ease of comparison with the rotation-dominated systems, we therefore report the dispersions from the thin disk model in Table \ref{tab:kin}.

We note that three of the dispersion-dominated galaxies have $v_{\rm rot}$/$\sigma_v$ ratios that are consistent with being greater than 1 within 1$\sigma$. These galaxies all have low inclinations, and could be rotation-dominated systems that are viewed face on. This number is roughly consistent with the assumption that galaxy inclinations are randomly distributed, since modeling of the data (see Section \ref{sec:pacontkin}) shows that below an inclination of $\sim$10\degr\ velocity gradients become difficult to discern. Consistent with the results in \citetalias{Neeleman2021}, we find that the average velocity dispersion of the dispersion-dominated sample (143 \kms) is larger than the rotation-dominated galaxies (120 \kms), although in this sample the difference is not statistically significant because of the large range of velocity dispersions within both sub-samples. If this higher velocity dispersion is driven by turbulence from interactions with outflows or shock-driven gas, not gravitational motion, then the dynamical mass estimates for these systems could be biased high.

\subsection{Companion Galaxies}
\label{sec:companions}

\begin{figure*}[!t]
\includegraphics[width=\textwidth]{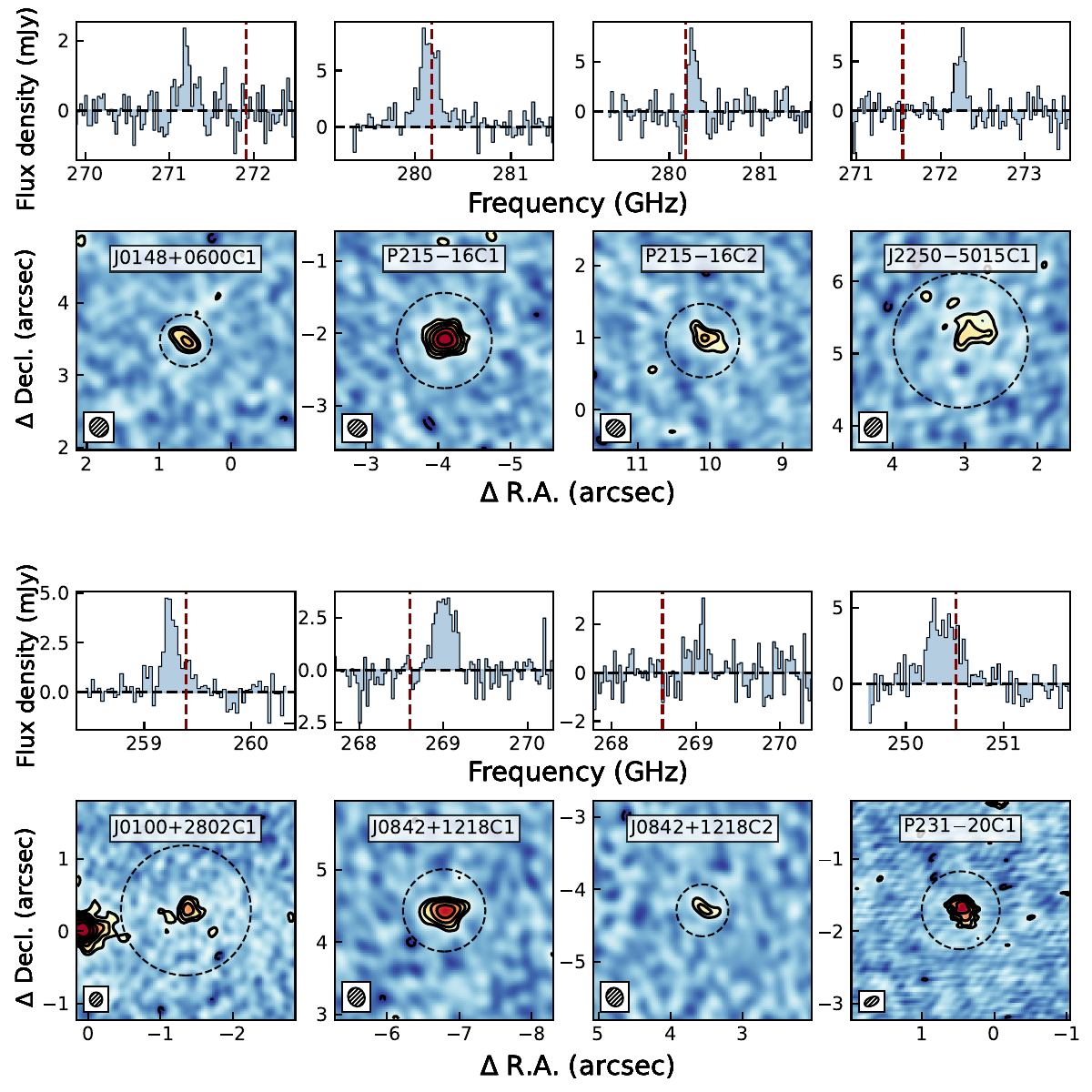}
\caption{Emission lines from galaxies within the field of the quasar. Top and third row shows the flux density spectrum of the emission line as measured from within the circular aperture (dashed circle in the moment-zero images). The red dashed line marks the frequency of the \CII\ emission line from the quasar. The second and fourth row show the moment-zero images of the companion galaxy's emission line created by imaging the emission line over 1.2$\times$ the FWHM of the line. Contours start at 3$\sigma$ and increase in powers of $\sqrt{2}$. Bottom left inset in each figure is the synthesized beam. 
\label{fig:cand}}
\end{figure*}

\begin{deluxetable*}{llrllllcrc}
\tablecaption{Companion sources in the sample
\label{tab:comp}}
\tabletypesize{\footnotesize}
\tablehead{
\colhead{Name} &
\colhead{R.A.$_{\rm CII}$} &
\colhead{Decl.$_{\rm CII}$} &
\colhead{$z_{\rm CII}$} &
\colhead{$F_{\rm CII}$} &
\colhead{FWHM$_{\rm CII}$} &
\colhead{$L_{\rm CII}$} &
\colhead{$b$} &
\colhead{$\Delta V$} &
\colhead{$M_{\rm dyn}$}\\
\colhead{} &
\colhead{(J2000)} &
\colhead{(J2000)} &
\colhead{} &
\colhead{(Jy \kms)} &
\colhead{(\kms)} &
\colhead{($10^9 L_\odot$)} &
\colhead{(kpc)} &
\colhead{(\kms)} &
\colhead{($10^{10} M_\odot$)}
}
\startdata
J0148+0600C1 & 01:48:37.682 & +06:00:23.54 & 6.0080(2) & $0.23\pm0.06$ & $ 99\pm 27$ & $0.22\pm0.05$ & $20.2$&$ 795$&\nodata\\
P215$-$16C1 & 14:20:36.055 & $-$16:02:32.32 & 5.7841(3) & $2.04\pm0.30$ & $225\pm 29$ & $1.82\pm0.27$ & $26.8$&$  34$&$0.37^{+0.19}_{-0.17}$\\
P215$-$16C2 & 14:20:37.039 & $-$16:02:29.26 & 5.7813(2) & $0.86\pm0.20$ & $128\pm 32$ & $0.77\pm0.18$ & $59.0$&$ -93$&\nodata\\
J2250$-$5015C1 & 22:50:02.329 & $-$50:15:36.99 & 5.9811(2) & $1.09\pm0.21$ & $165\pm 35$ & $1.03\pm0.20$ & $34.5$&$-758$&\nodata\\
J0100+2802C1$^{\rm a}$ & 01:00:12.916 & +28:02:26.12 & 6.3309(2) & $0.83\pm0.13$ & $139\pm 21$ & $0.85\pm0.14$ & $7.8$&$ 163$&\nodata\\
J0842+1218C1$^{\rm b}$ & 08:42:28.965 & +12:18:55.02 & 6.0649(4) & $1.09\pm0.19$ & $368\pm 64$ & $1.04\pm0.19$ & $46.1$&$-452$&$1.0^{+0.8}_{-0.5}$\\
J0842+1218C2$^{\rm c}$ & 08:42:29.671 & +12:18:46.29 & 6.06312(16) & $0.17\pm0.06$ & $ 67\pm 27$ & $0.17\pm0.06$ & $31.5$&$-529$&\nodata\\
P231$-$20C1$^{\rm b}$ & 15:26:37.873 & $-$20:50:02.34 & 6.5907(5) & $1.96\pm0.30$ & $431\pm 57$ & $2.13\pm0.32$ & $9.4$&$ 160$&$1.5^{+0.8}_{-0.5}$
\enddata
\tablecomments{The right ascension, declination and frequency are taken from the LineSeeker results. The impact parameter, $b$, is the on-sky distance between the companion galaxy and the quasar at the redshift of the quasar, whereas $\Delta V$, is the velocity difference between the peak emission of the quasar and the companion. Positive $\Delta V$ corresponds to a companion at a higher redshift than the quasar. $^{\rm a}$Galaxy was first analyzed in \citet{Tripodi2024a}, $^{\rm b}$Galaxy was first analyzed in \citet{Decarli2017}, $^{\rm c}$Galaxy was first analyzed in \citet{Neeleman2019}.}
\end{deluxetable*}

The ALMA primary beam at frequencies of $\approx$270 GHz covers a circular region with a radius of about 10$''$  around the quasar host galaxy. In this section, we report on the search for additional line emitters within this region. As we are particularly interested in companion galaxies of the quasar, we focus on only the 2 or 4 GHz ($\sim$1600 and $\sim$3200 \kms, respectively) contiguous spectral range covering the \CII\ emission line from the quasar. The other baseband, which covers a 4 GHz region that is at least 6000 \kms\ offset from the \CII\ emission line redshift of the quasar, was not searched for companion galaxies. 

To perform the search, we used the line-finding algorithm, LineSeeker \citep{Gonzalez-Lopez2017,Gonzalez-Lopez2019}. This search algorithm has been previously used to find emission lines in ALMA data \citep[e.g.,][]{Gonzalez-Lopez2019,Neeleman2025}. We refer the reader to these papers for a detailed description of the code and tests for completeness and fidelity, but, in short, the code convolves the cube in the spectral direction with different kernels and looks for spectral features on these scales. Previous testing with mock spectra found that the code has a greater than 99 \% fidelity for emission lines detected at $>$6$\sigma$ \citep{Neeleman2025}. 

In Table \ref{tab:comp}, we report on all the $>$6$\sigma$ emission lines that were detected in the 41 data cubes of this survey and their position and spectra are shown in Figure \ref{fig:cand}. Of the eight detections, four are in fields of newly observed quasars. The excellent agreement between the frequency of the emission line and the frequency of the \CII\ emission line from the quasar, as well as the relatively small impact parameter (i.e., on-sky separation) strongly suggests these are \CII-emitting companion galaxies to the quasar. We therefore assert that these are all \CII\ companions, but note that it is possible that these are emission lines from galaxies at a different redshift. Previous studies have shown an interloper rate of $\sim$ 1 galaxy per 50 GHz bandwidth searched with ALMA at these frequencies \citep{Decarli2020,Venemans2020,Loiacono2021,Neeleman2025}, therefore we expect at most one to two interlopers in this sample. In particular, P215$-$16C2 shows strong continuum emission at optical wavelengths that suggest a lower redshift origin \citep{Mazzucchelli2017a}. For one of the companion galaxies, J0148+0600C1, the redshift is confirmed though JWST/NIRCam grism observations \citep{Eilers2024,Kashino2026} that reveal [\ion{O}{3}] emission from the companion galaxy. For the four galaxies previously reported, the companion galaxy for J0100$+$2802C1 was discussed in \citep{Tripodi2024a}. Interestingly, we find that this companion galaxy sits within a region of low surface brightness \CII\ emission that extends from the quasar to about 3$''$ to the west and is detected at 5$\sigma$ significance. This low-level emission could be tidally stripped gas from the interaction between the companion and the quasar host galaxy. The remaining three companions, J0842+1218C1, J0842+1218C2 and P231$-$20C1 were first reported in \citet{Decarli2017} and \citet{Neeleman2019}. 

Under the assumption that these are \CII\ emission lines, we have calculated the redshift, $F_{\rm CII}$, FWHM and $L_{\rm CII}$ for all companion galaxies in Table \ref{tab:comp}. We have also included the impact parameter (i.e., the on-sky separation at the redshift of the quasar, $b$) between the quasar and companion galaxies, as well as the velocity difference between the peaks of the \CII\ emission, $\Delta V$. In most cases the velocity difference and on-sky separation between the companion galaxy and quasar host galaxy is sufficiently small that it is likely these systems will eventually merge. Remarkably, none of the systems at lower redshifts than the quasar are seen in absorption in the XSHOOTER quasar spectrum \citep[e.g., as \ion{O}{1} absorbers or higher ionization absorption systems;][]{Davies2023,Sodini2024}.

A full comparison of the expected number of companion \CII\ emitters based on clustering analysis is beyond the scope of this work, but we have compared our results with the recent analysis of \CII\ emitters in the field of 25 high-redshift quasars \citep{Wang2026}. They find a total of 17 \CII\ emitters within their sample, but this number reduces to seven, if we only consider companions within $\Delta V < 1000$ \kms\  and within $b < 65$ kpc, which are the search limits of our survey. If we further exclude the three galaxies that have $\Delta V < 300$ \kms\  and  $b < 8$ kpc, which would have been labeled mergers in this work, then the number of companion \CII\ emitters would have been four for a total of 25 quasars. This ratio is consistent with the number of companions sources that we detect in our sample.

\section{Discussion}
\label{sec:discussion}

In this section, we discuss the ALMA observations of the sample, focusing in particular on relating the different properties to each other and how these differ among the different kinematic classifications of the \CII\ emission line. In Section \ref{sec:extent}, we examine the difference between the extent of the dust emission and \CII\ emission. In Section \ref{sec:ciideficit}, we discuss the \CII\ deficit. In Section \ref{sec:pacontkin}, we introduce a new parameter that compares the position angle of the \CII\ and continuum emission to the position angle found from the kinematic fitting in order to look for non-standard rotation. In Section \ref{sec:mdynmbh} we compare the dynamical mass to the black hole mass for the galaxies in the enlarged XQR30 sample.  

\subsection{Extent of dust and [C II] emission}
\label{sec:extent}

\begin{figure}[!t]
\includegraphics[width=0.47\textwidth]{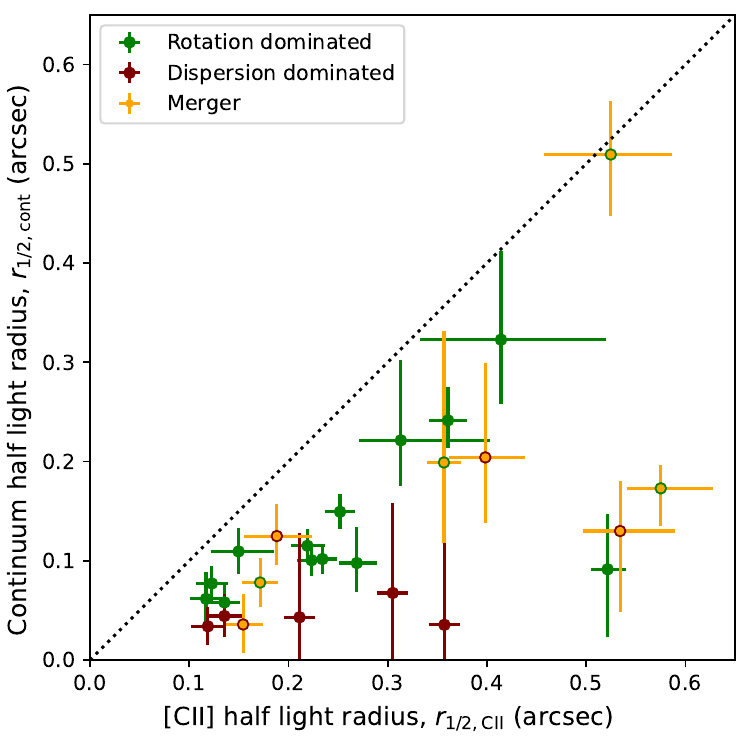}
\caption{Comparison between the extent of the \CII\ and dust continuum emission. The sample, excluding those sources with too low S/N, is sub-divided by their morphological classification. The black dotted line shows the one-to-one correlation. All galaxies have more extended \CII\ emission compared to the dust continuum independent of their morphology. We further find that dispersion-dominated systems have on average smaller continuum half-light radii than rotation-dominated systems.
\label{fig:extent}}
\end{figure}

Previous works have shown that dust emission is more compact than the \CII\ emission in $z \gtrsim 6$ quasars \citep[e.g.,][]{Venemans2020,Wang2024}. This has been attributed to the following two causes. First, dust temperatures are higher in the center of the galaxy  \citep{Walter2022,Meyer2025b}. Since FIR emission scales stronger with temperature than the \CII\ emission, the FIR emission will be relatively brighter in the center of the galaxy than the \CII\ emission \citep{Walter2022}. The resultant \CII\ emission profile will be less steep than the FIR emission profile, and therefore the \CII\ emission will have a larger half-light radius and observed extent. Second, \CII\ emission is expected to trace a range of gas phases, from molecular to ionized gas, this includes the gas that is tidally stripped from the galaxy during recent mergers. The larger extent of the \CII\ emission could be driven by this tidally disturbed gas. In this section, we try to differentiate between these two possible causes by looking at the morphology of the sources. If the second cause dominates, we would expect to see stronger size differences among those galaxies that are undergoing mergers and interactions.

In Figure \ref{fig:extent}, we plot the fitted half-light radii for both the dust and \CII\ emission for our sample of galaxies where we can constrain both measurements. Consistent with previous results  \citep[e.g.,][]{Venemans2020,Wang2024}, all of the galaxies fall below the one-to-one relation between these quantities, indicating that the dust emission is more compact for the galaxies in this sample. To understand the effect of morphology, we have divided the sample by morphological type. As a reminder, the morphological classification is based on the kinematics of the \CII\ emission line, so is mostly independent of the size of the emission. We find that those systems with little velocity gradient across the galaxy (i.e., dispersion-dominated systems) have very compact dust continuum emission compared to rotation-dominated galaxies, yet have a similar extent in their \CII\ emission. This strengthens the assertion made in \citetalias{Neeleman2021} that dispersion-dominated systems are physically distinct from rotation-dominated galaxies. One possible hypothesis that could explain these observations is that dispersion-dominated galaxies are galaxies that exhibit strong quasar feedback. In this scenario, the quasar shock-heats the gas and dust in its immediate surrounding creating turbulence, which prevents the gas from settling into a disk. The heated dust in the center also creates a large dust gradient, which results in more compact dust continuum sizes \citep{Meyer2025b}. Resolved dust temperature measurements would be able to  test this hypothesis.

We also find that the ratio between the continuum size and the \CII\ size is relatively constant with a value of $0.55 \pm 0.17$ for the majority of rotation-dominated galaxies. A similar ratio is also seen in a large fraction of mergers, suggesting that tidally disturbed gas in these mergers is not the primary cause for the increase size in \CII\ compared to the dust continuum size. This hints at an internal origin such as dust temperature gradients, which is consistent with the measured radial dust temperature gradients in $z \gtrsim 6$ quasar host galaxies \citep{Meyer2025b}. Of the three galaxies that have significantly larger \CII\ sizes compared to their dust, two are classified as mergers and do show extended \CII\ emission, PSO~J025$-$11 and PSO~J308$-$27. The remaining galaxy is PSO~J158$-$14. This galaxy shows disturbed, rotation-dominated kinematics, which, as we show in Section \ref{sec:pacontkin}, could be the result of a merger. These observations indicate that tidally disturbed gas can increase the size of the \CII\ emission in galaxies, but the main driver of the observed difference in sizes between the dust continuum and the \CII\ emission is likely internal.

 \subsection{[C II] deficit and Morphology}
\label{sec:ciideficit}

\begin{figure}[!t]
\includegraphics[width=0.47\textwidth]{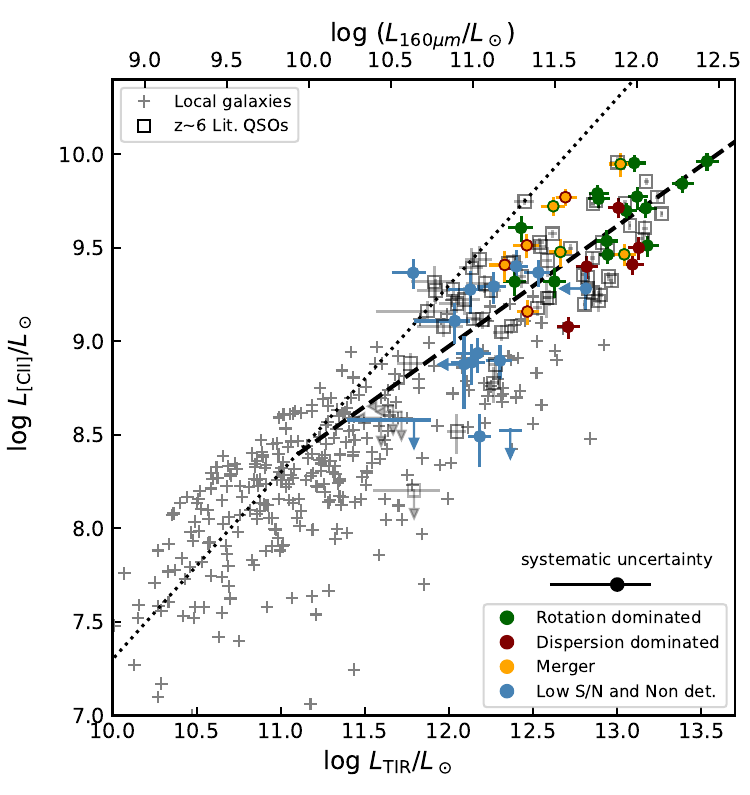}
\caption{\CII\ luminosity versus the total far-infrared luminosity for the quasars in the XQR30 sample. Also shown are a collection of local galaxies \citep{Malhotra2001,Farrah2013,DiazSantos2013}, as well as a literature search for other $z \gtrsim 6$ galaxies \citep[e.g.,][]{Venemans2020,Wang2024}. The dotted line is the linear correlation between these quantities and the dashed line is a fit to the high redshift quasars showing a clear deviation from this linear relation, the \CII\ deficit. For the majority of galaxies the same dust temperature and $\beta$ are used and for these sources the total far-infrared luminosity scales with the monochromatic 160$\mu$m continuum emission, $L_{\rm 160\mu m}$. We find that dispersion-dominated galaxies have on-average larger \CII\ deficits compared to mergers and rotation-dominated galaxies.
\label{fig:ciifir}}
\end{figure}

Many previous papers have noted that the one-to-one correlation between \CII\ luminosity and total far-infrared luminosity breaks down for galaxies with strong $L_{\rm TIR}$ \citep[e.g.,][]{Malhotra2001,Carilli2013}. These galaxies exhibit less $L_{\rm [CII]}$  than expected based on a simple extrapolation of the one-to-one relation. Many hypotheses have been put forth for the cause of this \CII\ deficit \citep{Malhotra2001,Sargsyan2012,Appleton2013,Goicoechea2015,Decarli2025}, but a likely contributor to the \CII\ deficit seen in $z \sim 6$ quasar host galaxies is that increased dust temperatures in the centers of these hosts are driving the increased total infrared luminosity, whilst only mildly increasing the \CII\ luminosity. This results in the centers of galaxies showing a larger \CII\ deficit than the outskirts, which has been observed in a range of galaxies both at high and low redshift \citep[e.g.,][]{DiazSantos2013,DiazSantos2017,Gullberg2018,Rybak2019}, and has been observed in detail using spatially resolved observations of high redshift quasar host galaxies \citep[e.g.,][]{Walter2022,Neeleman2023}.

In Figure \ref{fig:ciifir}, we plot $L_{\rm [CII]}$ as a function of $L_{\rm TIR}$ for the quasar host galaxies of our sample. These galaxies are systematically below the linear correlation, as shown by the fit to the data. This is in agreement with previous observations of $z \gtrsim 6$ quasar host galaxies \citep{Venemans2020,Wang2024}. We further subdivide the sample based on the morphology of the quasar host. Here we find that dispersion-dominated galaxies show the largest \CII\ deficit. This is consistent with the result from Section \ref{sec:extent}, which show that dispersion-dominated host galaxies have very compact dust emission. These galaxies therefore have the largest contribution from the central regions, which show the largest deficits. Unlike the dispersion-dominated systems, both rotation-dominated systems as well as mergers show slightly milder \CII\ deficits. This can be explained by their larger sizes, since emission further from the center shows less of a \CII\ deficit. We caution, however, that there are large systematic uncertainties involved in measuring $L_{\rm TIR}$ and only resolved, multi-wavelength observations of the dust continuum can help alleviate these systematics \citep{Tripodi2023,Tripodi2024b,Meyer2025b}.

\subsection{Evidence for Perturbed Kinematics}
\label{sec:pacontkin}

\begin{figure}[!t]
\includegraphics[width=0.45\textwidth]{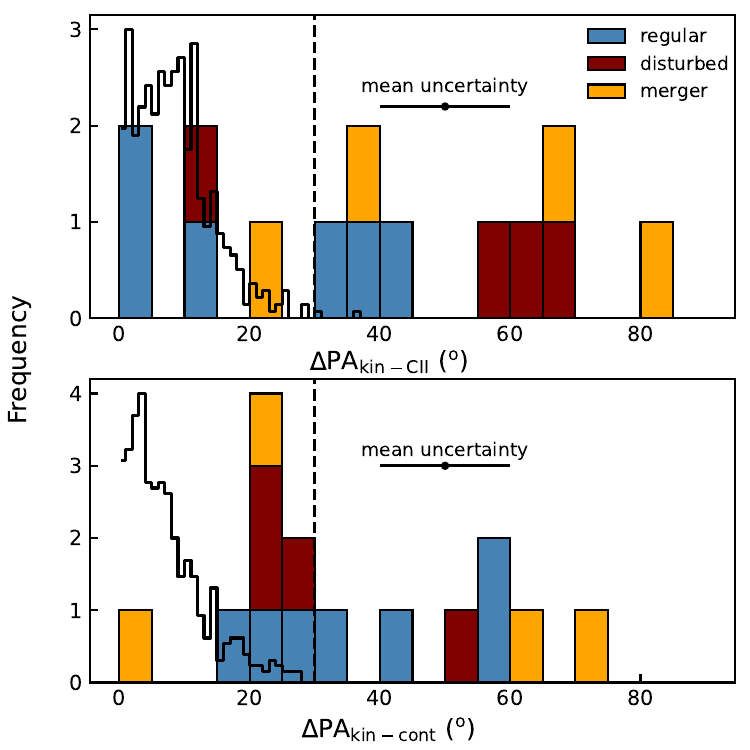}
\caption{Histograms of the kinematic position angle alignment parameters, $\Delta$PA$_{\rm kin-CII}$ and $\Delta$PA$_{\rm kin-cont}$. Only quasars with an ellipticity greater than 0.2 are shown in order to have reliable estimates of the position angle of the emission. The sample is subdivided by their morphology.  The black line shows the distribution of the kinematic position angle alignment parameters for a sample of mock disk galaxies at similar resolution and S/N as the data, showing that $\Delta$PA$_{\rm kin-CII}$ and $\Delta$PA$_{\rm kin-cont}$ are below 30\degr\ (dashed line) for pure disk rotation. This figure shows that $\Delta$PA$_{\rm kin-CII}$ is a good indicator of potential disk disturbances without performing a full kinematic analysis. 
\label{fig:pahist}}
\end{figure}

In this section, we explore if we can identify galaxies that show perturbations beyond simple disk rotation from differences between the kinematic major axis and the major axis of the \CII\ and dust continuum emission. Classification of the kinematics of the \CII\ emission line (Section \ref{sec:ana_kin}) shows that 18 out of 41 galaxies have a clear velocity gradient, and are therefore classified as rotation-dominated galaxies. Out of these 18 galaxies, we find that four of the galaxies are incompatible with the simple, uniform disk model used for fitting the emission. These four galaxies are also among the galaxies with the highest S/N observations; suggesting that deviations can only be observed at high S/N, and sensitive observations of quasar host galaxies are needed to reveal details of the kinematics that go beyond uniform disk rotation. 

There are many possible causes for the \CII\ emission to deviate from a uniform disk model: \CII\ could be tracing inflows or outflows, the galaxy could have a bar or spiral arms, the galaxy could have a physical warp or tidal structure from a recent merger, the \CII\ could trace clumpy star formation, etc. However, distinguishing between these different scenarios will require higher resolution observations. Nevertheless, we can try to quantify the number of galaxies that show potential deviations from a uniform disk model using the current observations by comparing the kinematics of the galaxies with the morphology of the emission. Although a full analysis of the kinematics would reveal any deviations from uniform disk emission, this can only be done for those galaxies with sufficient S/N. In this section, we therefore introduce a statistic that can be reliably obtained for the majority of galaxies without performing a detailed analysis of the kinematics, the kinematic position angle alignment parameter, $\Delta$PA$_{\rm kin-CII}$. 

We define $\Delta$PA$_{\rm kin-CII}$ as the absolute deviation between the positional angle of the kinematic axis, which is defined as the position angle of the gradient in the velocity field at the peak of the \CII\ emission, and the position angle of the morphological major axis of the \CII\ emission (Section \ref{sec:ana_cii}). A similar statistic can be defined for the dust continuum emission, $\Delta$PA$_{\rm kin-cont}$. Both $\Delta$PA$_{\rm kin-CII}$ and $\Delta$PA$_{\rm kin-cont}$ can vary between 0 and 90\degr, where 0\degr\ corresponds to the case where the kinematic axis is aligned with the major axis of the \CII\ or dust continuum emission, which is the expected geometry for a rotating disk galaxy. Therefore, any significant difference from 0\degr\ would indicate kinematics that are inconsistent with uniform disk rotation. To model this, we have generated a sample of 500 mock observations of pure disk galaxies, matching the distribution of S/N, resolution and inclination of the data using \texttt{QubeFit}. We find that $\Delta$PA$_{\rm kin-CII}$ and $\Delta$PA$_{\rm kin-cont}$ rarely exceeds 30\degr\ (Figure \ref{fig:pahist}), and therefore set this as our threshold for determining if the emission shows irregular disk rotation.

In Figure \ref{fig:pahist}, we plot the histogram of $\Delta$PA$_{\rm kin-CII}$ subdivided into those galaxies that were classified as regular rotation-dominated galaxies and those that show either evidence of mergers or disturbances. We only include those rotation-dominated galaxies with an ellipticity greater than 0.2, because more circular sources have ambiguous position angle measurements. {We find that the majority of galaxies classified as mergers or disturbed kinematics have $\Delta$PA$_{\rm kin-CII} > 30\degr$, consistent with the expectation that these should have a misaligned kinematic axis. A notable exception is PSO~J158$-$14. This galaxy has excess emission along the major axis, which the $\Delta$PA$_{\rm kin-CII}$ statistic cannot detect. This quasar has a nearby Ly$\alpha$ emitter \citep{Durovcikova2025}, and it is possible that a recent interaction could have triggered a UV-luminous phase of the quasar and resulted in disturbed kinematics. In general, our $\Delta$PA$_{\rm kin-CII}$ distribution is substantially skewed to higher values compared to the expectation of pure rotation. This suggests that some of the rotation-dominated systems classified as `regular' show disturbed kinematics that remains difficult to identify at this S/N, and that disturbances and mergers are common among high redshift quasar host galaxies.

Unlike $\Delta$PA$_{\rm kin-CII}$, the $\Delta$PA$_{\rm kin-cont}$ statistic reveals that the kinematic axis is much better aligned with the dust continuum major axis. $\Delta$PA$_{\rm kin-cont}$ is therefore a much less effective tool to look for disturbances. The likely reason for this is that the dust continuum is more compact (Section \ref{sec:extent}), and therefore is closer aligned with the central velocity gradient. This suggests that in galaxies where the kinematic axis is aligned with the dust but not the \CII\ emission, the disturbances are caused by external forces such as mergers and inflows \citep[e.g., PSO~J036+03;][]{Neeleman2023}. In contrast, galaxies that show misalignments in both the dust and \CII, the disturbances are caused by internal forces, such as bars, one such an example is SDSS~J2310+1855. We finally note that this kinematic analysis suggests that a large fraction of quasar host galaxies, but not all \citep[e.g.,][]{Walter2022}, will show deviations from a uniform disk model in higher resolution observations. Therefore, sensitive high-resolution \CII\ observations have the exciting potential to uncover the detailed structure of the interstellar medium of these high redshift galaxies.

\subsection{Dynamical Mass Estimates}
\label{sec:mdynmbh}

\begin{figure*}[!t]
\includegraphics[width=0.49\textwidth]{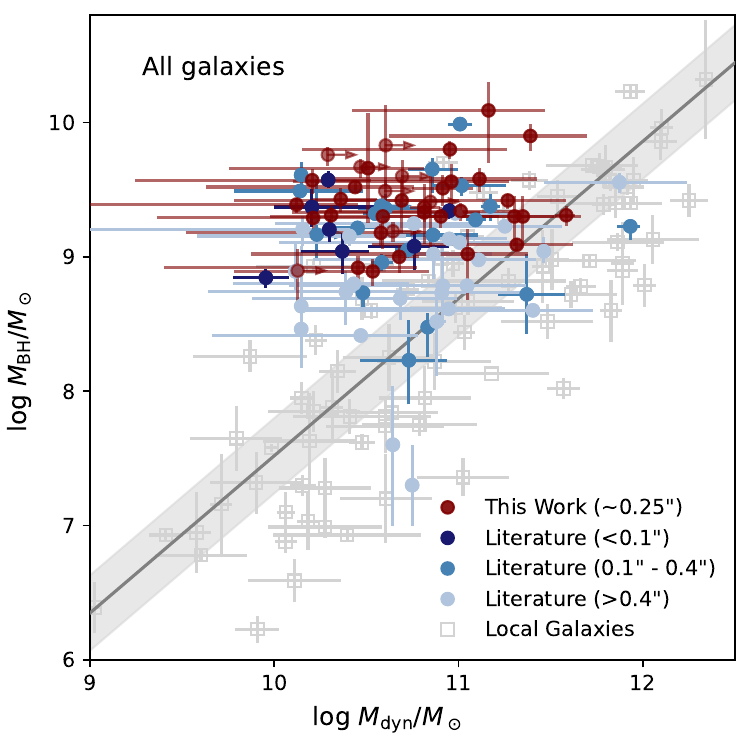}
\includegraphics[width=0.49\textwidth]{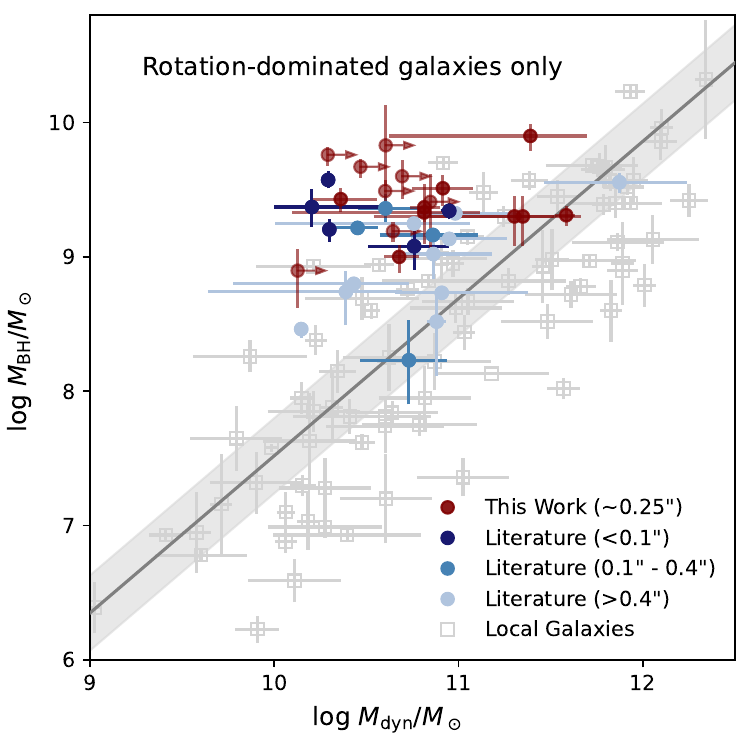}
\caption{Correlation between the mass of the SMBH and the mass of the galaxy. The gray points are local galaxies and the black line and shaded region show the observed correlation and scatter for local galaxies. The blue points are estimates for $z \gtrsim 6$ quasars from the literature (see Section \ref{sec:mdynmbh} for references), shaded by the resolution of the millimeter observations. The red points are the quasar host galaxies of the sample in this paper. In the right panel, we only show high redshift quasar host galaxies that are classified as rotation-dominated. Uncertainties on the black hole mass have a systematic uncertainty of $\sim$0.55 dex due to the scaling relation used. We find that our sample is fully consistent with previous results, and that low-resolution observations tend to overestimate the dynamical mass because of an overestimate of the extent of the \CII\ emission.
\label{fig:mbhmdyn}}
\end{figure*}

\begin{figure}[!t]
\includegraphics[width=0.47\textwidth]{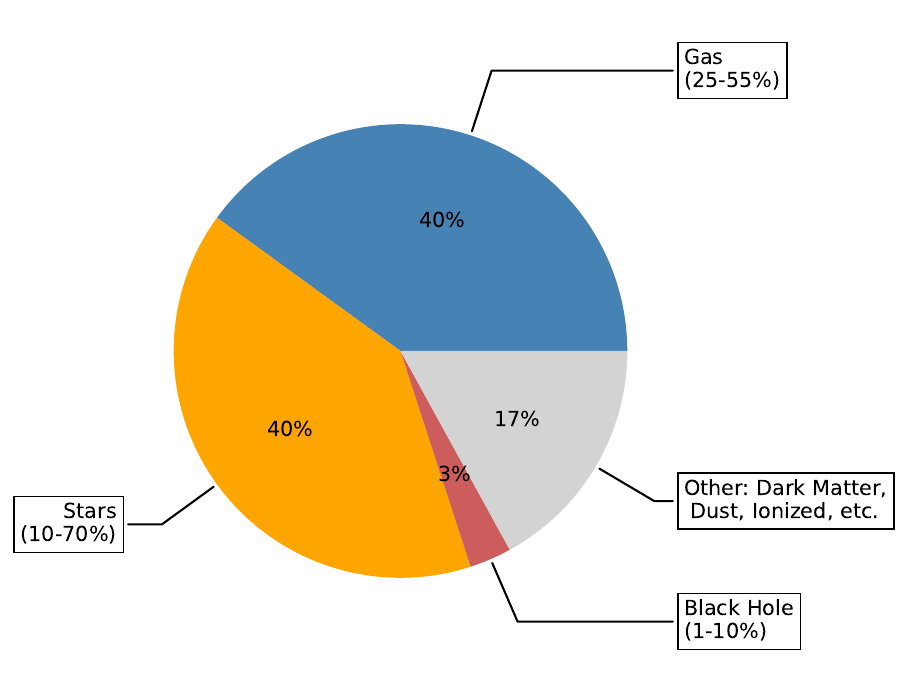}
\caption{Typical mass budget of $z \gtrsim 6$ quasar host galaxies within twice the \CII\ half-light radius ($\sim$ 2 kpc). The gas mass is estimated from the dust continuum, the stellar mass from JWST imaging and the mass of the SMBH is estimated from the quasar spectrum. This rough estimate reveals that the dynamical mass estimates are consistent with the estimates for the mass of gas, stars and the SMBH even leaving some room for other mass components. 
\label{fig:massbudget}}
\end{figure}

Using the \CII\ emission line, we can provide an estimate of the dynamical mass of the galaxy. The procedure is described in detail in \citetalias{Neeleman2021} including an in-depth analysis of the systematic uncertainties involved in these estimations. In this paper, we adopt a similar approach as \citetalias{Neeleman2021}. For the 14 rotation-dominated galaxies that were modeled with a thin disk model, the dynamical mass is calculated using equations 10 and 13 in \citetalias{Neeleman2021}, which takes into account that the galaxies are likely pressure supported.
\begin{equation}
M_{\rm dyn}(R) = 2.33 \times 10^5 (v_{\rm rot}^2 + 3.4 \sigma_v^2) R
\label{eq:mdyn}
\end{equation}
 For eight of these rotation-dominated galaxies, we have only lower limits on the rotation velocity, and for these galaxies we conservatively take the limit derived from the velocity gradient. For the other galaxies in the sample, we use equation 15 in \citetalias{Neeleman2021}. 
 \begin{equation}
\left(\frac{M_{\rm dyn}}{M_\odot}\right) = (1.9^{+1.5}_{-0.8} (^{+1.1}_{-1.3})) \times 10^5~\left(\frac{\text{FWHM}}{\text{km s}^{-1}}\right)^2\left(\frac{R}{\text{kpc}}\right).
\label{eq:mdyn2}
\end{equation}
 This equation relies on the size of the emitting region as well as the FWHM of the emission line. Only the two quasars not detected in \CII, VST~ATLAS~J029$-$36 and SDSS~J0836+0054, the quasar observed at low resolution, SDSS~J1030+0524 and the lensed quasar, WISEA~J0439+1634, have no constraints on their dynamical mass. All dynamical mass estimates are listed in Table \ref{tab:kin}, and are calculated within twice the effective radii of the \CII\ emission. We also provide the dynamical masses for the three companion galaxies that have sufficient signal to reliably constrain their size. These are listed in Table \ref{tab:comp}. 

A final word of caution on dynamical mass estimates, the uncertainties on these dynamical mass estimates are large, and are derived under the assumption that the kinematics of the \CII\ emission line are dominated by the gravitational potential of the galaxy. This can easily break down in these galaxies due to recent mergers and/or significant feedback from the quasar. In most cases, these effects will lower the dynamical mass estimates, and the quoted values can therefore be taken as upper limits to the dynamical mass.

The dynamical mass estimates of quasar host galaxies are often compared to other properties of the galaxy as well as of the quasar. One comparison of particular interest is the comparison with the mass of the SMBH, which is estimated from the rest-frame UV/optical light from the quasar, often the \ion{Mg}{2} line. Black hole mass estimates were previously derived for all of these quasars \citep{Mazzucchelli2017a,Mazzucchelli2023,DOdorico2023} and have a systematic uncertainty of $\sim$0.55 dex due to the scaling relation used. We show a plot of the black hole mass versus dynamical mass in Figure \ref{fig:mbhmdyn}. As this figure shows, these quasar host galaxies fall within the same parameter space as previous observations of high redshift quasar host galaxies \citep{Izumi2018,Izumi2021,Neeleman2021,Walter2022,Wang2024,Neeleman2023,Yue2024,Wilde2026}. Here, we have sub-divided the literature sample by the resolution of the \CII\ emission line observations, because higher resolution will provide more accurate estimates of the dynamical mass. Indeed, we see an observational trend that lower resolution data are closer to the local relation. We expect this is partly due to an overestimate of the galaxy size in these lower resolution data, which increases the dynamical mass estimate. In the right panel of Figure \ref{fig:mbhmdyn} we only show rotation-dominated galaxies, because these morphologies are most strongly associated with systems whose kinematics are dominated by the gravitational potential of the galaxy. In both cases, our results are fully consistent with the literature sample.

Many previous works \citep[e.g.,][]{Wang2013,Pensabene2020,Neeleman2021} have discussed the offset between high redshift quasar host galaxies and the correlation between black hole mass and galaxy mass seen in the local universe \citep{Kormendy2013}. This offset is in large part driven by selection effects \citep{Lauer2007,Li2022} and possible systematics in the galaxy mass determination (\citetalias{Neeleman2021}), and should not be interpreted as an indication that all high-redshift quasar host galaxies (or galaxies) have SMBHs that are over-massive compared to their galaxy mass \citep[see e.g.,][for counterexamples]{Izumi2018,Silverman2025}. However, it is interesting that the majority of luminous quasars reside in galaxies with relatively low dynamical masses, such that their black hole mass makes up on-average 3 \%, and as high as $>$10 \%, of their total mass within twice the effective radius of the \CII\ emission. Despite this high mass fraction, searches for signatures of the black hole on the gas kinematics (the black hole's sphere of influence), have only yielded non-detections \citep{Walter2022,Meyer2023}. These non-detections suggest that either the mass of the black hole is overestimated in these galaxies, or the \CII\ kinematics are not driven by the gravitational potential of the black hole. Distinguishing between these scenarios requires more precise black hole mass estimates, possible via JWST measurements of additional rest-frame optical tracers such H$\beta$ \citep{Yang2023, Yue2024}, and larger samples of high-resolution observations of the \CII\ kinematics to better characterize the overall kinematics within these systems. 

We finally try to account for the full mass budget of the quasar host galaxy by including estimates of the black hole mass, gas mass and stellar mass for these systems. The gas mass was estimated from the dust continuum measurement following \citet{Venemans2016}. To be specific, we calculated the dust mass under the assumption that the dust continuum emission follows a modified black body with a dust temperature of 46 K and an emissivity index of 1.6 \citep{Beelen2006}. We then use a dust-to-gas ratio of 100 to convert the dust mass to a gas mass\footnote{In practice, this is roughly equivalent to rescaling $L_{\rm TIR}$ by a conversion factor of $\alpha_{\rm TIR} = 290~L_\odot / M_\odot$.}. Under these assumptions, we find that the gas masses range between $10^9$ and $10^{11}~M_\odot$. As mentioned in Section \ref{sec:ana_cont}, these parameters are not constrained with the current single continuum measurement, and any gas measurements will have significant systematic uncertainties. We therefore opted to not report on individual gas mass measurements for these systems. We do however compare this to other scaling relations that compare the \CII\ emission to the molecular gas \citep[e.g.,][]{Zanella2018,Salvestrini2025}, and we find that our gas mass estimates are consistent with the scaling relation reported in \citet{Salvestrini2025}. The derived gas masses make up, on-average, about $40 \pm 15$ \% of the dynamical mass. 

Despite the advent of JWST, the stellar mass remains difficult to constrain, and only a handful of galaxies have stellar mass determination \citep{Ding2023,Yue2023,Stone2024,Yue2024,Ding2025,Marshall2025a,Marshall2025b,Onoue2025,Stone2025}, although this number is rapidly increasing. These measurements suggests that the stellar masses are comparable to the gas mass, and we therefore assume a mass fraction of $40 \pm 30$ \%, where the large uncertainty is due to the limited sample size. The average mass budget for luminous quasar host galaxies at $z \sim 6$ is shown in Figure \ref{fig:massbudget}. We find that the average dynamical mass is compatible with measurements of the black hole, gas and stellar mass, even leaving some room for additional mass components such as dark matter, dust and atomic/ionized gas. This rough mass budget indicates that the dynamical mass estimates derived from \CII\ are at least consistent with the mass estimates from stars, gas and the black hole. 

\section{Summary and Conclusions}
\label{sec:conclusions}
In this paper, we present the \CII\ and dust continuum emission for 40 quasars that make up the bulk of the enlarged XQR30 Sample \citep{DOdorico2023}, as well as one additional quasar PSO~J215$-$16. New observations are reported for 24 quasars, which are supplemented with a homogenized reduction of archival data for 17 additional quasars. Using this sample of 41 quasars, we find the following results.

\begin{itemize}
\item We detect \CII\ emission from the host galaxy of 39 quasars, resulting in a detection rate of 95~\%. We also detect dust continuum emission from 39 quasar host galaxies, but the two galaxies that are not detected in dust are different from the two not detected in \CII\ emission. The resulting \CII\ luminosities range between (0.3 and 9.2) $\times 10^9 L_\odot$ and dust continuum flux densities between 0.22 and 16.3 mJy. Results are tabulated in Tables \ref{tab:cii} and \ref{tab:cont}. 

\item We classify the morphology of the 39 galaxies that are resolved in \CII, and find that 18 galaxies show kinematics that are rotation-dominated, 9 galaxies that are dispersion-dominated and 12 galaxies that have too low S/N to constrain their kinematics. We further sub-classify the sample based on evidence of mergers, lensing or disturbed kinematics. All dispersion-dominated and rotation-dominated galaxies were further modeled using the fitting code \texttt{QubeFit} \citepalias{Neeleman2021} to constrain their gas kinematics. We find that nine rotation-dominated galaxies further satisfy the more stringent disk criterion described in \citet{Burkert2010} and are labeled with an asterisk in Table \ref{tab:kin}.

 \item Using the source finding algorithm, LineSeeker \citep{Gonzalez-Lopez2017,Gonzalez-Lopez2019}, we detect a total of eight companion galaxies in the fields of the 41 quasar host galaxies of which four are new detections. The field surrounding the brightest \CII-emitter in the sample, the quasar host galaxy of P215$-$16, has two possible close companions. This could provide a natural explanation why the disk seen in this quasar host galaxy shows a disturbance, since tidal forces can distort the disk of this galaxy.

\item Consistent with previous results, we find that all quasar host galaxies that have reliable estimates of their dust and \CII\ size are less extended in their dust continuum emission compared to their \CII\ emission. This difference is independent of the morphology of the system, which suggests that the cause of the difference is intrinsic to the galaxy. A possible cause that is consistent with these observations is that the dust temperature is warmer in the center of the galaxy, which results in a steeper dust continuum profile compared to the \CII\ profile \citep{Walter2022,Meyer2025b}, and therefore more compact dust continuum emission. Interestingly, dispersion-dominated systems have very compact dust continuum emission compared to rotation-dominated galaxies while having similar \CII\ emission extents. This strengthens the assertion made in \citetalias{Neeleman2021} that dispersion-dominated systems are distinct from rotation-dominated galaxies.

\item We explore the ratio of the \CII\ luminosity to total far-infrared luminosity for the galaxies in the sample and find that most galaxies are consistent with the deficit in \CII\ emission that is found in other luminous galaxies both locally and at high redshift \citep{Malhotra2001,DiazSantos2013,Venemans2020}. This deficit is largest for dispersion-dominated galaxies, which are more compact, consistent with the result in \citet{Venemans2020} and observations of dusty star-forming galaxies that show the center of galaxies have stronger \CII\ deficits \citep[e.g.,][]{DiazSantos2013,DiazSantos2017,Gullberg2018,Rybak2019}.

\item We introduce a new kinematic statistic, the kinematic position angle alignment parameter, $\Delta$PA$_{\rm kin-CII}$. This parameter is defined as the angle between the direction of the velocity gradient at the peak of the \CII\ emission and the positional angle of the major axis of the \CII\ emission from a 2D Gaussian fit to the \CII\ emission line. Values away from 0\degr\  indicate non-uniform disk rotation. We find that this parameter robustly predicts non-uniform disk rotation in 6 out of the 8 quasar host galaxies that show evidence of mergers or disturbances. In addition, the full distribution of galaxies is shifted towards higher values. This indicates that mergers and disturbances are relatively common features of $z \gtrsim 6$ quasar host galaxies.

\item We estimate the dynamical mass of the galaxies in our sample and find that they agree with previous observations of luminous $z \gtrsim 6$ quasar host galaxies. Subdividing the sample by resolution reveals a bias for lower resolutions to overestimate the dynamical mass due to an overestimation of their size. Using rough estimates of the mass of the SMBH, stars and gas, we find that the average dynamical mass estimates are consistent with these estimates, showing no evidence for a systematic under- or overestimation of the dynamical mass.

\end{itemize}

The ALMA observations discussed in this paper provide a resolved view of the cold ISM of massive, $z \gtrsim 6$ galaxies hosting some of the most luminous quasars in our universe. With these observations, we can study the kinematics of the gas within these galaxies, and we can explore how the distribution of gas affects the global properties of the galaxy as well as the properties of the SMBH at its center. Together with near-infrared spectra and the capabilities of JWST, these observations allow us to study the detailed baryonic budget of the most massive galaxies at cosmic dawn. Looking beyond these observations, a particularly exciting prospect is to go beyond galaxy-wide kinematics, and use higher resolution ALMA observations of the \CII\ and dust continuum emission to study the detailed structure of the ISM in these galaxies in order to refine our understanding of the interplay between the quasar and its host galaxy.

\begin{acknowledgments}
We would like to thank the referee for their helpful comments. VD acknowledges financial support from the Bando Ricerca Fondamentale INAF 2022 Large Grant“XQR-30”. RD acknowledges support from the INAF 2024 minigrant ``The interstellar medium at high redshift.'' CM acknowledges support from Fondecyt Iniciacion grant 11240336 and the ANID BASAL project FB210003. RAM acknowledges support from the Swiss National Science Foundation (SNSF) through project grant 200020\_207349. J.-T.S. acknowledges funding by the Deutsche Forschungsgemeinschaft (DFG, German Research Foundation) - Project number 518006966 (J.-T.S.) and support by the DFG under Germany’s Excellence Strategy – EXC 2121 ``Quantum Universe'' – 390833306. GK is supported by the Department of Atomic Energy (Government of India) research project with Project Identification Number RTI 4002. This paper makes use of the following ALMA data: \#2012.1.00240.S, \#2013.1.00591.S, \#2015.1.00399.S, \#2015.1.00692.S,  \#2015.1.01115.S,   \#2016.1.00544.S, \#2017.1.01301.S, \#2018.1.00566.S, \#2018.1.00908.S, \#2018.1.01188.S, \#2019.1.00080.S, \#2019.1.00111.S, \#2019.1.00661.S, \#2021.1.00211.S, \#2021.1.01320.S, \#2021.1.01018.S, \#2021.1.00934.S, \#2022.1.00300.S, \#2022.1.00998.S. ALMA is a partnership of ESO (representing its member states), NSF (USA) and NINS (Japan), together with NRC (Canada), MOST and ASIAA (Taiwan), and KASI (Republic of Korea), in cooperation with the Republic of Chile. The Joint ALMA Observatory is operated by ESO, AUI/NRAO and NAOJ. The National Radio Astronomy Observatory is a facility of the National Science Foundation operated under cooperative agreement by Associated Universities, Inc. 
\end{acknowledgments}

\facilities{ALMA}
\software{Astropy \citep{Astropy2013,Astropy2018}, \texttt{CASA} \citep{CASA2022},  Matplotlib \citep{Hunter2007}, Numpy \citep{Harris2020}, \texttt{QubeFit} \citepalias{Neeleman2021}}

\bibliography{bib}

@article{Burkert2010,
	adsurl = {https://ui.adsabs.harvard.edu/abs/2010ApJ...725.2324B},
	archiveprefix = {arXiv},
	author = {{Burkert}, A. and {Genzel}, R. and {Bouch{\'e}}, N. and {Cresci}, G. and {Khochfar}, S. and {Sommer-Larsen}, J. and {Sternberg}, A. and {Naab}, T. and {F{\"o}rster Schreiber}, N. and {Tacconi}, L. and {Shapiro}, K. and {Hicks}, E. and {Lutz}, D. and {Davies}, R. and {Buschkamp}, P. and {Genel}, S.},
	doi = {10.1088/0004-637X/725/2/2324},
	eprint = {0907.4777},
	journal = {\apj},
	month = dec,
	number = {2},
	pages = {2324-2332},
	primaryclass = {astro-ph.CO},
	title = {{High-redshift Star-forming Galaxies: Angular Momentum and Baryon Fraction, Turbulent Pressure Effects, and the Origin of Turbulence}},
	volume = {725},
	year = 2010}

@article{Rybak2019,
	adsurl = {https://ui.adsabs.harvard.edu/abs/2019ApJ...876..112R},
	archiveprefix = {arXiv},
	author = {{Rybak}, Matus and {Calistro Rivera}, G. and {Hodge}, J.~A. and {Smail}, Ian and {Walter}, F. and {van der Werf}, P. and {da Cunha}, E. and {Chen}, Chian-Chou and {Dannerbauer}, H. and {Ivison}, R.~J. and {Karim}, A. and {Simpson}, J.~M. and {Swinbank}, A.~M. and {Wardlow}, J.~L.},
	doi = {10.3847/1538-4357/ab0e0f},
	eid = {112},
	eprint = {1901.10027},
	journal = {\apj},
	month = may,
	number = {2},
	pages = {112},
	primaryclass = {astro-ph.GA},
	title = {{Strong Far-ultraviolet Fields Drive the [C II]/Far-infrared Deficit in z {\ensuremath{\sim}} 3 Dusty, Star-forming Galaxies}},
	volume = {876},
	year = 2019}

@article{Gullberg2018,
	adsurl = {https://ui.adsabs.harvard.edu/abs/2018ApJ...859...12G},
	archiveprefix = {arXiv},
	author = {{Gullberg}, B. and {Swinbank}, A.~M. and {Smail}, I. and {Biggs}, A.~D. and {Bertoldi}, F. and {De Breuck}, C. and {Chapman}, S.~C. and {Chen}, C.-C. and {Cooke}, E.~A. and {Coppin}, K.~E.~K. and {Cox}, P. and {Dannerbauer}, H. and {Dunlop}, J.~S. and {Edge}, A.~C. and {Farrah}, D. and {Geach}, J.~E. and {Greve}, T.~R. and {Hodge}, J. and {Ibar}, E. and {Ivison}, R.~J. and {Karim}, A. and {Schinnerer}, E. and {Scott}, D. and {Simpson}, J.~M. and {Stach}, S.~M. and {Thomson}, A.~P. and {van der Werf}, P. and {Walter}, F. and {Wardlow}, J.~L. and {Weiss}, A.},
	doi = {10.3847/1538-4357/aabe8c},
	eid = {12},
	eprint = {1804.03663},
	journal = {\apj},
	month = may,
	number = {1},
	pages = {12},
	primaryclass = {astro-ph.GA},
	title = {{The Dust and [C II] Morphologies of Redshift {\ensuremath{\sim}}4.5 Sub-millimeter Galaxies at {\ensuremath{\sim}}200 pc Resolution: The Absence of Large Clumps in the Interstellar Medium at High-redshift}},
	volume = {859},
	year = 2018}

@article{Decarli2023,
	adsurl = {https://ui.adsabs.harvard.edu/abs/2023A&A...673A.157D},
	archiveprefix = {arXiv},
	author = {{Decarli}, Roberto and {Pensabene}, Antonio and {Diaz-Santos}, Tanio and {Ferkinhoff}, Carl and {Strauss}, Michael A. and {Venemans}, Bram P. and {Walter}, Fabian and {Ba{\~n}ados}, Eduardo and {Bertoldi}, Frank and {Fan}, Xiaohui and {Farina}, Emanuele Paolo and {Riechers}, Dominik A. and {Rix}, Hans-Walter and {Wang}, Ran},
	doi = {10.1051/0004-6361/202245674},
	eid = {A157},
	eprint = {2302.04312},
	journal = {\aap},
	month = may,
	pages = {A157},
	primaryclass = {astro-ph.GA},
	title = {{A comprehensive view of the interstellar medium in a quasar host galaxy at z {\ensuremath{\approx}} 6.4}},
	volume = {673},
	year = 2023}

@article{Carniani2019,
	adsurl = {https://ui.adsabs.harvard.edu/abs/2019MNRAS.489.3939C},
	archiveprefix = {arXiv},
	author = {{Carniani}, S. and {Gallerani}, S. and {Vallini}, L. and {Pallottini}, A. and {Tazzari}, M. and {Ferrara}, A. and {Maiolino}, R. and {Cicone}, C. and {Feruglio}, C. and {Neri}, R. and {D'Odorico}, V. and {Wang}, R. and {Li}, J.},
	doi = {10.1093/mnras/stz2410},
	eprint = {1902.01413},
	journal = {\mnras},
	month = nov,
	number = {3},
	pages = {3939-3952},
	primaryclass = {astro-ph.GA},
	title = {{Constraints on high-J CO emission lines in z {\ensuremath{\sim}} 6 quasars}},
	volume = {489},
	year = 2019}

@misc{Mason2013,
	adsurl = {https://ui.adsabs.harvard.edu/abs/2013nrao.reptE...2M},
	author = {{Mason}, Brian and {Brogan}, Crystal L.},
	howpublished = {NAASC Memo 113/ALMA Memo 598, 8 pages, July 2013},
	month = jul,
	pages = {E2},
	title = {{Relative Integration Times for the ALMA Cycle 1 12-m, 7-m, and Total Power Arrays}},
	year = 2013}

@article{Sodini2024,
	adsurl = {https://ui.adsabs.harvard.edu/abs/2024A&A...687A.314S},
	archiveprefix = {arXiv},
	author = {{Sodini}, Alessio and {D'Odorico}, Valentina and {Salvadori}, Stefania and {Vanni}, Irene and {Bischetti}, Manuela and {Cupani}, Guido and {Davies}, Rebecca and {Becker}, George D. and {Ba{\~n}ados}, Eduardo and {Bosman}, Sarah and {Davies}, Frederick and {Paolo Farina}, Emanuele and {Ferrara}, Andrea and {Keating}, Laura and {Kulkarni}, Girish and {Lai}, Samuel and {Ryan-Weber}, Emma and {Maria Sebastian}, Alma and {Walter}, Fabian},
	doi = {10.1051/0004-6361/202349062},
	eid = {A314},
	eprint = {2404.10722},
	journal = {\aap},
	month = jul,
	pages = {A314},
	primaryclass = {astro-ph.GA},
	title = {{Evidence of Pop III stars' chemical signature in neutral gas at z {\ensuremath{\sim}} 6. A study based on the E-XQR-30 spectroscopic sample}},
	volume = {687},
	year = 2024}

@article{Davies2023,
	adsurl = {https://ui.adsabs.harvard.edu/abs/2023MNRAS.521..289D},
	archiveprefix = {arXiv},
	author = {{Davies}, Rebecca L. and {Ryan-Weber}, E. and {D'Odorico}, V. and {Bosman}, S.~E.~I. and {Meyer}, R.~A. and {Becker}, G.~D. and {Cupani}, G. and {Bischetti}, M. and {Sebastian}, A.~M. and {Eilers}, A.-C. and {Farina}, E.~P. and {Wang}, F. and {Yang}, J. and {Zhu}, Y.},
	doi = {10.1093/mnras/stac3662},
	eprint = {2211.15816},
	journal = {\mnras},
	month = may,
	number = {1},
	pages = {289-313},
	primaryclass = {astro-ph.GA},
	title = {{The XQR-30 metal absorber catalogue: 778 absorption systems spanning 2 {\ensuremath{\lesssim}} z {\ensuremath{\lesssim}} 6.5}},
	volume = {521},
	year = 2023}

@article{Loiacono2021,
	adsurl = {https://ui.adsabs.harvard.edu/abs/2021A&A...646A..76L},
	archiveprefix = {arXiv},
	author = {{Loiacono}, Federica and {Decarli}, Roberto and {Gruppioni}, Carlotta and {Talia}, Margherita and {Cimatti}, Andrea and {Zamorani}, Gianni and {Pozzi}, Francesca and {Yan}, Lin and {Lemaux}, Brian C. and {Riechers}, Dominik A. and {Le F{\`e}vre}, Olivier and {B{\`e}thermin}, Matthieu and {Capak}, Peter and {Cassata}, Paolo and {Faisst}, Andreas and {Schaerer}, Daniel and {Silverman}, John D. and {Bardelli}, Sandro and {Boquien}, M{\'e}d{\'e}ric and {Burkutean}, Sandra and {Dessauges-Zavadsky}, Miroslava and {Fudamoto}, Yoshinobu and {Fujimoto}, Seiji and {Ginolfi}, Michele and {Hathi}, Nimish P. and {Jones}, Gareth C. and {Khusanova}, Yana and {Koekemoer}, Anton M. and {Lagache}, Guilaine and {Lubin}, Lori M. and {Massardi}, Marcella and {Oesch}, Pascal and {Romano}, Michael and {Vallini}, Livia and {Vergani}, Daniela and {Zucca}, Elena},
	doi = {10.1051/0004-6361/202038607},
	eid = {A76},
	eprint = {2006.04837},
	journal = {\aap},
	month = feb,
	pages = {A76},
	primaryclass = {astro-ph.GA},
	title = {{The ALPINE-ALMA [C II] survey. Luminosity function of serendipitous [C II] line emitters at z {\ensuremath{\sim}} 5}},
	volume = {646},
	year = 2021}

@article{Decarli2020,
	adsurl = {https://ui.adsabs.harvard.edu/abs/2020ApJ...902..110D},
	archiveprefix = {arXiv},
	author = {{Decarli}, Roberto and {Aravena}, Manuel and {Boogaard}, Leindert and {Carilli}, Chris and {Gonz{\'a}lez-L{\'o}pez}, Jorge and {Walter}, Fabian and {Cortes}, Paulo C. and {Cox}, Pierre and {da Cunha}, Elisabete and {Daddi}, Emanuele and {D{\'\i}az-Santos}, Tanio and {Hodge}, Jacqueline A. and {Inami}, Hanae and {Neeleman}, Marcel and {Novak}, Mladen and {Oesch}, Pascal and {Popping}, Gerg{\"o} and {Riechers}, Dominik and {Smail}, Ian and {Uzgil}, Bade and {van der Werf}, Paul and {Wagg}, Jeff and {Weiss}, Axel},
	doi = {10.3847/1538-4357/abaa3b},
	eid = {110},
	eprint = {2009.10744},
	journal = {\apj},
	month = oct,
	number = {2},
	pages = {110},
	primaryclass = {astro-ph.GA},
	title = {{The ALMA Spectroscopic Survey in the Hubble Ultra Deep Field: Multiband Constraints on Line-luminosity Functions and the Cosmic Density of Molecular Gas}},
	volume = {902},
	year = 2020}

@article{Durovcikova2025,
	author = {{\v D}urov{\v c}{\'\i}kov{\'a}, Dominika and Eilers, Anna-Christina and Meyer, Romain A. and Farina, Emanuele Paolo and Ba{\~n}ados, Eduardo and Davies, Frederick B. and Hennawi, Joseph F. and Mazzucchelli, Chiara and Simcoe, Robert A. and Walter, Fabian},
	doi = {10.3847/1538-4357/adf6dd},
	issn = {1538-4357},
	journal = {The Astrophysical Journal},
	month = sep,
	number = {2},
	pages = {174},
	publisher = {American Astronomical Society},
	title = {Quasar Lifetime Measurements from Extended Lyα Nebulae at z ∼ 6},
	url = {http://dx.doi.org/10.3847/1538-4357/adf6dd},
	volume = {990},
	year = {2025}}

@article{Banados2024,
	adsurl = {https://ui.adsabs.harvard.edu/abs/2024ApJ...977L..46B},
	archiveprefix = {arXiv},
	author = {{Ba{\~n}ados}, Eduardo and {Khusanova}, Yana and {Decarli}, Roberto and {Momjian}, Emmanuel and {Walter}, Fabian and {Connor}, Thomas and {Carilli}, Christopher L. and {Mazzucchelli}, Chiara and {Rojas-Ruiz}, Sof{\'\i}a and {Venemans}, Bram P.},
	doi = {10.3847/2041-8213/ad823b},
	eid = {L46},
	eprint = {2408.12299},
	journal = {\apjl},
	month = dec,
	number = {2},
	pages = {L46},
	primaryclass = {astro-ph.GA},
	title = {{[C II] Properties and Far-infrared Variability of a z = 7 Blazar}},
	volume = {977},
	year = 2024}

@article{Banados2015,
	author = {Ba{\~n}ados, E. and Venemans, B. P. and Morganson, E. and Hodge, J. and Decarli, R. and Walter, F. and Stern, D. and Schlafly, E. and Farina, E. P. and Greiner, J. and Chambers, K. C. and Fan, X. and Rix, H-W. and Burgett, W. S. and Draper, P. W. and Flewelling, J. and Kaiser, N. and Metcalfe, N. and Morgan, J. S. and Tonry, J. L. and Wainscoat, R. J.},
	doi = {10.1088/0004-637x/804/2/118},
	issn = {1538-4357},
	journal = {The Astrophysical Journal},
	month = may,
	number = {2},
	pages = {118},
	publisher = {American Astronomical Society},
	title = {CONSTRAINING THE RADIO-LOUD FRACTION OF QUASARS ATz&gt; 5.5},
	url = {http://dx.doi.org/10.1088/0004-637X/804/2/118},
	volume = {804},
	year = {2015}}

@article{Wang2007,
	adsurl = {https://ui.adsabs.harvard.edu/abs/2007AJ....134..617W},
	archiveprefix = {arXiv},
	author = {{Wang}, Ran and {Carilli}, Chris L. and {Beelen}, Alexandre and {Bertoldi}, Frank and {Fan}, Xiaohui and {Walter}, Fabian and {Menten}, Karl M. and {Omont}, Alain and {Cox}, Pierre and {Strauss}, Michael A. and {Jiang}, Linhua},
	doi = {10.1086/518867},
	eprint = {0704.2053},
	journal = {\aj},
	month = aug,
	number = {2},
	pages = {617-627},
	primaryclass = {astro-ph},
	title = {{Millimeter and Radio Observations of z \raisebox{-0.5ex}\textasciitilde 6 Quasars}},
	volume = {134},
	year = 2007}

@article{Morganson2012,
	adsurl = {https://ui.adsabs.harvard.edu/abs/2012AJ....143..142M},
	archiveprefix = {arXiv},
	author = {{Morganson}, Eric and {De Rosa}, Gisella and {Decarli}, Roberto and {Walter}, Fabian and {Chambers}, Ken and {McGreer}, Ian and {Fan}, Xiaohui and {Burgett}, William and {Flewelling}, Heather and {Greiner}, Jochen and {Hodapp}, Klaus and {Kaiser}, Nick and {Magnier}, Eugene and {Price}, Paul and {Rix}, Hans-Walter and {Sweeney}, Bill and {Waters}, Christopher},
	doi = {10.1088/0004-6256/143/6/142},
	eid = {142},
	eprint = {1109.6241},
	journal = {\aj},
	month = jun,
	number = {6},
	pages = {142},
	primaryclass = {astro-ph.CO},
	title = {{The First High-redshift Quasar from Pan-STARRS}},
	volume = {143},
	year = 2012}

@article{Walter2025,
	adsurl = {https://ui.adsabs.harvard.edu/abs/2025ApJ...983L...8W},
	archiveprefix = {arXiv},
	author = {{Walter}, Fabian and {Ba{\~n}ados}, Eduardo and {Carilli}, Chris and {Neeleman}, Marcel and {Connor}, Thomas and {Decarli}, Roberto and {Farina}, Emanuele Paulo and {Khusanova}, Yana and {Mazzucchelli}, Chiara and {Meyer}, Romain and {Momjian}, Emmanuel and {Rix}, Hans─Walter and {Rojas-Ruiz}, Sof{\'\i}a and {Venemans}, Bram},
	doi = {10.3847/2041-8213/adc2f5},
	eid = {L8},
	eprint = {2503.23797},
	journal = {\apjl},
	month = apr,
	number = {1},
	pages = {L8},
	primaryclass = {astro-ph.GA},
	title = {{Kiloparsec-scale Alignment of a Radio Jet with Cool Gas and Dust in a z {\ensuremath{\sim}} 6 Quasar}},
	volume = {983},
	year = 2025}

@article{Mazzucchelli2017a,
	adsurl = {https://ui.adsabs.harvard.edu/abs/2017ApJ...834...83M},
	archiveprefix = {arXiv},
	author = {{Mazzucchelli}, C. and {Ba{\~n}ados}, E. and {Decarli}, R. and {Farina}, E.~P. and {Venemans}, B.~P. and {Walter}, F. and {Overzier}, R.},
	doi = {10.3847/1538-4357/834/1/83},
	eid = {83},
	eprint = {1611.02870},
	journal = {\apj},
	month = jan,
	number = {1},
	pages = {83},
	primaryclass = {astro-ph.GA},
	title = {{No Overdensity of Lyman-Alpha Emitting Galaxies around a Quasar at z {\ensuremath{\sim}} 5.7}},
	volume = {834},
	year = 2017}

@article{Yang2023,
	adsurl = {https://ui.adsabs.harvard.edu/abs/2023ApJ...951L...5Y},
	archiveprefix = {arXiv},
	author = {{Yang}, Jinyi and {Wang}, Feige and {Fan}, Xiaohui and {Hennawi}, Joseph F. and {Barth}, Aaron J. and {Ba{\~n}ados}, Eduardo and {Sun}, Fengwu and {Liu}, Weizhe and {Cai}, Zheng and {Jiang}, Linhua and {Li}, Zihao and {Onoue}, Masafusa and {Schindler}, Jan-Torge and {Shen}, Yue and {Wu}, Yunjing and {Bhowmick}, Aklant K. and {Bieri}, Rebekka and {Blecha}, Laura and {Bosman}, Sarah and {Champagne}, Jaclyn B. and {Colina}, Luis and {Connor}, Thomas and {Costa}, Tiago and {Davies}, Frederick B. and {Decarli}, Roberto and {De Rosa}, Gisella and {Drake}, Alyssa B. and {Egami}, Eiichi and {Eilers}, Anna-Christina and {Evans}, Analis E. and {Farina}, Emanuele Paolo and {Habouzit}, Melanie and {Haiman}, Zoltan and {Jin}, Xiangyu and {Jun}, Hyunsung D. and {Kakiichi}, Koki and {Khusanova}, Yana and {Kulkarni}, Girish and {Loiacono}, Federica and {Lupi}, Alessandro and {Mazzucchelli}, Chiara and {Pan}, Zhiwei and {Rojas-Ruiz}, Sof{\'\i}a and {Strauss}, Michael A. and {Tee}, Wei Leong and {Trakhtenbrot}, Benny and {Trebitsch}, Maxime and {Venemans}, Bram and {Vestergaard}, Marianne and {Volonteri}, Marta and {Walter}, Fabian and {Xie}, Zhang-Liang and {Yue}, Minghao and {Zhang}, Haowen and {Zhang}, Huanian and {Zou}, Siwei},
	doi = {10.3847/2041-8213/acc9c8},
	eid = {L5},
	eprint = {2304.09888},
	journal = {\apjl},
	month = jul,
	number = {1},
	pages = {L5},
	primaryclass = {astro-ph.GA},
	title = {{A SPectroscopic Survey of Biased Halos in the Reionization Era (ASPIRE): A First Look at the Rest-frame Optical Spectra of z > 6.5 Quasars Using JWST}},
	volume = {951},
	year = 2023}

@article{Mazzucchelli2025,
	adsurl = {https://ui.adsabs.harvard.edu/abs/2025A&A...694A.171M},
	archiveprefix = {arXiv},
	author = {{Mazzucchelli}, C. and {Decarli}, R. and {Belladitta}, S. and {Ba{\~n}ados}, E. and {Meyer}, R.~A. and {Connor}, T. and {Momjian}, E. and {Rojas-Ruiz}, S. and {Eilers}, A.-C. and {Khusanova}, Y. and {Farina}, E.~P. and {Drake}, A.~B. and {Walter}, F. and {Wang}, F. and {Onoue}, M. and {Venemans}, B.~P.},
	doi = {10.1051/0004-6361/202451290},
	eid = {A171},
	eprint = {2411.11952},
	journal = {\aap},
	month = feb,
	pages = {A171},
	primaryclass = {astro-ph.GA},
	title = {{The host galaxies of radio-loud quasars at z > 5 with ALMA}},
	volume = {694},
	year = 2025}

@article{Farina2022,
	adsurl = {https://ui.adsabs.harvard.edu/abs/2022ApJ...941..106F},
	archiveprefix = {arXiv},
	author = {{Farina}, Emanuele Paolo and {Schindler}, Jan-Torge and {Walter}, Fabian and {Ba{\~n}ados}, Eduardo and {Davies}, Frederick B. and {Decarli}, Roberto and {Eilers}, Anna-Christina and {Fan}, Xiaohui and {Hennawi}, Joseph F. and {Mazzucchelli}, Chiara and {Meyer}, Romain A. and {Trakhtenbrot}, Benny and {Volonteri}, Marta and {Wang}, Feige and {Worseck}, G{\'a}bor and {Yang}, Jinyi and {Gutcke}, Thales A. and {Venemans}, Bram P. and {Bosman}, Sarah E.~I. and {Costa}, Tiago and {De Rosa}, Gisella and {Drake}, Alyssa B. and {Onoue}, Masafusa},
	doi = {10.3847/1538-4357/ac9626},
	eid = {106},
	eprint = {2207.05113},
	journal = {\apj},
	month = dec,
	number = {2},
	pages = {106},
	primaryclass = {astro-ph.GA},
	title = {{The X-shooter/ALMA Sample of Quasars in the Epoch of Reionization. II. Black Hole Masses, Eddington Ratios, and the Formation of the First Quasars}},
	volume = {941},
	year = 2022}

@article{Kashino2026,
	adsurl = {https://ui.adsabs.harvard.edu/abs/2026ApJ...997..280K},
	archiveprefix = {arXiv},
	author = {{Kashino}, Daichi and {Lilly}, Simon J. and {Matthee}, Jorryt and {Mackenzie}, Ruari and {Eilers}, Anna-Christina and {Bordoloi}, Rongmon and {Simcoe}, Robert A. and {Naidu}, Rohan P. and {Yue}, Minghao and {Liu}, Bin},
	doi = {10.3847/1538-4357/ae2799},
	eid = {280},
	eprint = {2506.03121},
	journal = {\apj},
	month = feb,
	number = {2},
	pages = {280},
	primaryclass = {astro-ph.GA},
	title = {{EIGER. VII. The Evolving Relationship between Galaxies and the Intergalactic Medium in the Final Stages of Reionization}},
	volume = {997},
	year = 2026}

@article{Eilers2024,
	adsurl = {https://ui.adsabs.harvard.edu/abs/2024ApJ...974..275E},
	archiveprefix = {arXiv},
	author = {{Eilers}, Anna-Christina and {Mackenzie}, Ruari and {Pizzati}, Elia and {Matthee}, Jorryt and {Hennawi}, Joseph F. and {Zhang}, Haowen and {Bordoloi}, Rongmon and {Kashino}, Daichi and {Lilly}, Simon J. and {Naidu}, Rohan P. and {Simcoe}, Robert A. and {Yue}, Minghao and {Frenk}, Carlos S. and {Helly}, John C. and {Schaller}, Matthieu and {Schaye}, Joop},
	doi = {10.3847/1538-4357/ad778b},
	eid = {275},
	eprint = {2403.07986},
	journal = {\apj},
	month = oct,
	number = {2},
	pages = {275},
	primaryclass = {astro-ph.GA},
	title = {{EIGER. VI. The Correlation Function, Host Halo Mass, and Duty Cycle of Luminous Quasars at z {\ensuremath{\gtrsim}} 6}},
	volume = {974},
	year = 2024}

@article{Wang2026,
	adsurl = {https://ui.adsabs.harvard.edu/abs/2026arXiv260204979W},
	archiveprefix = {arXiv},
	author = {{Wang}, Feige and {Champagne}, Jaclyn B. and {Huang}, Jiamu and {Yang}, Jinyi and {Hennawi}, Joseph F. and {Fan}, Xiaohui and {Zhang}, Haowen and {Costa}, Tiago and {Decarli}, Roberto and {Habouzit}, Melanie and {Sun}, Fengwu and {Banados}, Eduardo and {Jin}, Xiangyu and {Kakiichi}, Koki and {Meyer}, Romain A. and {Wu}, Yunjing and {Belladitta}, Silvia and {Blecha}, Laura and {Bosman}, Sarah E.~I. and {Cai}, Zheng and {Connor}, Thomas and {Davies}, Frederick B. and {Eilers}, Anna-Christina and {Haiman}, Zoltan and {Jun}, Hyunsung D. and {Li}, Mingyu and {Li}, Zihao and {Liu}, Weizhe and {Lupi}, Alessandro and {Lyu}, Jianwei and {Mazzucchelli}, Chiara and {Onoue}, Masafusa and {Pudoka}, Maria and {Rojas-Ruiz}, Sofia and {Schindler}, Jan-Torge and {Shen}, Yue and {Tee}, Wei Leong and {Trakhtenbrot}, Benny and {Trebitsch}, Maxime and {Vestergaard}, Marianne and {Volonteri}, Marta and {Walter}, Fabian and {Zhang}, Huanian and {Zou}, Siwei},
	doi = {10.48550/arXiv.2602.04979},
	eid = {arXiv:2602.04979},
	eprint = {2602.04979},
	journal = {arXiv e-prints},
	month = feb,
	pages = {arXiv:2602.04979},
	primaryclass = {astro-ph.GA},
	title = {{ASPIRE: The Environments and Dark Matter Halos of Luminous Quasars in the Epoch of Reionization}},
	year = 2026}

@article{DiazSantos2017,
	adsurl = {https://ui.adsabs.harvard.edu/abs/2017ApJ...846...32D},
	archiveprefix = {arXiv},
	author = {{D{\'\i}az-Santos}, T. and {Armus}, L. and {Charmandaris}, V. and {Lu}, N. and {Stierwalt}, S. and {Stacey}, G. and {Malhotra}, S. and {van der Werf}, P.~P. and {Howell}, J.~H. and {Privon}, G.~C. and {Mazzarella}, J.~M. and {Goldsmith}, P.~F. and {Murphy}, E.~J. and {Barcos-Mu{\~n}oz}, L. and {Linden}, S.~T. and {Inami}, H. and {Larson}, K.~L. and {Evans}, A.~S. and {Appleton}, P. and {Iwasawa}, K. and {Lord}, S. and {Sanders}, D.~B. and {Surace}, J.~A.},
	doi = {10.3847/1538-4357/aa81d7},
	eid = {32},
	eprint = {1705.04326},
	journal = {\apj},
	month = sep,
	number = {1},
	pages = {32},
	primaryclass = {astro-ph.GA},
	title = {{A Herschel/PACS Far-infrared Line Emission Survey of Local Luminous Infrared Galaxies}},
	volume = {846},
	year = 2017}

@article{Witstok2023,
	adsurl = {https://ui.adsabs.harvard.edu/abs/2023MNRAS.523.3119W},
	archiveprefix = {arXiv},
	author = {{Witstok}, Joris and {Jones}, Gareth C. and {Maiolino}, Roberto and {Smit}, Renske and {Schneider}, Raffaella},
	doi = {10.1093/mnras/stad1470},
	eprint = {2305.09714},
	journal = {\mnras},
	month = aug,
	number = {2},
	pages = {3119-3132},
	primaryclass = {astro-ph.GA},
	title = {{An empirical study of dust properties at the earliest epochs}},
	volume = {523},
	year = 2023}

@article{ALMA2025,
	author = {Cortes, P. C. and Carpenter, J. and Kameno, S. and Loomis, R. and Vila-Vilaro, B. and Immer, K. and Vlahakis, C. and Law, J. and Stoehr, F. and Saini, K. and Hales, A.},
	doi = {10.5281/zenodo.14933753},
	journal = {ALMA Technical Handbook},
	pages = {ver. 1.5},
	title = {ALMA Technical Handbook},
	volume = {ALMA Doc 12.3},
	year = {2025}}

@article{Decarli2025,
	adsurl = {https://ui.adsabs.harvard.edu/abs/2025A&ARv..33....4D},
	archiveprefix = {arXiv},
	author = {{Decarli}, Roberto and {D{\'\i}az-Santos}, Tanio},
	doi = {10.1007/s00159-025-00162-7},
	eid = {4},
	eprint = {2509.19444},
	journal = {\aapr},
	month = sep,
	number = {1},
	pages = {4},
	primaryclass = {astro-ph.GA},
	title = {{Infrared fine-structure lines at high redshift}},
	volume = {33},
	year = 2025}

@article{Zanella2018,
	adsurl = {https://ui.adsabs.harvard.edu/abs/2018MNRAS.481.1976Z},
	archiveprefix = {arXiv},
	author = {{Zanella}, A. and {Daddi}, E. and {Magdis}, G. and {Diaz Santos}, T. and {Cormier}, D. and {Liu}, D. and {Cibinel}, A. and {Gobat}, R. and {Dickinson}, M. and {Sargent}, M. and {Popping}, G. and {Madden}, S.~C. and {Bethermin}, M. and {Hughes}, T.~M. and {Valentino}, F. and {Rujopakarn}, W. and {Pannella}, M. and {Bournaud}, F. and {Walter}, F. and {Wang}, T. and {Elbaz}, D. and {Coogan}, R.~T.},
	doi = {10.1093/mnras/sty2394},
	eprint = {1808.10331},
	journal = {\mnras},
	month = dec,
	number = {2},
	pages = {1976-1999},
	primaryclass = {astro-ph.GA},
	title = {{The [C II] emission as a molecular gas mass tracer in galaxies at low and high redshifts}},
	volume = {481},
	year = 2018}

@article{Salvestrini2025,
	adsurl = {https://ui.adsabs.harvard.edu/abs/2025A&A...695A..23S},
	archiveprefix = {arXiv},
	author = {{Salvestrini}, Francesco and {Feruglio}, Chiara and {Tripodi}, Roberta and {Fontanot}, Fabio and {Bischetti}, Manuela and {De Lucia}, Gabriella and {Fiore}, Fabrizio and {Hirschmann}, Michaela and {Maio}, Umberto and {Piconcelli}, Enrico and {Saccheo}, Ivano and {Tortosa}, Alessia and {Valiante}, Rosa and {Xie}, Lizhi and {Zappacosta}, Luca},
	doi = {10.1051/0004-6361/202453226},
	eid = {A23},
	eprint = {2412.02688},
	journal = {\aap},
	month = mar,
	pages = {A23},
	primaryclass = {astro-ph.GA},
	title = {{Molecular gas and dust properties in z > 7 quasar hosts}},
	volume = {695},
	year = 2025}

@article{Bischetti2025,
	adsurl = {https://ui.adsabs.harvard.edu/abs/2025ApJ...990L..31B},
	archiveprefix = {arXiv},
	author = {{Bischetti}, Manuela and {Feruglio}, Chiara and {Carniani}, Stefano and {D'Odorico}, Valentina and {Salvestrini}, Francesco and {Fiore}, Fabrizio},
	doi = {10.3847/2041-8213/ade6ff},
	eid = {L31},
	eprint = {2504.15357},
	journal = {\apjl},
	month = sep,
	number = {2},
	pages = {L31},
	primaryclass = {astro-ph.GA},
	title = {{ALMA Reveals Bright Circumgalactic Emission and a Biconical Outflow in z {\ensuremath{\sim}} 6.4 Quasar PSOJ183+05}},
	volume = {990},
	year = 2025}

@article{Bischetti2024,
	adsurl = {https://ui.adsabs.harvard.edu/abs/2024ApJ...970....9B},
	archiveprefix = {arXiv},
	author = {{Bischetti}, Manuela and {Choi}, Hyunseop and {Fiore}, Fabrizio and {Feruglio}, Chiara and {Carniani}, Stefano and {D'Odorico}, Valentina and {Ba{\~n}ados}, Eduardo and {Chen}, Huanqing and {Decarli}, Roberto and {Gallerani}, Simona and {Hlavacek-Larrondo}, Julie and {Lai}, Samuel and {Leighly}, Karen M. and {Mazzucchelli}, Chiara and {Perreault-Levasseur}, Laurence and {Tripodi}, Roberta and {Walter}, Fabian and {Wang}, Feige and {Yang}, Jinyi and {Zanchettin}, Maria Vittoria and {Zhu}, Yongda},
	doi = {10.3847/1538-4357/ad4a77},
	eid = {9},
	eprint = {2404.12443},
	journal = {\apj},
	month = jul,
	number = {1},
	pages = {9},
	primaryclass = {astro-ph.GA},
	title = {{Multiphase Black Hole Feedback and a Bright [C II] Halo in a LoBAL Quasar at z {\ensuremath{\sim}} 6.6}},
	volume = {970},
	year = 2024}

@article{Tripodi2024b,
	adsurl = {https://ui.adsabs.harvard.edu/abs/2024A&A...689A.220T},
	archiveprefix = {arXiv},
	author = {{Tripodi}, R. and {Feruglio}, C. and {Fiore}, F. and {Zappacosta}, L. and {Piconcelli}, E. and {Bischetti}, M. and {Bongiorno}, A. and {Carniani}, S. and {Civano}, F. and {Chen}, C.-C. and {Cristiani}, S. and {Cupani}, G. and {Di Mascia}, F. and {D'Odorico}, V. and {Fan}, X. and {Ferrara}, A. and {Gallerani}, S. and {Ginolfi}, M. and {Maiolino}, R. and {Mainieri}, V. and {Marconi}, A. and {Saccheo}, I. and {Salvestrini}, F. and {Tortosa}, A. and {Valiante}, R.},
	doi = {10.1051/0004-6361/202349054},
	eid = {A220},
	eprint = {2401.04211},
	journal = {\aap},
	month = sep,
	pages = {A220},
	primaryclass = {astro-ph.GA},
	title = {{HYPERION: Coevolution of supermassive black holes and galaxies at z > 6 and the build-up of massive galaxies}},
	volume = {689},
	year = 2024}

@article{Costa2025,
	adsurl = {https://ui.adsabs.harvard.edu/abs/2025arXiv251203143C},
	archiveprefix = {arXiv},
	author = {{Costa}, M. and {Decarli}, R. and {Pozzi}, F. and {Cox}, P. and {Meyer}, R.~A. and {Pensabene}, A. and {Venemans}, B.~P. and {Walter}, F. and {Xu}, F.},
	doi = {10.48550/arXiv.2512.03143},
	eid = {arXiv:2512.03143},
	eprint = {2512.03143},
	journal = {arXiv e-prints},
	month = dec,
	pages = {arXiv:2512.03143},
	primaryclass = {astro-ph.GA},
	title = {{Infrared emission from $z \sim 6.5$ quasar host galaxies: a direct estimate of dust physical properties}},
	year = 2025}

@article{Bertoldi2003,
	adsurl = {https://ui.adsabs.harvard.edu/abs/2003A&A...406L..55B},
	archiveprefix = {arXiv},
	author = {{Bertoldi}, F. and {Carilli}, C.~L. and {Cox}, P. and {Fan}, X. and {Strauss}, M.~A. and {Beelen}, A. and {Omont}, A. and {Zylka}, R.},
	doi = {10.1051/0004-6361:20030710},
	eprint = {astro-ph/0305116},
	journal = {\aap},
	month = jul,
	pages = {L55-L58},
	primaryclass = {astro-ph},
	title = {{Dust emission from the most distant quasars}},
	volume = {406},
	year = 2003}

@article{Maiolino2005,
	adsurl = {https://ui.adsabs.harvard.edu/abs/2005A&A...440L..51M},
	archiveprefix = {arXiv},
	author = {{Maiolino}, R. and {Cox}, P. and {Caselli}, P. and {Beelen}, A. and {Bertoldi}, F. and {Carilli}, C.~L. and {Kaufman}, M.~J. and {Menten}, K.~M. and {Nagao}, T. and {Omont}, A. and {Wei{\ss}}, A. and {Walmsley}, C.~M. and {Walter}, F.},
	doi = {10.1051/0004-6361:200500165},
	eprint = {astro-ph/0508064},
	journal = {\aap},
	month = sep,
	number = {2},
	pages = {L51-L54},
	primaryclass = {astro-ph},
	title = {{First detection of [CII]158 {\ensuremath{\mu}}m at high redshift: vigorous star formation in the early universe}},
	volume = {440},
	year = 2005}

@article{Onoue2025,
	adsurl = {https://ui.adsabs.harvard.edu/abs/2025NatAs...9.1541O},
	archiveprefix = {arXiv},
	author = {{Onoue}, Masafusa and {Ding}, Xuheng and {Silverman}, John D. and {Matsuoka}, Yoshiki and {Izumi}, Takuma and {Strauss}, Michael A. and {Ward}, Charlotte and {Phillips}, Camryn L. and {Ito}, Kei and {Andika}, Irham T. and {Aoki}, Kentaro and {Arita}, Junya and {Baba}, Shunsuke and {Bieri}, Rebekka and {Bosman}, Sarah E.~I. and {Eilers}, Anna-Christina and {Fujimoto}, Seiji and {Habouzit}, Melanie and {Haiman}, Zoltan and {Imanishi}, Masatoshi and {Inayoshi}, Kohei and {Iwasawa}, Kazushi and {Jahnke}, Knud and {Kashikawa}, Nobunari and {Kawaguchi}, Toshihiro and {Kohno}, Kotaro and {Lee}, Chien-Hsiu and {Li}, Junyao and {Lupi}, Alessandro and {Lyu}, Jianwei and {Nagao}, Tohru and {Overzier}, Roderik and {Schindler}, Jan-Torge and {Schramm}, Malte and {Scoggins}, Matthew T. and {Shimasaku}, Kazuhiro and {Toba}, Yoshiki and {Trakhtenbrot}, Benny and {Trebitsch}, Maxime and {Treu}, Tommaso and {Umehata}, Hideki and {Venemans}, Bram and {Vestergaard}, Marianne and {Volonteri}, Marta and {Walter}, Fabian and {Wang}, Feige and {Yang}, Jinyi and {Zhang}, Haowen},
	doi = {10.1038/s41550-025-02628-1},
	eprint = {2409.07113},
	journal = {Nature Astronomy},
	month = aug,
	pages = {1541-1552},
	primaryclass = {astro-ph.GA},
	title = {{A post-starburst pathway for the formation of massive galaxies and black holes at z > 6}},
	volume = {9},
	year = 2025}

@article{Silverman2025,
	adsurl = {https://ui.adsabs.harvard.edu/abs/2025ApJ...995L..67S},
	archiveprefix = {arXiv},
	author = {{Silverman}, John David and {Li}, Junyao and {Ding}, Xuheng and {Onoue}, Masafusa and {Strauss}, Michael A. and {Matsuoka}, Yoshiki and {Izumi}, Takuma and {Jahnke}, Knud and {Treu}, Tommaso and {Volonteri}, Marta and {Phillips}, Camryn L. and {Andika}, Irham T. and {Aoki}, Kentaro and {Arita}, Junya and {Baba}, Shunsuke and {Bosman}, Sarah E.~I. and {Eilers}, Anna-Christina and {Fan}, Xiaohui and {Fujimoto}, Seiji and {Habouzit}, Melanie and {Haiman}, Zoltan and {Imanishi}, Masatoshi and {Inayoshi}, Kohei and {Iwasawa}, Kazushi and {Kashikawa}, Nobunari and {Kawaguchi}, Toshihiro and {Lee}, Chien-Hsiu and {Lupi}, Alessandro and {Nagao}, Tohru and {Schindler}, Jan-Torge and {Schramm}, Malte and {Shimasaku}, Kazuhiro and {Toba}, Yoshiki and {Trakhtenbrot}, Benny and {Umehata}, Hideki and {Vestergaard}, Marianne and {Walter}, Fabian and {Wang}, Feige and {Yang}, Jinyi},
	doi = {10.3847/2041-8213/ae279c},
	eid = {L67},
	eprint = {2507.23066},
	journal = {\apjl},
	month = dec,
	number = {2},
	pages = {L67},
	primaryclass = {astro-ph.GA},
	title = {{SHELLQs─JWST Perspective on the Intrinsic Mass Relation between Supermassive Black Holes and Their Host Galaxies at z > 6}},
	volume = {995},
	year = 2025}

@article{Izumi2021,
	adsurl = {https://ui.adsabs.harvard.edu/abs/2021ApJ...908..235I},
	archiveprefix = {arXiv},
	author = {{Izumi}, Takuma and {Onoue}, Masafusa and {Matsuoka}, Yoshiki and {Strauss}, Michael A. and {Fujimoto}, Seiji and {Umehata}, Hideki and {Imanishi}, Masatoshi and {Kawamuro}, Taiki and {Nagao}, Tohru and {Toba}, Yoshiki and {Kohno}, Kotaro and {Kashikawa}, Nobunari and {Inayoshi}, Kohei and {Kawaguchi}, Toshihiro and {Iwasawa}, Kazushi and {Inoue}, Akio K. and {Goto}, Tomotsugu and {Baba}, Shunsuke and {Schramm}, Malte and {Suh}, Hyewon and {Harikane}, Yuichi and {Ueda}, Yoshihiro and {Silverman}, John D. and {Hashimoto}, Takuya and {Hashimoto}, Yasuhiro and {Ikarashi}, Soh and {Iono}, Daisuke and {Lee}, Chien-Hsiu and {Lee}, Kianhong and {Minezaki}, Takeo and {Nakanishi}, Kouichiro and {Nakano}, Suzuka and {Tamura}, Yoichi and {Tang}, Ji-Jia},
	doi = {10.3847/1538-4357/abd7ef},
	eid = {235},
	eprint = {2101.01199},
	journal = {\apj},
	month = feb,
	number = {2},
	pages = {235},
	primaryclass = {astro-ph.GA},
	title = {{Subaru High-z Exploration of Low-luminosity Quasars (SHELLQs). XII. Extended [C II] Structure (Merger or Outflow) in a z = 6.72 Red Quasar}},
	volume = {908},
	year = 2021}

@article{Yue2024,
	adsurl = {https://ui.adsabs.harvard.edu/abs/2024ApJ...966..176Y},
	archiveprefix = {arXiv},
	author = {{Yue}, Minghao and {Eilers}, Anna-Christina and {Simcoe}, Robert A. and {Mackenzie}, Ruari and {Matthee}, Jorryt and {Kashino}, Daichi and {Bordoloi}, Rongmon and {Lilly}, Simon J. and {Naidu}, Rohan P.},
	doi = {10.3847/1538-4357/ad3914},
	eid = {176},
	eprint = {2309.04614},
	journal = {\apj},
	month = may,
	number = {2},
	pages = {176},
	primaryclass = {astro-ph.GA},
	title = {{EIGER. V. Characterizing the Host Galaxies of Luminous Quasars at z {\ensuremath{\gtrsim}} 6}},
	volume = {966},
	year = 2024}

@article{DiazSantos2013,
	adsurl = {https://ui.adsabs.harvard.edu/abs/2013ApJ...774...68D},
	archiveprefix = {arXiv},
	author = {{D{\'\i}az-Santos}, T. and {Armus}, L. and {Charmandaris}, V. and {Stierwalt}, S. and {Murphy}, E.~J. and {Haan}, S. and {Inami}, H. and {Malhotra}, S. and {Meijerink}, R. and {Stacey}, G. and {Petric}, A.~O. and {Evans}, A.~S. and {Veilleux}, S. and {van der Werf}, P.~P. and {Lord}, S. and {Lu}, N. and {Howell}, J.~H. and {Appleton}, P. and {Mazzarella}, J.~M. and {Surace}, J.~A. and {Xu}, C.~K. and {Schulz}, B. and {Sanders}, D.~B. and {Bridge}, C. and {Chan}, B.~H.~P. and {Frayer}, D.~T. and {Iwasawa}, K. and {Melbourne}, J. and {Sturm}, E.},
	doi = {10.1088/0004-637X/774/1/68},
	eid = {68},
	eprint = {1307.2635},
	journal = {\apj},
	month = sep,
	number = {1},
	pages = {68},
	primaryclass = {astro-ph.CO},
	title = {{Explaining the [C II]157.7 {\ensuremath{\mu}}m Deficit in Luminous Infrared Galaxies{\textemdash}First Results from a Herschel/PACS Study of the GOALS Sample}},
	volume = {774},
	year = 2013}

@article{Farrah2013,
	adsurl = {https://ui.adsabs.harvard.edu/abs/2013ApJ...776...38F},
	archiveprefix = {arXiv},
	author = {{Farrah}, D. and {Lebouteiller}, V. and {Spoon}, H.~W.~W. and {Bernard-Salas}, J. and {Pearson}, C. and {Rigopoulou}, D. and {Smith}, H.~A. and {Gonz{\'a}lez-Alfonso}, E. and {Clements}, D.~L. and {Efstathiou}, A. and {Cormier}, D. and {Afonso}, J. and {Petty}, S.~M. and {Harris}, K. and {Hurley}, P. and {Borys}, C. and {Verma}, A. and {Cooray}, A. and {Salvatelli}, V.},
	doi = {10.1088/0004-637X/776/1/38},
	eid = {38},
	eprint = {1308.4165},
	journal = {\apj},
	month = oct,
	number = {1},
	pages = {38},
	primaryclass = {astro-ph.CO},
	title = {{Far-infrared Fine-structure Line Diagnostics of Ultraluminous Infrared Galaxies}},
	volume = {776},
	year = 2013}

@article{Goicoechea2015,
	adsurl = {https://ui.adsabs.harvard.edu/abs/2015ApJ...812...75G},
	archiveprefix = {arXiv},
	author = {{Goicoechea}, Javier R. and {Teyssier}, D. and {Etxaluze}, M. and {Goldsmith}, P.~F. and {Ossenkopf}, V. and {Gerin}, M. and {Bergin}, E.~A. and {Black}, J.~H. and {Cernicharo}, J. and {Cuadrado}, S. and {Encrenaz}, P. and {Falgarone}, E. and {Fuente}, A. and {Hacar}, A. and {Lis}, D.~C. and {Marcelino}, N. and {Melnick}, G.~J. and {M{\"u}ller}, H.~S.~P. and {Persson}, C. and {Pety}, J. and {R{\"o}llig}, M. and {Schilke}, P. and {Simon}, R. and {Snell}, R.~L. and {Stutzki}, J.},
	doi = {10.1088/0004-637X/812/1/75},
	eid = {75},
	eprint = {1508.03801},
	journal = {\apj},
	month = oct,
	number = {1},
	pages = {75},
	primaryclass = {astro-ph.GA},
	title = {{Velocity-resolved [CII] Emission and [CII]/FIR Mapping along Orion with Herschel}},
	volume = {812},
	year = 2015}

@article{Appleton2013,
	adsurl = {https://ui.adsabs.harvard.edu/abs/2013ApJ...777...66A},
	archiveprefix = {arXiv},
	author = {{Appleton}, P.~N. and {Guillard}, P. and {Boulanger}, F. and {Cluver}, M.~E. and {Ogle}, P. and {Falgarone}, E. and {Pineau des For{\^e}ts}, G. and {O'Sullivan}, E. and {Duc}, P.-A. and {Gallagher}, S. and {Gao}, Y. and {Jarrett}, T. and {Konstantopoulos}, I. and {Lisenfeld}, U. and {Lord}, S. and {Lu}, N. and {Peterson}, B.~W. and {Struck}, C. and {Sturm}, E. and {Tuffs}, R. and {Valchanov}, I. and {van der Werf}, P. and {Xu}, K.~C.},
	doi = {10.1088/0004-637X/777/1/66},
	eid = {66},
	eprint = {1309.1525},
	journal = {\apj},
	month = nov,
	number = {1},
	pages = {66},
	primaryclass = {astro-ph.CO},
	title = {{Shock-enhanced C$^{+}$ Emission and the Detection of H$_{2}$O from the Stephan's Quintet Group-wide Shock Using Herschel}},
	volume = {777},
	year = 2013}

@article{Sargsyan2012,
	adsurl = {https://ui.adsabs.harvard.edu/abs/2012ApJ...755..171S},
	archiveprefix = {arXiv},
	author = {{Sargsyan}, L. and {Lebouteiller}, V. and {Weedman}, D. and {Spoon}, H. and {Bernard-Salas}, J. and {Engels}, D. and {Stacey}, G. and {Houck}, J. and {Barry}, D. and {Miles}, J. and {Samsonyan}, A.},
	doi = {10.1088/0004-637X/755/2/171},
	eid = {171},
	eprint = {1206.5435},
	journal = {\apj},
	month = aug,
	number = {2},
	pages = {171},
	primaryclass = {astro-ph.CO},
	title = {{[C II] 158 {\ensuremath{\mu}}m Luminosities and Star Formation Rate in Dusty Starbursts and Active Galactic Nuclei}},
	volume = {755},
	year = 2012}

@article{Malhotra2001,
	adsurl = {https://ui.adsabs.harvard.edu/abs/2001ApJ...561..766M},
	archiveprefix = {arXiv},
	author = {{Malhotra}, S. and {Kaufman}, M.~J. and {Hollenbach}, D. and {Helou}, G. and {Rubin}, R.~H. and {Brauher}, J. and {Dale}, D. and {Lu}, N.~Y. and {Lord}, S. and {Stacey}, G. and {Contursi}, A. and {Hunter}, D.~A. and {Dinerstein}, H.},
	doi = {10.1086/323046},
	eprint = {astro-ph/0106485},
	journal = {\apj},
	month = nov,
	number = {2},
	pages = {766-786},
	primaryclass = {astro-ph},
	title = {{Far-Infrared Spectroscopy of Normal Galaxies: Physical Conditions in the Interstellar Medium}},
	volume = {561},
	year = 2001}

@article{Walter1997,
	adsurl = {https://ui.adsabs.harvard.edu/abs/1997AJ....113.2031W},
	author = {{Walter}, Fabian and {Brinks}, Elias and {Duric}, Neb and {Klein}, Ulrich},
	doi = {10.1086/118415},
	journal = {\aj},
	month = jun,
	pages = {2031-2045},
	title = {{A Dynamical Analysis of the HII Galaxy II Zwicky 33 and Its Low Surface Brightness Companion}},
	volume = {113},
	year = 1997}

@article{Appleton2023,
	adsurl = {https://ui.adsabs.harvard.edu/abs/2023ApJ...951..104A},
	archiveprefix = {arXiv},
	author = {{Appleton}, P.~N. and {Guillard}, P. and {Emonts}, Bjorn and {Boulanger}, Francois and {Togi}, Aditya and {Reach}, William T. and {Alatalo}, Katherine and {Cluver}, M. and {Diaz Santos}, T. and {Duc}, P.-A. and {Gallagher}, S. and {Ogle}, P. and {O'Sullivan}, E. and {Voggel}, K. and {Xu}, C.~K.},
	doi = {10.3847/1538-4357/accc2a},
	eid = {104},
	eprint = {2301.02928},
	journal = {\apj},
	month = jul,
	number = {2},
	pages = {104},
	primaryclass = {astro-ph.GA},
	title = {{Multiphase Gas Interactions on Subarcsec Scales in the Shocked Intergalactic Medium of Stephan's Quintet with JWST and ALMA}},
	volume = {951},
	year = 2023}

@article{Alatalo2011,
	adsurl = {https://ui.adsabs.harvard.edu/abs/2011ApJ...735...88A},
	archiveprefix = {arXiv},
	author = {{Alatalo}, K. and {Blitz}, L. and {Young}, L.~M. and {Davis}, T.~A. and {Bureau}, M. and {Lopez}, L.~A. and {Cappellari}, M. and {Scott}, N. and {Shapiro}, K.~L. and {Crocker}, A.~F. and {Mart{\'\i}n}, S. and {Bois}, M. and {Bournaud}, F. and {Davies}, R.~L. and {de Zeeuw}, P.~T. and {Duc}, P.-A. and {Emsellem}, E. and {Falc{\'o}n-Barroso}, J. and {Khochfar}, S. and {Krajnovi{\'c}}, D. and {Kuntschner}, H. and {Lablanche}, P.-Y. and {McDermid}, R.~M. and {Morganti}, R. and {Naab}, T. and {Oosterloo}, T. and {Sarzi}, M. and {Serra}, P. and {Weijmans}, A.},
	doi = {10.1088/0004-637X/735/2/88},
	eid = {88},
	eprint = {1104.2326},
	journal = {\apj},
	month = jul,
	number = {2},
	pages = {88},
	primaryclass = {astro-ph.CO},
	title = {{Discovery of an Active Galactic Nucleus Driven Molecular Outflow in the Local Early-type Galaxy NGC 1266}},
	volume = {735},
	year = 2011}

@article{Jorsater1995,
	adsurl = {https://ui.adsabs.harvard.edu/abs/1995AJ....110.2037J},
	author = {{Jorsater}, Steven and {van Moorsel}, Gustaaf A.},
	doi = {10.1086/117668},
	journal = {\aj},
	month = nov,
	pages = {2037},
	title = {{High Resolution Neutral Hydrogen Observations of the Barred Spiral Galaxy NGC 1365}},
	volume = {110},
	year = 1995}

@article{Czekala2021,
	adsurl = {https://ui.adsabs.harvard.edu/abs/2021ApJS..257....2C},
	archiveprefix = {arXiv},
	author = {{Czekala}, Ian and {Loomis}, Ryan A. and {Teague}, Richard and {Booth}, Alice S. and {Huang}, Jane and {Cataldi}, Gianni and {Ilee}, John D. and {Law}, Charles J. and {Walsh}, Catherine and {Bosman}, Arthur D. and {Guzm{\'a}n}, Viviana V. and {Le Gal}, Romane and {{\"O}berg}, Karin I. and {Yamato}, Yoshihide and {Aikawa}, Yuri and {Andrews}, Sean M. and {Bae}, Jaehan and {Bergin}, Edwin A. and {Bergner}, Jennifer B. and {Cleeves}, L. Ilsedore and {Kurtovic}, Nicolas T. and {M{\'e}nard}, Fran{\c{c}}ois and {Nomura}, Hideko and {P{\'e}rez}, Laura M. and {Qi}, Chunhua and {Schwarz}, Kamber R. and {Tsukagoshi}, Takashi and {Waggoner}, Abygail R. and {Wilner}, David J. and {Zhang}, Ke},
	doi = {10.3847/1538-4365/ac1430},
	eid = {2},
	eprint = {2109.06188},
	journal = {\apjs},
	month = nov,
	number = {1},
	pages = {2},
	primaryclass = {astro-ph.EP},
	title = {{Molecules with ALMA at Planet-forming Scales (MAPS). II. CLEAN Strategies for Synthesizing Images of Molecular Line Emission in Protoplanetary Disks}},
	volume = {257},
	year = 2021}

@article{Mazzucchelli2023,
	adsurl = {https://ui.adsabs.harvard.edu/abs/2023A&A...676A..71M},
	archiveprefix = {arXiv},
	author = {{Mazzucchelli}, C. and {Bischetti}, M. and {D'Odorico}, V. and {Feruglio}, C. and {Schindler}, J.-T. and {Onoue}, M. and {Ba{\~n}ados}, E. and {Becker}, G.~D. and {Bian}, F. and {Carniani}, S. and {Decarli}, R. and {Eilers}, A.-C. and {Farina}, E.~P. and {Gallerani}, S. and {Lai}, S. and {Meyer}, R.~A. and {Rojas-Ruiz}, S. and {Satyavolu}, S. and {Venemans}, B.~P. and {Wang}, F. and {Yang}, J. and {Zhu}, Y.},
	doi = {10.1051/0004-6361/202346317},
	eid = {A71},
	eprint = {2306.16474},
	journal = {\aap},
	month = aug,
	pages = {A71},
	primaryclass = {astro-ph.GA},
	title = {{XQR-30: Black hole masses and accretion rates of 42 z {\ensuremath{\gtrsim}} 6 quasars}},
	volume = {676},
	year = 2023}

@article{Vernet2011,
	adsurl = {https://ui.adsabs.harvard.edu/abs/2011A&A...536A.105V},
	archiveprefix = {arXiv},
	author = {{Vernet}, J. and {Dekker}, H. and {D'Odorico}, S. and {Kaper}, L. and {Kjaergaard}, P. and {Hammer}, F. and {Randich}, S. and {Zerbi}, F. and {Groot}, P.~J. and {Hjorth}, J. and {Guinouard}, I. and {Navarro}, R. and {Adolfse}, T. and {Albers}, P.~W. and {Amans}, J.-P. and {Andersen}, J.~J. and {Andersen}, M.~I. and {Binetruy}, P. and {Bristow}, P. and {Castillo}, R. and {Chemla}, F. and {Christensen}, L. and {Conconi}, P. and {Conzelmann}, R. and {Dam}, J. and {de Caprio}, V. and {de Ugarte Postigo}, A. and {Delabre}, B. and {di Marcantonio}, P. and {Downing}, M. and {Elswijk}, E. and {Finger}, G. and {Fischer}, G. and {Flores}, H. and {Fran{\c{c}}ois}, P. and {Goldoni}, P. and {Guglielmi}, L. and {Haigron}, R. and {Hanenburg}, H. and {Hendriks}, I. and {Horrobin}, M. and {Horville}, D. and {Jessen}, N.~C. and {Kerber}, F. and {Kern}, L. and {Kiekebusch}, M. and {Kleszcz}, P. and {Klougart}, J. and {Kragt}, J. and {Larsen}, H.~H. and {Lizon}, J.-L. and {Lucuix}, C. and {Mainieri}, V. and {Manuputy}, R. and {Martayan}, C. and {Mason}, E. and {Mazzoleni}, R. and {Michaelsen}, N. and {Modigliani}, A. and {Moehler}, S. and {M{\o}ller}, P. and {Norup S{\o}rensen}, A. and {N{\o}rregaard}, P. and {P{\'e}roux}, C. and {Patat}, F. and {Pena}, E. and {Pragt}, J. and {Reinero}, C. and {Rigal}, F. and {Riva}, M. and {Roelfsema}, R. and {Royer}, F. and {Sacco}, G. and {Santin}, P. and {Schoenmaker}, T. and {Spano}, P. and {Sweers}, E. and {Ter Horst}, R. and {Tintori}, M. and {Tromp}, N. and {van Dael}, P. and {van der Vliet}, H. and {Venema}, L. and {Vidali}, M. and {Vinther}, J. and {Vola}, P. and {Winters}, R. and {Wistisen}, D. and {Wulterkens}, G. and {Zacchei}, A.},
	doi = {10.1051/0004-6361/201117752},
	eid = {A105},
	eprint = {1110.1944},
	journal = {\aap},
	month = dec,
	pages = {A105},
	primaryclass = {astro-ph.IM},
	title = {{X-shooter, the new wide band intermediate resolution spectrograph at the ESO Very Large Telescope}},
	volume = {536},
	year = 2011}

@article{Li2022,
	adsurl = {https://ui.adsabs.harvard.edu/abs/2022ApJ...931L..11L},
	archiveprefix = {arXiv},
	author = {{Li}, Junyao and {Silverman}, John D. and {Izumi}, Takuma and {He}, Wanqiu and {Akiyama}, Masayuki and {Inayoshi}, Kohei and {Matsuoka}, Yoshiki and {Onoue}, Masafusa and {Toba}, Yoshiki},
	doi = {10.3847/2041-8213/ac6de8},
	eid = {L11},
	eprint = {2203.10663},
	journal = {\apjl},
	month = may,
	number = {1},
	pages = {L11},
	primaryclass = {astro-ph.GA},
	title = {{On the Connection between Supermassive Black Holes and Galaxy Growth in the Reionization Epoch}},
	volume = {931},
	year = 2022}

@article{Lauer2007,
	adsurl = {https://ui.adsabs.harvard.edu/abs/2007ApJ...670..249L},
	archiveprefix = {arXiv},
	author = {{Lauer}, Tod R. and {Tremaine}, Scott and {Richstone}, Douglas and {Faber}, S.~M.},
	doi = {10.1086/522083},
	eprint = {0705.4103},
	journal = {\apj},
	month = nov,
	number = {1},
	pages = {249-260},
	primaryclass = {astro-ph},
	title = {{Selection Bias in Observing the Cosmological Evolution of the M$_{{\textbullet}}$-{\ensuremath{\sigma}} and M$_{{\textbullet}}$-L Relationships}},
	volume = {670},
	year = 2007}

@article{Shao2017,
	adsurl = {https://ui.adsabs.harvard.edu/abs/2017ApJ...845..138S},
	archiveprefix = {arXiv},
	author = {{Shao}, Yali and {Wang}, Ran and {Jones}, Gareth C. and {Carilli}, Chris L. and {Walter}, Fabian and {Fan}, Xiaohui and {Riechers}, Dominik A. and {Bertoldi}, Frank and {Wagg}, Jeff and {Strauss}, Michael A. and {Omont}, Alain and {Cox}, Pierre and {Jiang}, Linhua and {Narayanan}, Desika and {Menten}, Karl M.},
	doi = {10.3847/1538-4357/aa826c},
	eid = {138},
	eprint = {1707.03078},
	journal = {\apj},
	month = aug,
	number = {2},
	pages = {138},
	primaryclass = {astro-ph.GA},
	title = {{Gas Dynamics of a Luminous z = 6.13 Quasar ULAS J1319+0950 Revealed by ALMA High-resolution Observations}},
	volume = {845},
	year = 2017}

@article{Willott2015,
	adsurl = {https://ui.adsabs.harvard.edu/abs/2015ApJ...801..123W},
	archiveprefix = {arXiv},
	author = {{Willott}, Chris J. and {Bergeron}, Jacqueline and {Omont}, Alain},
	doi = {10.1088/0004-637X/801/2/123},
	eid = {123},
	eprint = {1501.07538},
	journal = {\apj},
	month = mar,
	number = {2},
	pages = {123},
	primaryclass = {astro-ph.GA},
	title = {{Star Formation Rate and Dynamical Mass of {}10$^{8}$ Solar Mass Black Hole Host Galaxies At Redshift 6}},
	volume = {801},
	year = 2015}

@article{Venemans2016,
	adsurl = {https://ui.adsabs.harvard.edu/abs/2016ApJ...816...37V},
	archiveprefix = {arXiv},
	author = {{Venemans}, Bram P. and {Walter}, Fabian and {Zschaechner}, Laura and {Decarli}, Roberto and {De Rosa}, Gisella and {Findlay}, Joseph R. and {McMahon}, Richard G. and {Sutherland}, Will J.},
	doi = {10.3847/0004-637X/816/1/37},
	eid = {37},
	eprint = {1511.07432},
	journal = {\apj},
	month = jan,
	number = {1},
	pages = {37},
	primaryclass = {astro-ph.GA},
	title = {{Bright [C II] and Dust Emission in Three z > 6.6 Quasar Host Galaxies Observed by ALMA}},
	volume = {816},
	year = 2016}

@article{Pensabene2020,
	adsurl = {https://ui.adsabs.harvard.edu/abs/2020A&A...637A..84P},
	archiveprefix = {arXiv},
	author = {{Pensabene}, A. and {Carniani}, S. and {Perna}, M. and {Cresci}, G. and {Decarli}, R. and {Maiolino}, R. and {Marconi}, A.},
	doi = {10.1051/0004-6361/201936634},
	eid = {A84},
	eprint = {2002.00958},
	journal = {\aap},
	month = may,
	pages = {A84},
	primaryclass = {astro-ph.GA},
	title = {{The ALMA view of the high-redshift relation between supermassive black holes and their host galaxies}},
	volume = {637},
	year = 2020}

@article{Wilde2026,
	adsurl = {https://ui.adsabs.harvard.edu/abs/2026ApJ...996....8W},
	archiveprefix = {arXiv},
	author = {{Wilde}, Aurora and {Neeleman}, Marcel and {Meyer}, Romain A. and {Decarli}, Roberto and {Walter}, Fabian and {Frye}, Brenda L. and {Fan}, Xiaohui},
	doi = {10.3847/1538-4357/ae1d42},
	eid = {8},
	eprint = {2511.03659},
	journal = {\apj},
	month = jan,
	number = {1},
	pages = {8},
	primaryclass = {astro-ph.GA},
	title = {{ALMA and JWST Imaging of z > 6 Quasars: No Spatial Position Offset Observed between Quasars and Their Host Galaxies}},
	volume = {996},
	year = 2026}

@article{Meyer2023,
	adsurl = {https://ui.adsabs.harvard.edu/abs/2023ApJ...956..127M},
	archiveprefix = {arXiv},
	author = {{Meyer}, Romain A. and {Neeleman}, Marcel and {Walter}, Fabian and {Venemans}, Bram},
	doi = {10.3847/1538-4357/acf047},
	eid = {127},
	eprint = {2308.03477},
	journal = {\apj},
	month = oct,
	number = {2},
	pages = {127},
	primaryclass = {astro-ph.GA},
	title = {{ALMA 300 pc Resolution Imaging of a z = 6.79 Quasar: No Evidence for Supermassive Black Hole Influence on the C II Kinematics}},
	volume = {956},
	year = 2023}

@article{Willott2017,
	adsurl = {https://ui.adsabs.harvard.edu/abs/2017ApJ...850..108W},
	archiveprefix = {arXiv},
	author = {{Willott}, Chris J. and {Bergeron}, Jacqueline and {Omont}, Alain},
	doi = {10.3847/1538-4357/aa921b},
	eid = {108},
	eprint = {1710.02212},
	journal = {\apj},
	month = nov,
	number = {1},
	pages = {108},
	primaryclass = {astro-ph.GA},
	title = {{A Wide Dispersion in Star Formation Rate and Dynamical Mass of {}10$^{8}$ Solar Mass Black Hole Host Galaxies at Redshift 6}},
	volume = {850},
	year = 2017}

@article{Bouwens2025,
	adsurl = {https://ui.adsabs.harvard.edu/abs/2025arXiv250624128B},
	archiveprefix = {arXiv},
	author = {{Bouwens}, R.~J. and {Banados}, E. and {Decarli}, R. and {Hennawi}, J. and {Yang}, D. and {Algera}, H. and {Aravena}, M. and {Farina}, E. and {Gloudemans}, A. and {Hodge}, J. and {Inami}, H. and {Matthee}, J. and {Meyer}, R. and {Naidu}, R.~P. and {Oesch}, P. and {Rottgering}, H.~J.~A. and {Schouws}, S. and {Smit}, R. and {Stefanon}, M. and {van der Werf}, P. and {Venemans}, B. and {Walter}, F. and {Fudamoto}, Y.},
	doi = {10.48550/arXiv.2506.24128},
	eid = {arXiv:2506.24128},
	eprint = {2506.24128},
	journal = {arXiv e-prints},
	month = jun,
	pages = {arXiv:2506.24128},
	primaryclass = {astro-ph.GA},
	title = {{Beneath the Surface: >85\% of z>5.9 QSOs in Massive Host Galaxies are UV-Faint}},
	year = 2025}

@article{Izumi2018,
	adsurl = {https://ui.adsabs.harvard.edu/abs/2018PASJ...70...36I},
	archiveprefix = {arXiv},
	author = {{Izumi}, Takuma and {Onoue}, Masafusa and {Shirakata}, Hikari and {Nagao}, Tohru and {Kohno}, Kotaro and {Matsuoka}, Yoshiki and {Imanishi}, Masatoshi and {Strauss}, Michael A. and {Kashikawa}, Nobunari and {Schulze}, Andreas and {Silverman}, John D. and {Fujimoto}, Seiji and {Harikane}, Yuichi and {Toba}, Yoshiki and {Umehata}, Hideki and {Nakanishi}, Kouichiro and {Greene}, Jenny E. and {Tamura}, Yoichi and {Taniguchi}, Akio and {Yamaguchi}, Yuki and {Goto}, Tomotsugu and {Hashimoto}, Yasuhiro and {Ikarashi}, Soh and {Iono}, Daisuke and {Iwasawa}, Kazushi and {Lee}, Chien-Hsiu and {Makiya}, Ryu and {Minezaki}, Takeo and {Tang}, Ji-Jia},
	doi = {10.1093/pasj/psy026},
	eid = {36},
	eprint = {1802.05742},
	journal = {\pasj},
	month = jun,
	number = {3},
	pages = {36},
	primaryclass = {astro-ph.GA},
	title = {{Subaru High-z Exploration of Low-Luminosity Quasars (SHELLQs). III. Star formation properties of the host galaxies at z {\ensuremath{\gtrsim}} 6 studied with ALMA}},
	volume = {70},
	year = 2018}

@article{Wang2013,
	adsurl = {https://ui.adsabs.harvard.edu/abs/2013ApJ...773...44W},
	archiveprefix = {arXiv},
	author = {{Wang}, Ran and {Wagg}, Jeff and {Carilli}, Chris L. and {Walter}, Fabian and {Lentati}, Lindley and {Fan}, Xiaohui and {Riechers}, Dominik A. and {Bertoldi}, Frank and {Narayanan}, Desika and {Strauss}, Michael A. and {Cox}, Pierre and {Omont}, Alain and {Menten}, Karl M. and {Knudsen}, Kirsten K. and {Neri}, Roberto and {Jiang}, Linhua},
	doi = {10.1088/0004-637X/773/1/44},
	eid = {44},
	eprint = {1302.4154},
	journal = {\apj},
	month = aug,
	number = {1},
	pages = {44},
	primaryclass = {astro-ph.CO},
	title = {{Star Formation and Gas Kinematics of Quasar Host Galaxies at z \raisebox{-0.5ex}\textasciitilde 6: New Insights from ALMA}},
	volume = {773},
	year = 2013}

@article{Mechtley2012,
	adsurl = {https://ui.adsabs.harvard.edu/abs/2012ApJ...756L..38M},
	archiveprefix = {arXiv},
	author = {{Mechtley}, M. and {Windhorst}, R.~A. and {Ryan}, R.~E. and {Schneider}, G. and {Cohen}, S.~H. and {Jansen}, R.~A. and {Fan}, X. and {Hathi}, N.~P. and {Keel}, W.~C. and {Koekemoer}, A.~M. and {R{\"o}ttgering}, H. and {Scannapieco}, E. and {Schneider}, D.~P. and {Strauss}, M.~A. and {Yan}, H.~J.},
	doi = {10.1088/2041-8205/756/2/L38},
	eid = {L38},
	eprint = {1207.3283},
	journal = {\apjl},
	month = sep,
	number = {2},
	pages = {L38},
	primaryclass = {astro-ph.CO},
	title = {{Near-infrared Imaging of a z = 6.42 Quasar Host Galaxy with the Hubble Space Telescope Wide Field Camera 3}},
	volume = {756},
	year = 2012}

@article{Marshall2025a,
	adsurl = {https://ui.adsabs.harvard.edu/abs/2025A&A...702A..50M},
	archiveprefix = {arXiv},
	author = {{Marshall}, Madeline A. and {Yue}, Minghao and {Eilers}, Anna-Christina and {Scholtz}, Jan and {Perna}, Michele and {Willott}, Chris J. and {Maiolino}, Roberto and {{\"U}bler}, Hannah and {Arribas}, Santiago and {Bunker}, Andrew J. and {Charlot}, Stephane and {Rodr{\'\i}guez Del Pino}, Bruno and {B{\"o}ker}, Torsten and {Carniani}, Stefano and {Circosta}, Chiara and {Cresci}, Giovanni and {D'Eugenio}, Francesco and {Jones}, Gareth C. and {Venturi}, Giacomo and {Bordoloi}, Rongmon and {Kashino}, Daichi and {Mackenzie}, Ruari and {Matthee}, Jorryt and {Naidu}, Rohan and {Simcoe}, Robert A.},
	doi = {10.1051/0004-6361/202452650},
	eid = {A50},
	eprint = {2410.11035},
	journal = {\aap},
	month = oct,
	pages = {A50},
	primaryclass = {astro-ph.GA},
	title = {{GA-NIFS and EIGER: A merging quasar host at z = 7 with an overmassive black hole}},
	volume = {702},
	year = 2025}

@article{Marshall2025b,
	adsurl = {https://ui.adsabs.harvard.edu/abs/2025A&A...702A.174M},
	archiveprefix = {arXiv},
	author = {{Marshall}, Madeline A. and {Windhorst}, Rogier A. and {Ferrami}, Giovanni and {Willner}, S.~P. and {del Carmen Polletta}, Maria and {Keel}, William C. and {Fazio}, Giovanni G. and {Cohen}, Seth H. and {Carleton}, Timothy and {Jansen}, Rolf A. and {Honor}, Rachel and {Ortiz}, III, Rafael and {Summers}, Jake and {D'Silva}, Jordan C.~J. and {Koekemoer}, Anton M. and {Coe}, Dan and {Conselice}, Christopher J. and {Diego}, Jose M. and {Driver}, Simon P. and {Frye}, Brenda and {Grogin}, Norman A. and {Pirzkal}, Nor and {Robotham}, Aaron and {Ryan}, Jr., Russell E. and {Willmer}, Christopher N.~A. and {Yan}, Haojing and {Ricotti}, Massimo and {Zitrin}, Adi and {Adams}, Nathan J. and {Cheng}, Cheng and {Wyithe}, J. Stuart B. and {Lim}, Jeremy and {Perna}, Michele and {{\"U}bler}, Hannah and {Willott}, Chris J. and {Jones}, Gareth and {Scholtz}, Jan and {Mechtley}, Mira},
	doi = {10.1051/0004-6361/202554307},
	eid = {A174},
	eprint = {2502.20550},
	journal = {\aap},
	month = oct,
	pages = {A174},
	primaryclass = {astro-ph.GA},
	title = {{JWST's PEARLS: A z ≃ 6 quasar in a train-wreck galaxy merger system}},
	volume = {702},
	year = 2025}

@article{Stone2025,
	adsurl = {https://ui.adsabs.harvard.edu/abs/2025ApJ...993..168S},
	archiveprefix = {arXiv},
	author = {{Stone}, Meredith A. and {Rieke}, George H. and {Lyu}, Jianwei and {Florian}, Michael K. and {Hainline}, Kevin N. and {Sun}, Yang and {Zhu}, Yongda},
	doi = {10.3847/1538-4357/ae062a},
	eid = {168},
	eprint = {2507.13489},
	journal = {\apj},
	month = nov,
	number = {2},
	pages = {168},
	primaryclass = {astro-ph.GA},
	title = {{The z = 7.08 Quasar ULAS J1120+0641 May Never Reach a ``Normal'' Black Hole to Stellar Mass Ratio}},
	volume = {993},
	year = 2025}

@article{Stone2024,
	adsurl = {https://ui.adsabs.harvard.edu/abs/2024ApJ...964...90S},
	archiveprefix = {arXiv},
	author = {{Stone}, Meredith A. and {Lyu}, Jianwei and {Rieke}, George H. and {Alberts}, Stacey and {Hainline}, Kevin N.},
	doi = {10.3847/1538-4357/ad2a57},
	eid = {90},
	eprint = {2310.18395},
	journal = {\apj},
	month = mar,
	number = {1},
	pages = {90},
	primaryclass = {astro-ph.GA},
	title = {{Undermassive Host Galaxies of Five z {\ensuremath{\sim}} 6 Luminous Quasars Detected with JWST}},
	volume = {964},
	year = 2024}

@article{Yue2023,
	adsurl = {https://ui.adsabs.harvard.edu/abs/2023ApJ...950..105Y},
	archiveprefix = {arXiv},
	author = {{Yue}, Minghao and {Eilers}, Anna-Christina and {Simcoe}, Robert A. and {Belli}, Sirio and {Davies}, Frederick B. and {DePalma}, David and {Hennawi}, Joseph F. and {Mason}, Charlotte A. and {Mu{\~n}oz}, Julian B. and {Nelson}, Erica J. and {Tacchella}, Sandro},
	doi = {10.3847/1538-4357/accf20},
	eid = {105},
	eprint = {2304.09256},
	journal = {\apj},
	month = jun,
	number = {2},
	pages = {105},
	primaryclass = {astro-ph.GA},
	title = {{Detecting and Characterizing Young Quasars. III. The Impact of Gravitational Lensing Magnification}},
	volume = {950},
	year = 2023}

@article{Ding2025,
	adsurl = {https://ui.adsabs.harvard.edu/abs/2025ApJ...993...91D},
	archiveprefix = {arXiv},
	author = {{Ding}, Xuheng and {Onoue}, Masafusa and {Silverman}, John D. and {Matsuoka}, Yoshiki and {Izumi}, Takuma and {Strauss}, Michael A. and {Yang}, Lilan and {Jahnke}, Knud and {Phillips}, Camryn L. and {Treu}, Tommaso and {Andika}, Irham T. and {Aoki}, Kentaro and {Arita}, Junya and {Baba}, Shunsuke and {Bosman}, Sarah E.~I. and {Eilers}, Anna-Christina and {Fujimoto}, Seiji and {Haiman}, Zoltan and {Imanishi}, Masatoshi and {Inayoshi}, Kohei and {Iwasawa}, Kazushi and {Kartaltepe}, Jeyhan and {Kashikawa}, Nobunari and {Kawaguchi}, Toshihiro and {Li}, Junyao and {Lee}, Chien-Hsiu and {Lupi}, Alessandro and {Schindler}, Jan-Torge and {Schramm}, Malte and {Shimasaku}, Kazuhiro and {Shuntov}, Marko and {Tanaka}, Takumi S. and {Toba}, Yoshiki and {Trakhtenbrot}, Benny and {Umehata}, Hideki and {Vestergaard}, Marianne and {Wang}, Feige and {Yang}, Jinyi},
	doi = {10.3847/1538-4357/ae045b},
	eid = {91},
	eprint = {2505.03876},
	journal = {\apj},
	month = nov,
	number = {1},
	pages = {91},
	primaryclass = {astro-ph.GA},
	title = {{SHELLQs-JWST Unveils the Host Galaxies of 12 Quasars at z > 6}},
	volume = {993},
	year = 2025}

@article{Ding2023,
	adsurl = {https://ui.adsabs.harvard.edu/abs/2023Natur.621...51D},
	archiveprefix = {arXiv},
	author = {{Ding}, Xuheng and {Onoue}, Masafusa and {Silverman}, John D. and {Matsuoka}, Yoshiki and {Izumi}, Takuma and {Strauss}, Michael A. and {Jahnke}, Knud and {Phillips}, Camryn L. and {Li}, Junyao and {Volonteri}, Marta and {Haiman}, Zoltan and {Andika}, Irham Taufik and {Aoki}, Kentaro and {Baba}, Shunsuke and {Bieri}, Rebekka and {Bosman}, Sarah E.~I. and {Bottrell}, Connor and {Eilers}, Anna-Christina and {Fujimoto}, Seiji and {Habouzit}, Melanie and {Imanishi}, Masatoshi and {Inayoshi}, Kohei and {Iwasawa}, Kazushi and {Kashikawa}, Nobunari and {Kawaguchi}, Toshihiro and {Kohno}, Kotaro and {Lee}, Chien-Hsiu and {Lupi}, Alessandro and {Lyu}, Jianwei and {Nagao}, Tohru and {Overzier}, Roderik and {Schindler}, Jan-Torge and {Schramm}, Malte and {Shimasaku}, Kazuhiro and {Toba}, Yoshiki and {Trakhtenbrot}, Benny and {Trebitsch}, Maxime and {Treu}, Tommaso and {Umehata}, Hideki and {Venemans}, Bram P. and {Vestergaard}, Marianne and {Walter}, Fabian and {Wang}, Feige and {Yang}, Jinyi},
	doi = {10.1038/s41586-023-06345-5},
	eprint = {2211.14329},
	journal = {\nat},
	month = sep,
	number = {7977},
	pages = {51-55},
	primaryclass = {astro-ph.GA},
	title = {{Detection of stellar light from quasar host galaxies at redshifts above 6}},
	volume = {621},
	year = 2023}

@article{Fan2023,
	adsurl = {https://ui.adsabs.harvard.edu/abs/2023ARA&A..61..373F},
	archiveprefix = {arXiv},
	author = {{Fan}, Xiaohui and {Ba{\~n}ados}, Eduardo and {Simcoe}, Robert A.},
	doi = {10.1146/annurev-astro-052920-102455},
	eprint = {2212.06907},
	journal = {\araa},
	month = aug,
	pages = {373-426},
	primaryclass = {astro-ph.GA},
	title = {{Quasars and the Intergalactic Medium at Cosmic Dawn}},
	volume = {61},
	year = 2023}

@article{Habouzit2022,
	adsurl = {https://ui.adsabs.harvard.edu/abs/2022MNRAS.511.3751H},
	archiveprefix = {arXiv},
	author = {{Habouzit}, M{\'e}lanie and {Onoue}, Masafusa and {Ba{\~n}ados}, Eduardo and {Neeleman}, Marcel and {Angl{\'e}s-Alc{\'a}zar}, Daniel and {Walter}, Fabian and {Pillepich}, Annalisa and {Dav{\'e}}, Romeel and {Jahnke}, Knud and {Dubois}, Yohan},
	doi = {10.1093/mnras/stac225},
	eprint = {2201.09892},
	journal = {\mnras},
	month = apr,
	number = {3},
	pages = {3751-3767},
	primaryclass = {astro-ph.CO},
	title = {{Co-evolution of massive black holes and their host galaxies at high redshift: discrepancies from six cosmological simulations and the key role of JWST}},
	volume = {511},
	year = 2022}

@article{Kormendy2013,
	adsurl = {https://ui.adsabs.harvard.edu/abs/2013ARA&A..51..511K},
	archiveprefix = {arXiv},
	author = {{Kormendy}, John and {Ho}, Luis C.},
	doi = {10.1146/annurev-astro-082708-101811},
	eprint = {1304.7762},
	journal = {\araa},
	month = aug,
	number = {1},
	pages = {511-653},
	primaryclass = {astro-ph.CO},
	title = {{Coevolution (Or Not) of Supermassive Black Holes and Host Galaxies}},
	volume = {51},
	year = 2013}

@article{DeRosa2011,
	adsurl = {https://ui.adsabs.harvard.edu/abs/2011ApJ...739...56D},
	archiveprefix = {arXiv},
	author = {{De Rosa}, G. and {Decarli}, R. and {Walter}, F. and {Fan}, X. and {Jiang}, L. and {Kurk}, J. and {Pasquali}, A. and {Rix}, H.~W.},
	doi = {10.1088/0004-637X/739/2/56},
	eid = {56},
	eprint = {1106.5501},
	journal = {\apj},
	month = oct,
	number = {2},
	pages = {56},
	primaryclass = {astro-ph.CO},
	title = {{Evidence for Non-evolving Fe II/Mg II Ratios in Rapidly Accreting z \raisebox{-0.5ex}\textasciitilde 6 QSOs}},
	volume = {739},
	year = 2011}

@article{Mazzucchelli2017b,
	adsurl = {https://ui.adsabs.harvard.edu/abs/2017ApJ...849...91M},
	archiveprefix = {arXiv},
	author = {{Mazzucchelli}, C. and {Ba{\~n}ados}, E. and {Venemans}, B.~P. and {Decarli}, R. and {Farina}, E.~P. and {Walter}, F. and {Eilers}, A.-C. and {Rix}, H.-W. and {Simcoe}, R. and {Stern}, D. and {Fan}, X. and {Schlafly}, E. and {De Rosa}, G. and {Hennawi}, J. and {Chambers}, K.~C. and {Greiner}, J. and {Burgett}, W. and {Draper}, P.~W. and {Kaiser}, N. and {Kudritzki}, R.-P. and {Magnier}, E. and {Metcalfe}, N. and {Waters}, C. and {Wainscoat}, R.~J.},
	doi = {10.3847/1538-4357/aa9185},
	eid = {91},
	eprint = {1710.01251},
	journal = {\apj},
	month = nov,
	number = {2},
	pages = {91},
	primaryclass = {astro-ph.GA},
	title = {{Physical Properties of 15 Quasars at z {\ensuremath{\gtrsim}} 6.5}},
	volume = {849},
	year = 2017}

@article{Gebhardt2000,
	adsurl = {https://ui.adsabs.harvard.edu/abs/2000ApJ...539L..13G},
	archiveprefix = {arXiv},
	author = {{Gebhardt}, Karl and {Bender}, Ralf and {Bower}, Gary and {Dressler}, Alan and {Faber}, S.~M. and {Filippenko}, Alexei V. and {Green}, Richard and {Grillmair}, Carl and {Ho}, Luis C. and {Kormendy}, John and {Lauer}, Tod R. and {Magorrian}, John and {Pinkney}, Jason and {Richstone}, Douglas and {Tremaine}, Scott},
	doi = {10.1086/312840},
	eprint = {astro-ph/0006289},
	journal = {\apjl},
	month = aug,
	number = {1},
	pages = {L13-L16},
	primaryclass = {astro-ph},
	title = {{A Relationship between Nuclear Black Hole Mass and Galaxy Velocity Dispersion}},
	volume = {539},
	year = 2000}

@article{Ferrarese2000,
	adsurl = {https://ui.adsabs.harvard.edu/abs/2000ApJ...539L...9F},
	archiveprefix = {arXiv},
	author = {{Ferrarese}, Laura and {Merritt}, David},
	doi = {10.1086/312838},
	eprint = {astro-ph/0006053},
	journal = {\apjl},
	month = aug,
	number = {1},
	pages = {L9-L12},
	primaryclass = {astro-ph},
	title = {{A Fundamental Relation between Supermassive Black Holes and Their Host Galaxies}},
	volume = {539},
	year = 2000}

@article{Wang2021,
	adsurl = {https://ui.adsabs.harvard.edu/abs/2021ApJ...907L...1W},
	archiveprefix = {arXiv},
	author = {{Wang}, Feige and {Yang}, Jinyi and {Fan}, Xiaohui and {Hennawi}, Joseph F. and {Barth}, Aaron J. and {Banados}, Eduardo and {Bian}, Fuyan and {Boutsia}, Konstantina and {Connor}, Thomas and {Davies}, Frederick B. and {Decarli}, Roberto and {Eilers}, Anna-Christina and {Farina}, Emanuele Paolo and {Green}, Richard and {Jiang}, Linhua and {Li}, Jiang-Tao and {Mazzucchelli}, Chiara and {Nanni}, Riccardo and {Schindler}, Jan-Torge and {Venemans}, Bram and {Walter}, Fabian and {Wu}, Xue-Bing and {Yue}, Minghao},
	doi = {10.3847/2041-8213/abd8c6},
	eid = {L1},
	eprint = {2101.03179},
	journal = {\apjl},
	month = jan,
	number = {1},
	pages = {L1},
	primaryclass = {astro-ph.GA},
	title = {{A Luminous Quasar at Redshift 7.642}},
	volume = {907},
	year = 2021}

@article{Fan2001,
	adsurl = {https://ui.adsabs.harvard.edu/abs/2001AJ....122.2833F},
	archiveprefix = {arXiv},
	author = {{Fan}, Xiaohui and {Narayanan}, Vijay K. and {Lupton}, Robert H. and {Strauss}, Michael A. and {Knapp}, Gillian R. and {Becker}, Robert H. and {White}, Richard L. and {Pentericci}, Laura and {Leggett}, S.~K. and {Haiman}, Zolt{\'a}n and {Gunn}, James E. and {Ivezi{\'c}}, {\v{Z}}eljko and {Schneider}, Donald P. and {Anderson}, Scott F. and {Brinkmann}, J. and {Bahcall}, Neta A. and {Connolly}, Andrew J. and {Csabai}, Istv{\'a}n and {Doi}, Mamoru and {Fukugita}, Masataka and {Geballe}, Tom and {Grebel}, Eva K. and {Harbeck}, Daniel and {Hennessy}, Gregory and {Lamb}, Don Q. and {Miknaitis}, Gajus and {Munn}, Jeffrey A. and {Nichol}, Robert and {Okamura}, Sadanori and {Pier}, Jeffrey R. and {Prada}, Francisco and {Richards}, Gordon T. and {Szalay}, Alex and {York}, Donald G.},
	doi = {10.1086/324111},
	eprint = {astro-ph/0108063},
	journal = {\aj},
	month = dec,
	number = {6},
	pages = {2833-2849},
	primaryclass = {astro-ph},
	title = {{A Survey of z>5.8 Quasars in the Sloan Digital Sky Survey. I. Discovery of Three New Quasars and the Spatial Density of Luminous Quasars at z\raisebox{-0.5ex}\textasciitilde6}},
	volume = {122},
	year = 2001}

@article{Banados2018,
	adsurl = {https://ui.adsabs.harvard.edu/abs/2018Natur.553..473B},
	archiveprefix = {arXiv},
	author = {{Ba{\~n}ados}, Eduardo and {Venemans}, Bram P. and {Mazzucchelli}, Chiara and {Farina}, Emanuele P. and {Walter}, Fabian and {Wang}, Feige and {Decarli}, Roberto and {Stern}, Daniel and {Fan}, Xiaohui and {Davies}, Frederick B. and {Hennawi}, Joseph F. and {Simcoe}, Robert A. and {Turner}, Monica L. and {Rix}, Hans-Walter and {Yang}, Jinyi and {Kelson}, Daniel D. and {Rudie}, Gwen C. and {Winters}, Jan Martin},
	doi = {10.1038/nature25180},
	eprint = {1712.01860},
	journal = {\nat},
	month = jan,
	number = {7689},
	pages = {473-476},
	primaryclass = {astro-ph.GA},
	title = {{An 800-million-solar-mass black hole in a significantly neutral Universe at a redshift of 7.5}},
	volume = {553},
	year = 2018}

@article{Casey2014,
	adsurl = {https://ui.adsabs.harvard.edu/abs/2014PhR...541...45C},
	archiveprefix = {arXiv},
	author = {{Casey}, Caitlin M. and {Narayanan}, Desika and {Cooray}, Asantha},
	doi = {10.1016/j.physrep.2014.02.009},
	eprint = {1402.1456},
	journal = {\physrep},
	month = aug,
	number = {2},
	pages = {45-161},
	primaryclass = {astro-ph.CO},
	title = {{Dusty star-forming galaxies at high redshift}},
	volume = {541},
	year = 2014}

@article{DeGraaff2025,
	adsurl = {https://ui.adsabs.harvard.edu/abs/2025NatAs...9..280D},
	archiveprefix = {arXiv},
	author = {{de Graaff}, Anna and {Setton}, David J. and {Brammer}, Gabriel and {Cutler}, Sam and {Suess}, Katherine A. and {Labb{\'e}}, Ivo and {Leja}, Joel and {Weibel}, Andrea and {Maseda}, Michael V. and {Whitaker}, Katherine E. and {Bezanson}, Rachel and {Boogaard}, Leindert A. and {Cleri}, Nikko J. and {De Lucia}, Gabriella and {Franx}, Marijn and {Greene}, Jenny E. and {Hirschmann}, Michaela and {Matthee}, Jorryt and {McConachie}, Ian and {Naidu}, Rohan P. and {Oesch}, Pascal A. and {Price}, Sedona H. and {Rix}, Hans-Walter and {Valentino}, Francesco and {Wang}, Bingjie and {Williams}, Christina C.},
	doi = {10.1038/s41550-024-02424-3},
	eprint = {2404.05683},
	journal = {Nature Astronomy},
	month = feb,
	pages = {280-292},
	primaryclass = {astro-ph.GA},
	title = {{Efficient formation of a massive quiescent galaxy at redshift 4.9}},
	volume = {9},
	year = 2025}

@article{Cattaneo2006,
	adsurl = {https://ui.adsabs.harvard.edu/abs/2006MNRAS.370.1651C},
	archiveprefix = {arXiv},
	author = {{Cattaneo}, A. and {Dekel}, A. and {Devriendt}, J. and {Guiderdoni}, B. and {Blaizot}, J.},
	doi = {10.1111/j.1365-2966.2006.10608.x},
	eprint = {astro-ph/0601295},
	journal = {\mnras},
	month = aug,
	number = {4},
	pages = {1651-1665},
	primaryclass = {astro-ph},
	title = {{Modelling the galaxy bimodality: shutdown above a critical halo mass}},
	volume = {370},
	year = 2006}

@article{Costa2014,
	adsurl = {https://ui.adsabs.harvard.edu/abs/2014MNRAS.439.2146C},
	archiveprefix = {arXiv},
	author = {{Costa}, Tiago and {Sijacki}, Debora and {Trenti}, Michele and {Haehnelt}, Martin G.},
	doi = {10.1093/mnras/stu101},
	eprint = {1307.5854},
	journal = {\mnras},
	month = apr,
	number = {2},
	pages = {2146-2174},
	primaryclass = {astro-ph.CO},
	title = {{The environment of bright QSOs at z {\ensuremath{\sim}} 6: star-forming galaxies and X-ray emission}},
	volume = {439},
	year = 2014}

@article{Volonteri2006,
	adsurl = {https://ui.adsabs.harvard.edu/abs/2006ApJ...650..669V},
	archiveprefix = {arXiv},
	author = {{Volonteri}, Marta and {Rees}, Martin J.},
	doi = {10.1086/507444},
	eprint = {astro-ph/0607093},
	journal = {\apj},
	month = oct,
	number = {2},
	pages = {669-678},
	primaryclass = {astro-ph},
	title = {{Quasars at z=6: The Survival of the Fittest}},
	volume = {650},
	year = 2006}

@article{Efstathiou1988,
	adsurl = {https://ui.adsabs.harvard.edu/abs/1988MNRAS.230P...5E},
	author = {{Efstathiou}, G. and {Rees}, M.~J.},
	doi = {10.1093/mnras/230.1.5P},
	journal = {\mnras},
	month = feb,
	pages = {5p-11p},
	title = {{High-redshift quasars in the Cold Dark Matter cosmogony}},
	volume = {230},
	year = 1988}

@article{Meyer2025a,
	adsurl = {https://ui.adsabs.harvard.edu/abs/2025ApJ...980...20M},
	archiveprefix = {arXiv},
	author = {{Meyer}, Romain A. and {Venemans}, Bram and {Neeleman}, Marcel and {Decarli}, Roberto and {Walter}, Fabian},
	doi = {10.3847/1538-4357/ada351},
	eid = {20},
	eprint = {2412.07474},
	journal = {\apj},
	month = feb,
	number = {1},
	pages = {20},
	primaryclass = {astro-ph.GA},
	title = {{Pushing the Atacama Large Millimeter/submillimeter Array to the Limit: 140 pc Resolution Observations of a z = 6.6 Quasar{\textendash}Galaxy Merger Resolve Strikingly Different Morphologies of Dust Continuum and [C II] 158 {\ensuremath{\mu}}m Emission}},
	volume = {980},
	year = 2025}

@article{Meyer2025b,
	adsurl = {https://ui.adsabs.harvard.edu/abs/2025A&A...695L..18M},
	archiveprefix = {arXiv},
	author = {{Meyer}, Romain A. and {Walter}, Fabian and {Di Mascia}, Fabio and {Decarli}, Roberto and {Neeleman}, Marcel and {Venemans}, Bram},
	doi = {10.1051/0004-6361/202453279},
	eid = {L18},
	eprint = {2502.14539},
	journal = {\aap},
	month = mar,
	pages = {L18},
	primaryclass = {astro-ph.GA},
	title = {{ALMA 360 parsec, high-frequency observations reveal warm dust in the center of a z = 6.9 quasar}},
	volume = {695},
	year = 2025}

@article{Walter2022,
	adsurl = {https://ui.adsabs.harvard.edu/abs/2022ApJ...927...21W},
	archiveprefix = {arXiv},
	author = {{Walter}, Fabian and {Neeleman}, Marcel and {Decarli}, Roberto and {Venemans}, Bram and {Meyer}, Romain and {Weiss}, Axel and {Ba{\~n}ados}, Eduardo and {Bosman}, Sarah E.~I. and {Carilli}, Chris and {Fan}, Xiaohui and {Riechers}, Dominik and {Rix}, Hans-Walter and {Thompson}, Todd A.},
	doi = {10.3847/1538-4357/ac49e8},
	eid = {21},
	eprint = {2201.06396},
	journal = {\apj},
	month = mar,
	number = {1},
	pages = {21},
	primaryclass = {astro-ph.GA},
	title = {{ALMA 200 pc Imaging of a z 7 Quasar Reveals a Compact, Disk-like Host Galaxy}},
	volume = {927},
	year = 2022}

@article{Neeleman2019,
	adsurl = {https://ui.adsabs.harvard.edu/abs/2019ApJ...882...10N},
	archiveprefix = {arXiv},
	author = {{Neeleman}, Marcel and {Ba{\~n}ados}, Eduardo and {Walter}, Fabian and {Decarli}, Roberto and {Venemans}, Bram P. and {Carilli}, Chris L. and {Fan}, Xiaohui and {Farina}, Emanuele P. and {Mazzucchelli}, Chiara and {Novak}, Mladen and {Riechers}, Dominik A. and {Rix}, Hans-Walter and {Wang}, Ran},
	doi = {10.3847/1538-4357/ab2ed3},
	eid = {10},
	eprint = {1907.02536},
	journal = {\apj},
	month = sep,
	number = {1},
	pages = {10},
	primaryclass = {astro-ph.GA},
	title = {{Resolved [C II] Emission from z > 6 Quasar Host-Companion Galaxy Pairs}},
	volume = {882},
	year = 2019}

@article{Decarli2017,
	adsurl = {https://ui.adsabs.harvard.edu/abs/2017Natur.545..457D},
	archiveprefix = {arXiv},
	author = {{Decarli}, R. and {Walter}, F. and {Venemans}, B.~P. and {Ba{\~n}ados}, E. and {Bertoldi}, F. and {Carilli}, C. and {Fan}, X. and {Farina}, E.~P. and {Mazzucchelli}, C. and {Riechers}, D. and {Rix}, H. -W. and {Strauss}, M.~A. and {Wang}, R. and {Yang}, Y.},
	doi = {10.1038/nature22358},
	eprint = {1705.08662},
	journal = {\nat},
	month = may,
	number = {7655},
	pages = {457-461},
	primaryclass = {astro-ph.GA},
	title = {{Rapidly star-forming galaxies adjacent to quasars at redshifts exceeding 6}},
	volume = {545},
	year = 2017}

@article{Neeleman2025,
	adsurl = {https://ui.adsabs.harvard.edu/abs/2025ApJ...983...26N},
	archiveprefix = {arXiv},
	author = {{Neeleman}, Marcel and {Kanekar}, Nissim and {Prochaska}, J. Xavier and {Rafelski}, Marc A. and {Kahinga}, Lordrick A.},
	doi = {10.3847/1538-4357/adbbd1},
	eid = {26},
	eprint = {2504.03828},
	journal = {\apj},
	month = apr,
	number = {1},
	pages = {26},
	primaryclass = {astro-ph.GA},
	title = {{A [C II] 158 {\ensuremath{\mu}}m Survey of Damped Ly{\ensuremath{\alpha}} Absorber Galaxies at z {\ensuremath{\sim}} 4: Observations of Dense and Metal-enriched Neutral Gas within the Circumgalactic Medium of Star-forming Galaxies}},
	volume = {983},
	year = 2025}

@article{Gonzalez-Lopez2017,
	adsurl = {https://ui.adsabs.harvard.edu/abs/2017A&A...608A.138G},
	archiveprefix = {arXiv},
	author = {{Gonz{\'a}lez-L{\'o}pez}, J. and {Bauer}, F.~E. and {Aravena}, M. and {Laporte}, N. and {Bradley}, L. and {Carrasco}, M. and {Carvajal}, R. and {Demarco}, R. and {Infante}, L. and {Kneissl}, R. and {Koekemoer}, A.~M. and {Mu{\~n}oz Arancibia}, A.~M. and {Troncoso}, P. and {Villard}, E. and {Zitrin}, A.},
	doi = {10.1051/0004-6361/201730961},
	eid = {A138},
	eprint = {1704.03007},
	journal = {\aap},
	month = dec,
	pages = {A138},
	primaryclass = {astro-ph.GA},
	title = {{The ALMA Frontier Fields Survey. III. 1.1 mm emission line identifications in Abell 2744, MACSJ 0416.1-2403, MACSJ 1149.5+2223, Abell 370, and Abell S1063}},
	volume = {608},
	year = 2017}

@article{Gonzalez-Lopez2019,
	adsurl = {https://ui.adsabs.harvard.edu/abs/2019ApJ...882..139G},
	archiveprefix = {arXiv},
	author = {{Gonz{\'a}lez-L{\'o}pez}, Jorge and {Decarli}, Roberto and {Pavesi}, Riccardo and {Walter}, Fabian and {Aravena}, Manuel and {Carilli}, Chris and {Boogaard}, Leindert and {Popping}, Gerg{\"o} and {Weiss}, Axel and {Assef}, Roberto J. and {Bauer}, Franz Erik and {Bertoldi}, Frank and {Bouwens}, Richard and {Contini}, Thierry and {Cortes}, Paulo C. and {Cox}, Pierre and {da Cunha}, Elisabete and {Daddi}, Emanuele and {D{\'\i}az-Santos}, Tanio and {Inami}, Hanae and {Hodge}, Jacqueline and {Ivison}, Rob and {Le F{\`e}vre}, Olivier and {Magnelli}, Benjamin and {Oesch}, Pascal and {Riechers}, Dominik and {Rix}, Hans-Walter and {Smail}, Ian and {Swinbank}, A.~M. and {Somerville}, Rachel S. and {Uzgil}, Bade and {van der Werf}, Paul},
	doi = {10.3847/1538-4357/ab3105},
	eid = {139},
	eprint = {1903.09161},
	journal = {\apj},
	month = sep,
	number = {2},
	pages = {139},
	primaryclass = {astro-ph.GA},
	title = {{The Atacama Large Millimeter/submillimeter Array Spectroscopic Survey in the Hubble Ultra Deep Field: CO Emission Lines and 3 mm Continuum Sources}},
	volume = {882},
	year = 2019}

@article{Harris2020,
	author = {Charles R. Harris and K. Jarrod Millman and St{\'{e}}fan J. van der Walt and Ralf Gommers and Pauli Virtanen and David Cournapeau and Eric Wieser and Julian Taylor and Sebastian Berg and Nathaniel J. Smith and Robert Kern and Matti Picus and Stephan Hoyer and Marten H. van Kerkwijk and Matthew Brett and Allan Haldane and Jaime Fern{\'{a}}ndez del R{\'{i}}o and Mark Wiebe and Pearu Peterson and Pierre G{\'{e}}rard-Marchant and Kevin Sheppard and Tyler Reddy and Warren Weckesser and Hameer Abbasi and Christoph Gohlke and Travis E. Oliphant},
	journal = {Nature},
	month = sep,
	number = {7825},
	pages = {357--362},
	publisher = {Springer Science and Business Media {LLC}},
	title = {Array programming with {NumPy}},
	volume = {585},
	year = {2020}}

@article{Hunter2007,
	adsurl = {https://ui.adsabs.harvard.edu/abs/2007CSE.....9...90H},
	author = {{Hunter}, John D.},
	journal = {Computing in Science and Engineering},
	month = may,
	number = {3},
	pages = {90-95},
	title = {{Matplotlib: A 2D Graphics Environment}},
	volume = {9},
	year = 2007}

@article{Astropy2018,
	adsurl = {https://ui.adsabs.harvard.edu/abs/2018AJ....156..123A},
	archiveprefix = {arXiv},
	author = {{Astropy Collaboration} and {Price-Whelan}, A.~M. and {Sip{\H{o}}cz}, B.~M. and {G{\"u}nther}, H.~M. and {Lim}, P.~L. and {Crawford}, S.~M. and {Conseil}, S. and {Shupe}, D.~L. and {Craig}, M.~W. and {Dencheva}, N. and {Ginsburg}, A. and {VanderPlas}, J.~T. and {Bradley}, L.~D. and {P{\'e}rez-Su{\'a}rez}, D. and {de Val-Borro}, M. and {Aldcroft}, T.~L. and {Cruz}, K.~L. and {Robitaille}, T.~P. and {Tollerud}, E.~J. and {Ardelean}, C. and {Babej}, T. and {Bach}, Y.~P. and {Bachetti}, M. and {Bakanov}, A.~V. and {Bamford}, S.~P. and {Barentsen}, G. and {Barmby}, P. and {Baumbach}, A. and {Berry}, K.~L. and {Biscani}, F. and {Boquien}, M. and {Bostroem}, K.~A. and {Bouma}, L.~G. and {Brammer}, G.~B. and {Bray}, E.~M. and {Breytenbach}, H. and {Buddelmeijer}, H. and {Burke}, D.~J. and {Calderone}, G. and {Cano Rodr{\'\i}guez}, J.~L. and {Cara}, M. and {Cardoso}, J.~V.~M. and {Cheedella}, S. and {Copin}, Y. and {Corrales}, L. and {Crichton}, D. and {D'Avella}, D. and {Deil}, C. and {Depagne}, {\'E}. and {Dietrich}, J.~P. and {Donath}, A. and {Droettboom}, M. and {Earl}, N. and {Erben}, T. and {Fabbro}, S. and {Ferreira}, L.~A. and {Finethy}, T. and {Fox}, R.~T. and {Garrison}, L.~H. and {Gibbons}, S.~L.~J. and {Goldstein}, D.~A. and {Gommers}, R. and {Greco}, J.~P. and {Greenfield}, P. and {Groener}, A.~M. and {Grollier}, F. and {Hagen}, A. and {Hirst}, P. and {Homeier}, D. and {Horton}, A.~J. and {Hosseinzadeh}, G. and {Hu}, L. and {Hunkeler}, J.~S. and {Ivezi{\'c}}, {\v{Z}}. and {Jain}, A. and {Jenness}, T. and {Kanarek}, G. and {Kendrew}, S. and {Kern}, N.~S. and {Kerzendorf}, W.~E. and {Khvalko}, A. and {King}, J. and {Kirkby}, D. and {Kulkarni}, A.~M. and {Kumar}, A. and {Lee}, A. and {Lenz}, D. and {Littlefair}, S.~P. and {Ma}, Z. and {Macleod}, D.~M. and {Mastropietro}, M. and {McCully}, C. and {Montagnac}, S. and {Morris}, B.~M. and {Mueller}, M. and {Mumford}, S.~J. and {Muna}, D. and {Murphy}, N.~A. and {Nelson}, S. and {Nguyen}, G.~H. and {Ninan}, J.~P. and {N{\"o}the}, M. and {Ogaz}, S. and {Oh}, S. and {Parejko}, J.~K. and {Parley}, N. and {Pascual}, S. and {Patil}, R. and {Patil}, A.~A. and {Plunkett}, A.~L. and {Prochaska}, J.~X. and {Rastogi}, T. and {Reddy Janga}, V. and {Sabater}, J. and {Sakurikar}, P. and {Seifert}, M. and {Sherbert}, L.~E. and {Sherwood-Taylor}, H. and {Shih}, A.~Y. and {Sick}, J. and {Silbiger}, M.~T. and {Singanamalla}, S. and {Singer}, L.~P. and {Sladen}, P.~H. and {Sooley}, K.~A. and {Sornarajah}, S. and {Streicher}, O. and {Teuben}, P. and {Thomas}, S.~W. and {Tremblay}, G.~R. and {Turner}, J.~E.~H. and {Terr{\'o}n}, V. and {van Kerkwijk}, M.~H. and {de la Vega}, A. and {Watkins}, L.~L. and {Weaver}, B.~A. and {Whitmore}, J.~B. and {Woillez}, J. and {Zabalza}, V. and {Astropy Contributors}},
	eid = {123},
	eprint = {1801.02634},
	journal = {\aj},
	month = sep,
	number = {3},
	pages = {123},
	primaryclass = {astro-ph.IM},
	title = {{The Astropy Project: Building an Open-science Project and Status of the v2.0 Core Package}},
	volume = {156},
	year = 2018}

@article{Astropy2013,
	adsurl = {https://ui.adsabs.harvard.edu/abs/2013A&A...558A..33A},
	archiveprefix = {arXiv},
	author = {{Astropy Collaboration} and {Robitaille}, Thomas P. and {Tollerud}, Erik J. and {Greenfield}, Perry and {Droettboom}, Michael and {Bray}, Erik and {Aldcroft}, Tom and {Davis}, Matt and {Ginsburg}, Adam and {Price-Whelan}, Adrian M. and {Kerzendorf}, Wolfgang E. and {Conley}, Alexander and {Crighton}, Neil and {Barbary}, Kyle and {Muna}, Demitri and {Ferguson}, Henry and {Grollier}, Fr{\'e}d{\'e}ric and {Parikh}, Madhura M. and {Nair}, Prasanth H. and {Unther}, Hans M. and {Deil}, Christoph and {Woillez}, Julien and {Conseil}, Simon and {Kramer}, Roban and {Turner}, James E.~H. and {Singer}, Leo and {Fox}, Ryan and {Weaver}, Benjamin A. and {Zabalza}, Victor and {Edwards}, Zachary I. and {Azalee Bostroem}, K. and {Burke}, D.~J. and {Casey}, Andrew R. and {Crawford}, Steven M. and {Dencheva}, Nadia and {Ely}, Justin and {Jenness}, Tim and {Labrie}, Kathleen and {Lim}, Pey Lian and {Pierfederici}, Francesco and {Pontzen}, Andrew and {Ptak}, Andy and {Refsdal}, Brian and {Servillat}, Mathieu and {Streicher}, Ole},
	eid = {A33},
	eprint = {1307.6212},
	journal = {\aap},
	month = oct,
	pages = {A33},
	primaryclass = {astro-ph.IM},
	title = {{Astropy: A community Python package for astronomy}},
	volume = {558},
	year = 2013}

@article{Fei2025,
	adsurl = {https://ui.adsabs.harvard.edu/abs/2025ApJ...980...84F},
	archiveprefix = {arXiv},
	author = {{Fei}, Qinyue and {Silverman}, John D. and {Fujimoto}, Seiji and {Wang}, Ran and {Ho}, Luis C. and {Bischetti}, Manuela and {Carniani}, Stefano and {Ginolfi}, Michele and {Jones}, Gareth and {Maiolino}, Roberto and {Rujopakarn}, Wiphu and {F{\"o}rster Schreiber}, N.~M. and {Espejo Salcedo}, Juan M. and {Lee}, L.~L.},
	eid = {84},
	eprint = {2501.09077},
	journal = {\apj},
	month = feb,
	number = {1},
	pages = {84},
	primaryclass = {astro-ph.GA},
	title = {{Assessing the Dark Matter Content of Two Quasar Host Galaxies at z {\ensuremath{\sim}} 6 through Gas Kinematics}},
	volume = {980},
	year = 2025}

@article{Neeleman2023,
	adsurl = {https://ui.adsabs.harvard.edu/abs/2023ApJ...958..132N},
	archiveprefix = {arXiv},
	author = {{Neeleman}, Marcel and {Walter}, Fabian and {Decarli}, Roberto and {Drake}, Alyssa B. and {Eilers}, Anna-Christina and {Meyer}, Romain A. and {Venemans}, Bram P.},
	eid = {132},
	eprint = {2311.00771},
	journal = {\apj},
	month = dec,
	number = {2},
	pages = {132},
	primaryclass = {astro-ph.GA},
	title = {{ALMA 400 pc Imaging of a z = 6.5 Massive Warped Disk Galaxy}},
	volume = {958},
	year = 2023}

@article{Guevara2020,
	adsurl = {https://ui.adsabs.harvard.edu/abs/2020A&A...636A..16G},
	archiveprefix = {arXiv},
	author = {{Guevara}, C. and {Stutzki}, J. and {Ossenkopf-Okada}, V. and {Simon}, R. and {P{\'e}rez-Beaupuits}, J.~P. and {Beuther}, H. and {Bihr}, S. and {Higgins}, R. and {Graf}, U. and {G{\"u}sten}, R.},
	eid = {A16},
	eprint = {2002.12692},
	journal = {\aap},
	month = apr,
	pages = {A16},
	primaryclass = {astro-ph.GA},
	title = {{[C II] 158 {\ensuremath{\mu}}m self-absorption and optical depth effects}},
	volume = {636},
	year = 2020}

@article{Goldsmith2012,
	adsurl = {https://ui.adsabs.harvard.edu/abs/2012ApJS..203...13G},
	archiveprefix = {arXiv},
	author = {{Goldsmith}, Paul F. and {Langer}, William D. and {Pineda}, Jorge L. and {Velusamy}, T.},
	eid = {13},
	eprint = {1209.4536},
	journal = {\apjs},
	month = nov,
	number = {1},
	pages = {13},
	primaryclass = {astro-ph.GA},
	title = {{Collisional Excitation of the [C II] Fine Structure Transition in Interstellar Clouds}},
	volume = {203},
	year = 2012}

@article{Novak2019,
	adsurl = {https://ui.adsabs.harvard.edu/abs/2019ApJ...881...63N},
	archiveprefix = {arXiv},
	author = {{Novak}, Mladen and {Ba{\~n}ados}, Eduardo and {Decarli}, Roberto and {Walter}, Fabian and {Venemans}, Bram and {Neeleman}, Marcel and {Farina}, Emanuele Paolo and {Mazzucchelli}, Chiara and {Carilli}, Chris and {Fan}, Xiaohui and {Rix}, Hans{\textendash}Walter and {Wang}, Feige},
	eid = {63},
	eprint = {1906.08569},
	journal = {\apj},
	month = aug,
	number = {1},
	pages = {63},
	primaryclass = {astro-ph.GA},
	title = {{An ALMA Multiline Survey of the Interstellar Medium of the Redshift 7.5 Quasar Host Galaxy J1342+0928}},
	volume = {881},
	year = 2019}

@article{Beelen2006,
	adsurl = {https://ui.adsabs.harvard.edu/abs/2006ApJ...642..694B},
	archiveprefix = {arXiv},
	author = {{Beelen}, Alexandre and {Cox}, Pierre and {Benford}, Dominic J. and {Dowell}, C. Darren and {Kov{\'a}cs}, Attila and {Bertoldi}, Frank and {Omont}, Alain and {Carilli}, Chris L.},
	eprint = {astro-ph/0603121},
	journal = {\apj},
	month = may,
	number = {2},
	pages = {694-701},
	primaryclass = {astro-ph},
	title = {{350 {\ensuremath{\mu}}m Dust Emission from High-Redshift Quasars}},
	volume = {642},
	year = 2006}

@article{Tripodi2023,
	adsurl = {https://ui.adsabs.harvard.edu/abs/2023ApJ...946L..45T},
	archiveprefix = {arXiv},
	author = {{Tripodi}, Roberta and {Feruglio}, Chiara and {Kemper}, Francisca and {Civano}, Francesca and {Costa}, Tiago and {Elvis}, Martin and {Bischetti}, Manuela and {Carniani}, Stefano and {Di Mascia}, Fabio and {D'Odorico}, Valentina and {Fiore}, Fabrizio and {Gallerani}, Simona and {Ginolfi}, Michele and {Maiolino}, Roberto and {Piconcelli}, Enrico and {Valiante}, Rosa and {Zappacosta}, Luca},
	eid = {L45},
	eprint = {2303.11961},
	journal = {\apjl},
	month = apr,
	number = {2},
	pages = {L45},
	primaryclass = {astro-ph.GA},
	title = {{Accurate Dust Temperature and Star Formation Rate in the Most Luminous z > 6 Quasar in the Hyperluminous Quasars at the Epoch of Reionization (HYPERION) Sample}},
	volume = {946},
	year = 2023}

@article{Rojas-Ruiz2021,
	adsurl = {https://ui.adsabs.harvard.edu/abs/2021ApJ...920..150R},
	archiveprefix = {arXiv},
	author = {{Rojas-Ruiz}, Sof{\'\i}a and {Ba{\~n}ados}, Eduardo and {Neeleman}, Marcel and {Connor}, Thomas and {Eilers}, Anna-Christina and {Venemans}, Bram P. and {Khusanova}, Yana and {Carilli}, Chris and {Mazzucchelli}, Chiara and {Decarli}, Roberto and {Momjian}, Emmanuel and {Novak}, Mladen},
	eid = {150},
	eprint = {2108.04257},
	journal = {\apj},
	month = oct,
	number = {2},
	pages = {150},
	primaryclass = {astro-ph.GA},
	title = {{The Impact of Powerful Jets on the Far-infrared Emission of an Extreme Radio Quasar at z {\ensuremath{\sim}} 6}},
	volume = {920},
	year = 2021}

@article{Khusanova2022,
	adsurl = {https://ui.adsabs.harvard.edu/abs/2022A&A...664A..39K},
	archiveprefix = {arXiv},
	author = {{Khusanova}, Y. and {Ba{\~n}ados}, E. and {Mazzucchelli}, C. and {Rojas-Ruiz}, S. and {Momjian}, E. and {Walter}, F. and {Decarli}, R. and {Venemans}, B. and {Farina}, E.~P. and {Meyer}, R. and {Wang}, F. and {Yang}, J.},
	eid = {A39},
	eprint = {2204.08973},
	journal = {\aap},
	month = aug,
	pages = {A39},
	primaryclass = {astro-ph.GA},
	title = {{The [CII] and FIR properties of z > 6 radio-loud quasars}},
	volume = {664},
	year = 2022}

@article{Foreman-Mackey2013,
	adsurl = {https://ui.adsabs.harvard.edu/abs/2013PASP..125..306F},
	archiveprefix = {arXiv},
	author = {{Foreman-Mackey}, Daniel and {Hogg}, David W. and {Lang}, Dustin and {Goodman}, Jonathan},
	eprint = {1202.3665},
	journal = {\pasp},
	month = mar,
	number = {925},
	pages = {306},
	primaryclass = {astro-ph.IM},
	title = {{emcee: The MCMC Hammer}},
	volume = {125},
	year = 2013}

@article{Neeleman2021,
	adsurl = {https://ui.adsabs.harvard.edu/abs/2021ApJ...911..141N},
	archiveprefix = {arXiv},
	author = {{Neeleman}, Marcel and {Novak}, Mladen and {Venemans}, Bram P. and {Walter}, Fabian and {Decarli}, Roberto and {Kaasinen}, Melanie and {Schindler}, Jan-Torge and {Ba{\~n}ados}, Eduardo and {Carilli}, Chris L. and {Drake}, Alyssa B. and {Fan}, Xiaohui and {Rix}, Hans-Walter},
	eid = {141},
	eprint = {2102.05679},
	journal = {\apj},
	month = apr,
	number = {2},
	pages = {141},
	primaryclass = {astro-ph.GA},
	title = {{The Kinematics of z {\ensuremath{\gtrsim}} 6 Quasar Host Galaxies}},
	volume = {911},
	year = 2021}

@article{Yue2021,
	adsurl = {https://ui.adsabs.harvard.edu/abs/2021ApJ...917...99Y},
	archiveprefix = {arXiv},
	author = {{Yue}, Minghao and {Yang}, Jinyi and {Fan}, Xiaohui and {Wang}, Feige and {Spilker}, Justin and {Georgiev}, Iskren Y. and {Keeton}, Charles R. and {Litke}, Katrina C. and {Marrone}, Daniel P. and {Walter}, Fabian and {Wang}, Ran and {Wu}, Xue-Bing and {Venemans}, Bram P. and {Zabludoff}, Ann},
	eid = {99},
	eprint = {2106.07429},
	journal = {\apj},
	month = aug,
	number = {2},
	pages = {99},
	primaryclass = {astro-ph.GA},
	title = {{ALMA Observations of the Sub-kpc Structure of the Host Galaxy of a z = 6.5 Lensed Quasar: A Rotationally Supported Hyper-Starburst System at the Epoch of Reionization}},
	volume = {917},
	year = 2021}

@article{Carilli2013,
	adsurl = {https://ui.adsabs.harvard.edu/abs/2013ARA&A..51..105C},
	archiveprefix = {arXiv},
	author = {{Carilli}, C.~L. and {Walter}, F.},
	eprint = {1301.0371},
	journal = {\araa},
	month = aug,
	number = {1},
	pages = {105-161},
	primaryclass = {astro-ph.CO},
	title = {{Cool Gas in High-Redshift Galaxies}},
	volume = {51},
	year = 2013}

@article{Lenz1992,
	adsurl = {https://ui.adsabs.harvard.edu/abs/1992PASP..104.1104L},
	author = {{Lenz}, Dawn D. and {Ayres}, Thomas R.},
	journal = {\pasp},
	month = nov,
	pages = {1104},
	title = {{Errors Associated with Fitting Gaussian Profiles to Noisy Emission-Line Spectra}},
	volume = {104},
	year = 1992}

@article{Tripodi2024a,
	adsurl = {https://ui.adsabs.harvard.edu/abs/2024A&A...682A..54T},
	archiveprefix = {arXiv},
	author = {{Tripodi}, R. and {Scholtz}, J. and {Maiolino}, R. and {Fujimoto}, S. and {Carniani}, S. and {Silverman}, J.~D. and {Feruglio}, C. and {Ginolfi}, M. and {Zappacosta}, L. and {Costa}, T. and {Jones}, G.~C. and {Piconcelli}, E. and {Bischetti}, M. and {Fiore}, F.},
	eid = {A54},
	eprint = {2306.01644},
	journal = {\aap},
	month = feb,
	pages = {A54},
	primaryclass = {astro-ph.GA},
	title = {{HYPERION. Interacting companion and outflow in the most luminous z > 6 quasar}},
	volume = {682},
	year = 2024}

@article{Novak2020,
	adsurl = {https://ui.adsabs.harvard.edu/abs/2020ApJ...904..131N},
	archiveprefix = {arXiv},
	author = {{Novak}, Mladen and {Venemans}, Bram P. and {Walter}, Fabian and {Neeleman}, Marcel and {Kaasinen}, Melanie and {Liang}, Lichen and {Feldmann}, Robert and {Ba{\~n}ados}, Eduardo and {Carilli}, Chris and {Decarli}, Roberto and {Drake}, Alyssa B. and {Fan}, Xiaohui and {Farina}, Emanuele P. and {Mazzucchelli}, Chiara and {Rix}, Hans-Walter and {Wang}, Ran},
	eid = {131},
	eprint = {2010.14875},
	journal = {\apj},
	month = dec,
	number = {2},
	pages = {131},
	primaryclass = {astro-ph.GA},
	title = {{No Evidence for [C II] Halos or High-velocity Outflows in z {\ensuremath{\gtrsim}} 6 Quasar Host Galaxies}},
	volume = {904},
	year = 2020}

@inproceedings{Petry2024,
	adsurl = {https://ui.adsabs.harvard.edu/abs/2024SPIE13098E..0PP},
	author = {{Petry}, Dirk and {D{\'\i}az Trigo}, Mar{\'\i}a. and {Kneissl}, R{\"u}diger and {Toledo}, Ignacio and {Miyazaki}, Atsushi and {Takagi}, Toshinobu and {Barnes}, Ashley and {Bonanomi}, Francesca},
	booktitle = {Observatory Operations: Strategies, Processes, and Systems X},
	editor = {{Benn}, Chris R. and {Chrysostomou}, Antonio and {Storrie-Lombardi}, Lisa J.},
	eid = {130980P},
	month = jul,
	pages = {130980P},
	series = {Society of Photo-Optical Instrumentation Engineers (SPIE) Conference Series},
	title = {{New methods for ALMA angular-scale-based observation scheduling, quality assessment, and beam shaping II: refinements}},
	volume = {13098},
	year = 2024}

@article{CASA2022,
	adsurl = {https://ui.adsabs.harvard.edu/abs/2022PASP..134k4501C},
	archiveprefix = {arXiv},
	author = {{CASA Team} and {Bean}, Ben and {Bhatnagar}, Sanjay and {Castro}, Sandra and {Donovan Meyer}, Jennifer and {Emonts}, Bjorn and {Garcia}, Enrique and {Garwood}, Robert and {Golap}, Kumar and {Gonzalez Villalba}, Justo and {Harris}, Pamela and {Hayashi}, Yohei and {Hoskins}, Josh and {Hsieh}, Mingyu and {Jagannathan}, Preshanth and {Kawasaki}, Wataru and {Keimpema}, Aard and {Kettenis}, Mark and {Lopez}, Jorge and {Marvil}, Joshua and {Masters}, Joseph and {McNichols}, Andrew and {Mehringer}, David and {Miel}, Renaud and {Moellenbrock}, George and {Montesino}, Federico and {Nakazato}, Takeshi and {Ott}, Juergen and {Petry}, Dirk and {Pokorny}, Martin and {Raba}, Ryan and {Rau}, Urvashi and {Schiebel}, Darrell and {Schweighart}, Neal and {Sekhar}, Srikrishna and {Shimada}, Kazuhiko and {Small}, Des and {Steeb}, Jan-Willem and {Sugimoto}, Kanako and {Suoranta}, Ville and {Tsutsumi}, Takahiro and {van Bemmel}, Ilse M. and {Verkouter}, Marjolein and {Wells}, Akeem and {Xiong}, Wei and {Szomoru}, Arpad and {Griffith}, Morgan and {Glendenning}, Brian and {Kern}, Jeff},
	eid = {114501},
	eprint = {2210.02276},
	journal = {\pasp},
	month = nov,
	number = {1041},
	pages = {114501},
	primaryclass = {astro-ph.IM},
	title = {{CASA, the Common Astronomy Software Applications for Radio Astronomy}},
	volume = {134},
	year = 2022}

@article{Hunter2023,
	adsurl = {https://ui.adsabs.harvard.edu/abs/2023PASP..135g4501H},
	archiveprefix = {arXiv},
	author = {{Hunter}, Todd R. and {Indebetouw}, Remy and {Brogan}, Crystal L. and {Berry}, Kristin and {Chang}, Chin-Shin and {Francke}, Harold and {Geers}, Vincent C. and {G{\'o}mez}, Laura and {Hibbard}, John E. and {Humphreys}, Elizabeth M. and {Kent}, Brian R. and {Kepley}, Amanda A. and {Kunneriath}, Devaky and {Lipnicky}, Andrew and {Loomis}, Ryan A. and {Mason}, Brian S. and {Masters}, Joseph S. and {Maud}, Luke T. and {Muders}, Dirk and {Sabater}, Jose and {Sugimoto}, Kanako and {Sz{\H{u}}cs}, L{\'a}szl{\'o} and {Vasiliev}, Eugene and {Videla}, Liza and {Villard}, Eric and {Williams}, Stewart J. and {Xue}, Rui and {Yoon}, Ilsang},
	eid = {074501},
	eprint = {2306.07420},
	journal = {\pasp},
	month = jul,
	number = {1049},
	pages = {074501},
	primaryclass = {astro-ph.IM},
	title = {{The ALMA Interferometric Pipeline Heuristics}},
	volume = {135},
	year = 2023}

@article{Decarli2018,
	adsurl = {https://ui.adsabs.harvard.edu/abs/2018ApJ...854...97D},
	archiveprefix = {arXiv},
	author = {{Decarli}, Roberto and {Walter}, Fabian and {Venemans}, Bram P. and {Ba{\~n}ados}, Eduardo and {Bertoldi}, Frank and {Carilli}, Chris and {Fan}, Xiaohui and {Farina}, Emanuele Paolo and {Mazzucchelli}, Chiara and {Riechers}, Dominik and {Rix}, Hans-Walter and {Strauss}, Michael A. and {Wang}, Ran and {Yang}, Yujin},
	eid = {97},
	eprint = {1801.02641},
	journal = {\apj},
	month = feb,
	number = {2},
	pages = {97},
	primaryclass = {astro-ph.GA},
	title = {{An ALMA [C II] Survey of 27 Quasars at z > 5.94}},
	volume = {854},
	year = 2018}

@article{Tripodi2022,
	adsurl = {https://ui.adsabs.harvard.edu/abs/2022A&A...665A.107T},
	archiveprefix = {arXiv},
	author = {{Tripodi}, R. and {Feruglio}, C. and {Fiore}, F. and {Bischetti}, M. and {D'Odorico}, V. and {Carniani}, S. and {Cristiani}, S. and {Gallerani}, S. and {Maiolino}, R. and {Marconi}, A. and {Pallottini}, A. and {Piconcelli}, E. and {Vallini}, L. and {Zana}, T.},
	eid = {A107},
	eprint = {2207.03314},
	journal = {\aap},
	month = sep,
	pages = {A107},
	primaryclass = {astro-ph.GA},
	title = {{Black hole and host galaxy growth in an isolated z {\ensuremath{\sim}} 6 QSO observed with ALMA}},
	volume = {665},
	year = 2022}

@article{Wang2024,
	adsurl = {https://ui.adsabs.harvard.edu/abs/2024ApJ...968....9W},
	archiveprefix = {arXiv},
	author = {{Wang}, Feige and {Yang}, Jinyi and {Fan}, Xiaohui and {Venemans}, Bram and {Decarli}, Roberto and {Ba{\~n}ados}, Eduardo and {Walter}, Fabian and {Barth}, Aaron J. and {Bian}, Fuyan and {Davies}, Frederick B. and {Eilers}, Anna-Christina and {Farina}, Emanuele Paolo and {Hennawi}, Joseph F. and {Li}, Jiang-Tao and {Mazzucchelli}, Chiara and {Wang}, Ran and {Wu}, Xue-Bing and {Yue}, Minghao},
	eid = {9},
	eprint = {2404.15413},
	journal = {\apj},
	month = jun,
	number = {1},
	pages = {9},
	primaryclass = {astro-ph.GA},
	title = {{A Spatially Resolved [C II] Survey of 31 z {\ensuremath{\sim}} 7 Massive Galaxies Hosting Luminous Quasars}},
	volume = {968},
	year = 2024}

@article{Venemans2020,
	adsurl = {https://ui.adsabs.harvard.edu/abs/2020ApJ...904..130V},
	archiveprefix = {arXiv},
	author = {{Venemans}, Bram P. and {Walter}, Fabian and {Neeleman}, Marcel and {Novak}, Mladen and {Otter}, Justin and {Decarli}, Roberto and {Ba{\~n}ados}, Eduardo and {Drake}, Alyssa and {Farina}, Emanuele Paolo and {Kaasinen}, Melanie and {Mazzucchelli}, Chiara and {Carilli}, Chris and {Fan}, Xiaohui and {Rix}, Hans-Walter and {Wang}, Ran},
	eid = {130},
	eprint = {2010.14874},
	journal = {\apj},
	month = dec,
	number = {2},
	pages = {130},
	primaryclass = {astro-ph.GA},
	title = {{Kiloparsec-scale ALMA Imaging of [C II] and Dust Continuum Emission of 27 Quasar Host Galaxies at z {\ensuremath{\sim}} 6}},
	volume = {904},
	year = 2020}

@article{DOdorico2023,
	adsurl = {https://ui.adsabs.harvard.edu/abs/2023MNRAS.523.1399D},
	archiveprefix = {arXiv},
	author = {{D'Odorico}, Valentina and {Ba{\~n}ados}, E. and {Becker}, G.~D. and {Bischetti}, M. and {Bosman}, S.~E.~I. and {Cupani}, G. and {Davies}, R. and {Farina}, E.~P. and {Ferrara}, A. and {Feruglio}, C. and {Mazzucchelli}, C. and {Ryan-Weber}, E. and {Schindler}, J. -T. and {Sodini}, A. and {Venemans}, B.~P. and {Walter}, F. and {Chen}, H. and {Lai}, S. and {Zhu}, Y. and {Bian}, F. and {Campo}, S. and {Carniani}, S. and {Cristiani}, S. and {Davies}, F. and {Decarli}, R. and {Drake}, A. and {Eilers}, A. -C. and {Fan}, X. and {Gaikwad}, P. and {Gallerani}, S. and {Greig}, B. and {Haehnelt}, M.~G. and {Hennawi}, J. and {Keating}, L. and {Kulkarni}, G. and {Mesinger}, A. and {Meyer}, R.~A. and {Neeleman}, M. and {Onoue}, M. and {Pallottini}, A. and {Qin}, Y. and {Rojas-Ruiz}, S. and {Satyavolu}, S. and {Sebastian}, A. and {Tripodi}, R. and {Wang}, F. and {Wolfson}, M. and {Yang}, J. and {Zanchettin}, M.~V.},
	eprint = {2305.05053},
	journal = {\mnras},
	month = jul,
	number = {1},
	pages = {1399-1420},
	primaryclass = {astro-ph.GA},
	title = {{XQR-30: The ultimate XSHOOTER quasar sample at the reionization epoch}},
	volume = {523},
	year = 2023}
\bibliographystyle{aasjournalv7}

\appendix

\section{Combining a heterogeneous set of ALMA observations}
\label{sec:comb_data}
Interferometric observations by their very design are sensitive to different spatial scales depending on the configurations that are used during the observations. By combining different configurations into a single data set, it is possible to recover emission across a wide range of spatial scales. This can be very important for correctly interpreting the data. For instance, high-resolution observations might be insensitive to extended emission, and by adding in low-resolution observations, it is possible to recover this emission. In practice, the goal is to fully sample the Fourier space ($uv$-plane) with the combined observations, which will minimize gaps in the data at specific spatial scales, and in turn will recover the best representation of the emission.

However, fully sampling the $uv$-plane is not sufficient, it is also important to optimally sample the $uv$-space. For instance, consider a data set where the short baselines (those near the center of the $uv$-plane) are far more numerous than the long baselines. This will result in two noticeable and important effects. First, the synthesized beam, which is directly calculated from the density and distribution of the points in the $uv$-plane, will have a distinct non-Gaussian shape, often described as a plateau \citep[e.g.,][]{Czekala2021}. This causes the clean beam, which is a Gaussian that is fit to the primary lobe of the dirty beam, to have a size that is different from the actual size of the primary lobe of the dirty beam. As a result, flux measurements from the observations will be incorrect, but can be corrected using the so-called Jorsater van Moorsel correction \citep[e.g.,][]{Jorsater1995,Novak2020}. Second, the image will have different sensitivities to different spatial scales. In practice, this will result in noise that is spatially correlated on larger scales than the clean beam; complicating identification of extended emission over this spatially correlated noise.

A conservative approach that alleviates both problems is to sample the $uv$-plane in such a way that the dirty beam is as Gaussian as possible. For dedicated, multi-configuration ALMA observations, theoretical time ratios of short versus long baseline configurations are used that try to achieve this sampling \citep{Mason2013}. However, when observations of different programs are combined, these time ratios can deviate significantly from the optimal values, often resulting in too many short baselines. Therefore, the different data needs to be scaled in order to achieve a correct sampling of the $uv$-plane. In this paper, we adapt the code discussed in \citep{Petry2024} to accomplish this task\footnote{https://github.com/mneeleman/ALMA-DR}. In words, the code creates a 1D histogram\footnote{Using a 1D histogram is an approximation since the $uv$-plane is 2D. This approximation is sufficient for our case, but could be generalized to 2D.} of the baseline distributions of the ALMA observations and compares this distribution to the theoretical baseline distribution of a tapered, scaled Gaussian that has a beam size of 0.25$''$. The code then flags any excess data above the theoretical baseline distribution. The resulting flagged data set is then processed as usual. 

The above approach provides an explanation why we did not process all archival data for the quasars. For some well-studied quasars (e.g., SDSS~J2310+1855), the distribution at short (or long) baselines is sufficiently good compared to intermediate baselines that the additional data sets would not have improved the coverage, and any additional baselines would have simply been flagged. As a result, we did not process these data sets.

\section{Sizes}
\label{sec:sizes}
 In this paper, we report on the extent of the \CII\ line and dust continuum emission for the quasar host galaxies in our sample. Previous papers have generally measured sizes by fitting 2D Gaussian profiles to the emission \citep{Venemans2020,Wang2024}. This approach assumes a Gaussian profile for the emission, which need not be the case for all quasar host galaxies. To explore this assumption, we fit a more generalized profile, i.e., a S\'{e}rsic profile, to the quasar host galaxies in our sample using the 2D fitting algorithm within \texttt{QubeFit}. For the data, we created a mask containing all contiguous 2$\sigma$ features surrounding the quasar. We then grew this mask to include sufficient area to properly sample the noise surrounding the emission. This masking approach will remove distinct secondary sources, such as the companion in PSO~J231$-$20, but will include emission from close mergers and extended emission for which it is difficult to distinguish between the galaxy and surrounding gas. The 2D S\'{e}rsic model that was used has seven parameters, the emission center ($x_0, y_0$), the amplitude of the emission ($A$), the ellipticity ($e$), the half-light radius ($R_{\rm 1/2}$), the position angle of the major axis ($\theta$) and the S\'{e}rsic index ($n$). The model was convolved with the ALMA synthesized beam and the parameter space was explored using a MCMC approach that is described in detail in the documentation for \texttt{QubeFit}. 

The results of fitting the \CII\ emission map are tabulated in Table \ref{tab:size}. Two of the galaxies detected in \CII\ emission were not fitted: SDSS~J1030+0524, because its emission was too faint to get a reliable size estimate, and WISEA~J0439+1634, because any size estimate would be unreliable due to the effect of gravitational lensing. For 13 of the galaxies, the \CII\ emission was not bright enough to constrain the S\'{e}rsic index. For these galaxies, we fitted the \CII\ emission using a standard 2D Gaussian model. For the remaining 24 galaxies, the S\'{e}rsic indices span a wide range of values from nearly Gaussian ($n_{\rm CII} \approx 0.5$) to profiles that are centrally peaked and show extended emission ($n_{\rm CII} > 2$). The latter is either driven by strongly peaked \CII\ emission in the center (e.g., PSO~J231$-$20) or extended emission in the outskirts (e.g., PSO~J025$-$11). Despite the varying S\'{e}rsic indices, the \CII\ half-light radii are consistent with the half-light radii obtained using 2D Gaussian fitting in the majority of galaxies. Only for those galaxies with very large S\'{e}rsic indices ($n_{\rm CII} \gtrsim 3.5$) are the half-light radii significantly larger, because these profiles capture the extended \CII\ emission in these galaxies. We find that there is only a slight degeneracy between the half-light radii and the S\'{e}rsic indices, especially for larger indices, but this is captured in the uncertainty on these values by the MCMC approach. We therefore believe that the resulting half-light radii accurately describe the extent of the emission. 

For the dust continuum emission, we employ the same strategy as for the \CII\ emission. In three of the galaxies, the dust continuum was too compact to get a reliable size estimate and 2$\sigma$ upper limits are reported in Table \ref{tab:size}. For the remaining galaxies, eight galaxies were fitted with a Gaussian profile and 27 with a S\'{e}rsic profile. Similar to the \CII\ emission, the dust emission also shows a wide range of S\'{e}rsic indices with half-light radii that are mostly smaller than the \CII\ half-light radii for the same galaxy. In general, the dust continuum fits are slightly worse than the fits to the \CII\ emission. This is driven by substructure within the dust emission that is not accurately described by the azimuthally invariant emission profile adopted in the fit. Finally, we see a rough correlation between the S\'{e}rsic indices of the \CII\ and dust continuum emission. This suggests that both profiles are driven by a similar distribution despite differences in the extent of the emission. 

\begin{deluxetable*}{lllllllllll}
\tablecaption{Results of quasar sizes fitting for the sample
\label{tab:size}}
\tabletypesize{\scriptsize}
\tablehead{
\colhead{QSO} &
\colhead{$r_{1/2,\rm CII}$} &
\colhead{$e_{\rm CII}$} &
\colhead{PA$_{\rm CII}$} &
\colhead{$n_{\rm CII}$} &
\colhead{reduced $\chi^2$} &
\colhead{$r_{1/2,\rm cont}$} &
\colhead{$e_{\rm cont}$} &
\colhead{PA$_{\rm cont}$} &
\colhead{$n_{\rm cont}$} &
\colhead{reduced $\chi^2$}\\
\colhead{} &
\colhead{($''$)} &
\colhead{} &
\colhead{\degr} &
\colhead{} &
\colhead{} &
\colhead{($''$)} &
\colhead{} &
\colhead{\degr} &
\colhead{} &
\colhead{}
}
\startdata
PSO J023$-$02 &   0.36 $\pm$   0.16 & $<$0.66 &   10 $\pm$ 60 & $<$4.3 & 1.54 &$<$0.15 & \nodata & \nodata & \nodata & 2.45\\
PSO J025$-$11 &   0.57 $\pm$   0.04 &   0.33 $\pm$   0.05 &   19 $\pm$  5 &   2.33 $\pm$   0.19 & 1.71 &  0.17 $\pm$   0.05 &   0.37 $\pm$   0.09 &  178 $\pm$  9 & $>$2.8 & 1.12\\
ULAS J0148+0600 &   0.31 $\pm$   0.09 & $<$0.33 &  155 $\pm$ 30 & $>$3.1 & 1.58 &  0.07 $\pm$   0.02 &   0.23 $\pm$   0.04 &  163 $\pm$  5 &   4.8 $\pm$   1.3 & 2.34\\
PSO J029$-$29 &   0.36 $\pm$   0.14 &   0.48 $\pm$   0.11 &   48 $\pm$  8 &   2.6 $\pm$   1.1 & 1.08 &  0.04 $\pm$   0.02 &   0.27 $\pm$   0.05 &   45 $\pm$  6 &   4.7 $\pm$   1.0 & 1.82\\
VST ATLAS J029$-$36 & \nodata & \nodata & \nodata & \nodata & \nodata &  0.05 $\pm$   0.02 & $<$0.66 &  170 $\pm$ 50 & \nodata & 2.03\\
PSO J060+24 &   0.43 $\pm$   0.12 & $<$0.6 &   20 $\pm$ 60 &   2.6 $\pm$   1.4 & 1.45 &  0.48 $\pm$   0.10 & $<$0.62 &  174 $\pm$ 21 &   2.8 $\pm$   1.1 & 2.73\\
VDES J0408$-$5632 &   0.13 $\pm$   0.14 &   0.55 $\pm$   0.34$^{\rm a}$ &   30 $\pm$ 70 & \nodata & 1.88 &  0.21 $\pm$   0.04 &   0.59 $\pm$   0.25$^{\rm a}$ &   63 $\pm$ 12 & \nodata & 1.03\\
PSO J065+01 &   0.14 $\pm$   0.02 & $<$0.42 &  164 $\pm$ 34 & \nodata & 1.02 &  0.06 $\pm$   0.02 &   0.38 $\pm$   0.03 &   27 $\pm$  3 &   3.0 $\pm$   0.6 & 1.41\\
PSO J089$-$15 &   0.19 $\pm$   0.05 &   0.51 $\pm$   0.31$^{\rm a}$ &   45 $\pm$ 27 & \nodata & 1.26 &  0.37 $\pm$   0.15 & $<$0.74 &  100 $\pm$ 40 &   2.6 $\pm$   1.7 & 1.37\\
PSO J108+08 &   0.17 $\pm$   0.03 & $<$0.53 &  115 $\pm$ 16 & \nodata & 1.35 &  0.08 $\pm$   0.02 & $<$0.43 &    2 $\pm$ 15 & $<$2.6 & 2.60\\
SDSS J0818+1722 &   0.15 $\pm$   0.02 & $<$0.5 &  160 $\pm$ 70 & \nodata & 1.32 &  0.11 $\pm$   0.04 & $<$0.55 &    1 $\pm$ 27 & $>$2.1 & 2.14\\
SDSS J0836+0054 & \nodata & \nodata & \nodata & \nodata & \nodata &$<$0.39 & \nodata & \nodata & \nodata & 3.15\\
DELS J0923+0402 &   0.51 $\pm$   0.09 & $<$0.66 &  115 $\pm$ 16 &   1.6 $\pm$   0.7 & 0.75 &  0.23 $\pm$   0.10 &   0.34 $\pm$   0.29$^{\rm a}$ &  100 $\pm$ 50 & \nodata & 0.64\\
PSO J158$-$14 &   0.52 $\pm$   0.07 &   0.28 $\pm$   0.05 &   13 $\pm$  7 &   4.0 $\pm$   0.4 & 1.31 &  0.09 $\pm$   0.02 &   0.23 $\pm$   0.06 &   22 $\pm$  8 &   4.7 $\pm$   1.4 & 1.93\\
PSO J215$-$16 &   0.22 $\pm$   0.02 &   0.26 $\pm$   0.03 &  169 $\pm$  4 &   0.78 $\pm$   0.09 & 1.26 &  0.10 $\pm$   0.01 &   0.27 $\pm$   0.01 &   19 $\pm$  2 &   1.49 $\pm$   0.09 & 11.3\\
PSO J217$-$16 &   0.38 $\pm$   0.14 & $<$0.75 &  170 $\pm$ 70 &   2.0 $\pm$   1.8 & 1.37 &  0.41 $\pm$   0.11 & $>$0.18 &   26 $\pm$ 12 & $<$5.6 & 2.07\\
PSO J217$-$07 &   0.25 $\pm$   0.02 &   0.17 $\pm$   0.04 &   86 $\pm$  8 &   0.67 $\pm$   0.10 & 0.96 &  0.15 $\pm$   0.02 &   0.14 $\pm$   0.03 &   81 $\pm$  7 &   0.78 $\pm$   0.11 & 1.20\\
CFHQS J1509$-$1749 &   0.12 $\pm$   0.03 & $<$0.61 &   60 $\pm$ 50 & \nodata & 1.38 &  0.06 $\pm$   0.02 & $<$0.33 &    0 $\pm$ 40 & \nodata & 2.17\\
DELS J1535+1943 &   0.15 $\pm$   0.03 & $<$0.76 &  105 $\pm$ 23 & \nodata & 1.36 &  0.04 $\pm$   0.02 & $<$0.62 &   53 $\pm$ 42 & $>$2.0 & 1.44\\
PSO J239$-$07 &   0.45 $\pm$   0.10 &   0.49 $\pm$   0.22 &   95 $\pm$ 13 & $<$1.7 & 1.27 &  0.10 $\pm$   0.05 &   0.53 $\pm$   0.34$^{\rm a}$ &   80 $\pm$ 50 & \nodata & 1.54\\
PSO J242$-$12 &   0.12 $\pm$   0.02 &   0.32 $\pm$   0.10 &  100 $\pm$ 10 & \nodata & 1.48 &  0.08 $\pm$   0.02 &   0.18 $\pm$   0.05 &  119 $\pm$  9 &   1.2 $\pm$   0.4 & 2.98\\
PSO J308$-$27 &   0.53 $\pm$   0.08 &   0.61 $\pm$   0.10 &   18 $\pm$  5 &   3.4 $\pm$   0.9 & 2.21 &  0.13 $\pm$   0.06 & $<$0.58 &   49 $\pm$ 28 & $>$2.1 & 1.29\\
VDES J2211$-$3206 &   0.15 $\pm$   0.11 &   0.54 $\pm$   0.34$^{\rm a}$ &  150 $\pm$ 90 & \nodata & 1.61 &$<$0.07 & \nodata & \nodata & \nodata & 2.44\\
VDES J2250$-$5015 &   0.14 $\pm$   0.02 & $<$0.46 &  100 $\pm$ 50 & \nodata & 1.33 &  0.04 $\pm$   0.02 &   0.44 $\pm$   0.11 &   82 $\pm$  9 &   2.2 $\pm$   1.1 & 3.97\\
\tableline
PSO J007+04 &   0.12 $\pm$   0.02 & $<$0.41 &  160 $\pm$ 60 & \nodata & 0.91 &  0.03 $\pm$   0.02 & $<$0.4 &  160 $\pm$ 40 &   3.5 $\pm$   1.5 & 1.37\\
PSO J009$-$10 &   0.52 $\pm$   0.06 &   0.30 $\pm$   0.10 &  139 $\pm$  9 &   1.09 $\pm$   0.20 & 1.78 &  0.51 $\pm$   0.07 &   0.59 $\pm$   0.06 &  137 $\pm$  4 &   1.5 $\pm$   0.4 & 1.88\\
SDSS J0100+2802 &   0.40 $\pm$   0.10 &   0.33 $\pm$   0.04 &   55 $\pm$  4 &   3.4 $\pm$   0.6 & 1.06 &  0.20 $\pm$   0.04 &   0.35 $\pm$   0.06 &   56 $\pm$  5 & $>$5.3 & 2.06\\
VST ATLAS J025$-$33 &   0.41 $\pm$   0.09 &   0.30 $\pm$   0.05 &   84 $\pm$  6 &   2.7 $\pm$   0.5 & 1.24 &  0.32 $\pm$   0.11 &   0.41 $\pm$   0.07 &  102 $\pm$  6 & $>$3.8 & 1.14\\
VDES J0224$-$4711 &   0.27 $\pm$   0.04 &   0.17 $\pm$   0.04 &  150 $\pm$  7 &   2.8 $\pm$   0.4 & 1.50 &  0.10 $\pm$   0.02 & $<$0.24 &  117 $\pm$ 18 & $>$3.9 & 1.76\\
PSO J036+03 &   0.22 $\pm$   0.02 &   0.33 $\pm$   0.03 &  143 $\pm$  4 &   0.58 $\pm$   0.08 & 1.98 &  0.12 $\pm$   0.02 &   0.19 $\pm$   0.04 &    5 $\pm$  8 &   1.86 $\pm$   0.33 & 3.65\\
PSO J065$-$26 &   0.46 $\pm$   0.11 & $<$0.51 &  170 $\pm$ 40 &   1.7 $\pm$   0.7 & 1.04 &  0.10 $\pm$   0.03 & $<$0.77 &  150 $\pm$ 80 & \nodata & 0.95\\
WISEA J0439+1634$^{\rm b}$ & \nodata & \nodata & \nodata & \nodata & \nodata &\nodata & \nodata & \nodata & \nodata & \nodata\\
SDSS J0842+1218 &   0.15 $\pm$   0.04 &   0.37 $\pm$   0.28$^{\rm a}$ &  144 $\pm$ 32 & \nodata & 1.25 &  0.09 $\pm$   0.03 &   0.65 $\pm$   0.33$^{\rm a}$ &  103 $\pm$ 19 & \nodata & 1.54\\
SDSS J1030+0524 & \nodata & \nodata & \nodata & \nodata & \nodata &\nodata & \nodata & \nodata & \nodata & \nodata\\
PSO J183+05 &   0.36 $\pm$   0.03 & $<$0.28 &    2 $\pm$ 12 &   0.89 $\pm$   0.17 & 1.14 &  0.24 $\pm$   0.02 &   0.23 $\pm$   0.04 &   14 $\pm$  6 &   1.10 $\pm$   0.18 & 1.22\\
SDSS J1306+0356 &   0.19 $\pm$   0.03 & $<$0.51 &   71 $\pm$ 30 & \nodata & 0.80 &  0.12 $\pm$   0.04 &   0.37 $\pm$   0.32$^{\rm a}$ &   90 $\pm$ 40 & \nodata & 1.27\\
ULAS J1319+0950 &   0.36 $\pm$   0.13 & $<$0.53 &   38 $\pm$ 23 &   1.5 $\pm$   0.8 & 1.12 &  0.20 $\pm$   0.02 &   0.17 $\pm$   0.06 &   12 $\pm$ 11 &   0.76 $\pm$   0.15 & 0.99\\
PSO J231$-$20 &   0.21 $\pm$   0.08 & $<$0.49 &   83 $\pm$ 13 & $>$1.9 & 0.74 &  0.04 $\pm$   0.02 &   0.20 $\pm$   0.06 &  109 $\pm$  9 &   2.6 $\pm$   0.9 & 4.58\\
PSO J323+12 &   0.43 $\pm$   0.12 &   0.66 $\pm$   0.17 &   63 $\pm$  7 & $<$1.8 & 0.94 &\nodata & \nodata & \nodata & \nodata & \nodata\\
SDSS J2310+1855 &   0.23 $\pm$   0.01 &   0.34 $\pm$   0.01 &  137 $\pm$  1 &   0.88 $\pm$   0.04 & 2.43 &  0.10 $\pm$   0.01 &   0.19 $\pm$   0.01 &  149 $\pm$  1 &   1.52 $\pm$   0.03 & 4.59\\
PSO J359$-$06 &   0.31 $\pm$   0.08 &   0.41 $\pm$   0.08 &  146 $\pm$  7 &   1.6 $\pm$   0.6 & 1.19 &  0.22 $\pm$   0.09 & $<$0.61 &  163 $\pm$ 21 &   1.4 $\pm$   1.3 & 0.74
\enddata
\tablecomments{Quasars without a S\'{e}rsic index are fitted using a two-dimensional Gaussian. PA$_{\rm CII}$ and PA$_{\rm cont}$ are the position angles of the major axes relative to due north. $n_{\rm CII}$ and $n_{\rm cont}$ are the S\'{e}rsic indices for the emission profiles. All uncertainties denote the 16$^{\rm th}$ and 84$^{\rm th}$ percentile ranges and limits denote the 2$\sigma$ ($\approx$2.3 or $\approx$97.7 percentile) ranges of the posterior distribution. $^{\rm a}$ Emission is unresolved along the minor axis. $^{\rm b}$Quasar is lensed and the resulting emission profile is not accurately described by a standard profile in the plane of the sky.}
\end{deluxetable*}

\end{document}